# Ultrathin Broadband Metasurface Superabsorbers from a van der Waals Semimetal

Adam D. Alfieri[1], Michael J. Motala[2], Michael Snure[3], Jason Lynch[1], Pawan Kumar[1,4], Huiqin Zhang[1], Susanna Post[5], Christopher Muratore[5], Joshua R. Hendrickson[3], Nicholas R. Glavin*[6], Deep Jariwala*[1].

[1]Department of Electrical and Systems Engineering, University of Pennsylvania, Philadelphia, PA 19104, USA
[2]UES Inc. Dayton, OH, 45432
[3]Air Force Research Laboratory, Sensors Directorate, Wright-Patterson Air Force Base, OH, 45433, USA
[4]Department of Materials Science and Engineering, University of Pennsylvania, Philadelphia, PA 19104, USA
[5]Department of Chemical and Materials Engineering, University of Dayton, Dayton, OH, 45469, USA
[6]Air Force Research Laboratory, Materials and Manufacturing Directorate, Wright-Patterson Air Force Base, OH, 45433, USA

*Corresponding Authors: dmj@seas.upenn.edu; nicholas.glavin.1@afrl.af.mil

**Abstract**
Metamaterials and metasurfaces operating in the visible and near-infrared (NIR) offer a promising route towards next-generation photodetectors and devices for solar energy harvesting. While numerous metamaterials and metasurfaces using metals and semiconductors have been demonstrated, semimetals-based metasurfaces in the vis-NIR range are notably missing. Here, we experimentally demonstrate a broadband metasurface superabsorber based on large area, semimetallic, van der Waals $PtSe_2$ thin films in agreement with electromagnetic simulations. Our results show that $PtSe_2$ is an ultrathin and scalable semimetal that concurrently possesses high index and high extinction across the vis-NIR range. Consequently, the thin-film PtSe2 on a reflector separated by a dielectric spacer can absorb > 85 % for the unpatterned case and ~97 % for the optimized 2D metasurface in the 400-900 nm range making it one of the strongest and thinnest broadband perfect absorbers to date. Our results present a scalable approach to photodetection and solar energy harvesting, demonstrating the practical utility of high index, high extinction semimetals for nanoscale optics.

## 1. Introduction

Optical metamaterials and their planar counterpart, metasurfaces, are artificial materials engineered using subwavelength features to exhibit a desired response to electromagnetic fields unachievable with natural materials[1,2]. Metamaterial perfect absorbers (MMPAs) have been designed for frequencies ranging from the visible to microwave[3]. Narrow bandwidth MMPAs are typically based on a frequency-dependent resonant response and have applications in spectral filtering[4] and sensing[5]. In the visible to near infrared (NIR), broadband MMPAs are important for photodetection and solar energy harvesting[6,7].

Broadband solar absorption has been achieved using anti-reflectance effects in superlattices of dielectrics and lossy metals[8], nanostructuring the surface of thick semiconductor wafers[9], light trapping in vertically aligned nanowire or nanotube arrays[10–13], and by light trapping in metamaterials with tapered geometries.[14,15] While effective, these approaches require relatively thick structures and/or complicated fabrication processes, limiting their practicality. Additionally, it is desirable for certain applications– such as photocatalysis– to localize EM waves at or near the surface[7]. Ultrathin metasurfaces exhibiting near-unity absorption and deep subwavelength field confinement are therefore ideal as electromagnetic energy is concentrated at the surface while the thin structure enables lightweight, low-cost devices.

A common approach to metasurface absorbers is using a patterned metal film on top of a lossless dielectric spacer layer on a metal reflector[5,6,16,17]. The dielectric layer recirculates light or can be chosen to produce a Fabry-Perot cavity resonance[4]. Periodic patterning of the top metal layer results in localized surface plasmon (LSP) modes excited in the top layer[18] and a gap surface plasmon (GSP) between the two metal films[19]. These plasmons – collective excitations of electrons at the interface between a metal and dielectric – enhance light absorption/emission and can localize electromagnetic fields to deep subwavelength mode volumes[18,20]. Further, hot electrons created by plasmons are particularly useful for photoelectrocatalysis[21], photodetection[22,23], and plasmonic solar cells[24].

The periodically patterned metal film exhibits a plasmonic response that is highly dependent on the structure geometry and the polarization and incident angle of the incident light. Moreover, plasmonic responses tend to be narrowband[6,25] and do not absorb off-resonance light or transverse electric (TE) modes. The polarization dependence can be significantly reduced by patterning elaborate nanoantenna structures[26,27], and broadband absorption can be achieved by resonance multiplexing[6,16,28,29] or by incorporating epsilon-near-zero materials in the dielectric spacer[30]. However, achieving polarization independent broadband absorption in purely plasmonic absorbers requires intricate geometries that make design and fabrication challenging. Using lossy plasmonic metals like chromium, titanium nitride, and tungsten has been shown to improve broadband absorption[31–38], but the ohmic loss results in energy being dissipated as heat, limiting potential applications.

Alternatively, two-dimensional metal dichalcogenides exhibit strong interaction with light despite ultrathin thicknesses[39–41]. Among them, noble metal dichalcogenides (NMDs) of form $MX_2$ (M = Pt, Pd; X = S, Se, Te) have only been recently explored[42–49]. $PtSe_2$ is an indirect-gap semiconductor in the monolayer limit but becomes a type-II Dirac semimetal beyond a few layers in thickness[50–52]. $PtSe_2$ has attracted the attention of researchers due to its broadband optical properties[53,54], saturable absorption[55], air stability[47], low temperature synthesis compatible with CMOS back end of line (BEOL) processing[56], and high mobilities in single crystalline flakes and films[57]. Additionally, $PtSe_2$ has demonstrated immense promise as a photo-electrocatalytic material[44,58–63]. Broadband absorption and strong catalytic properties make $PtSe_2$ an exciting material for broadband metasurface solar absorbers with direct applications in solar photocatalysis.

In this work, we demonstrate the scalable synthesis of $PtSe_2$ thin films with extraordinarily large refractive index and extinction, dominated by inter-band optical transitions, throughout the visible-NIR range. We then replace the commonly used top metal film in the metal-insulator-metal (MIM) metasurface structure with a semimetallic $PtSe_2$ film. We show that use of the dielectric spacer and back reflector are sufficient to achieve broadband absorption even in unpatterned films. By patterning the $PtSe_2$ to produce plasmons from the Ag and dielectric modes in the $PtSe_2$, broadband near unity absorption (97.0 %) is achieved from 400 nm to 900 nm. Remarkably, our $PtSe_2$-based approach stands out as simultaneously among the strongest and thinnest broadband thin-film or meta-absorbers known to date, exemplifying the broader value of taking a materials-first approach to nanophotonics: simple designs with emerging high-index materials can outperform more complex designs using conventional materials.

## 2. Results and Discussion
### 2.1. Synthesis and Characterization

PtSe$_2$ films were synthesized via thermally assisted conversion (TAC) of a sputtered Pt film[56,64] by annealing in H$_2$Se at 550°C (see Methods). A picture of a large area PtSe$_2$ grown by TAC on sapphire is shown in Figure 1a. Raman spectra (Figure 1b) collected at 5 spots on the sample, corresponding to locations at the center and at each edge, confirm the formation of PtSe$_2$ with high uniformity across the sample. The peaks observed at 178 cm$^{-1}$, 206 cm$^{-1}$, and ~237 cm$^{-1}$ correspond to the E$_g$, A$_{1g}$, and longitudinal optical (LO) modes of PtSe$_2$[65]. Cross-sectional transmission electron micrographs confirm the horizontally aligned, well-layered structure, and high angle annular dark field (HAADF) elemental mapping verifies the complete selenization of the film (supporting information section S1.1.). We synthesized PtSe$_2$ at temperatures ranging from 375°C to 650°C (supporting information section S1.1.) and found that annealing at 550°C produced films with the largest optical constants (supporting information section S1.2.1.), but the ability to selenize PtSe$_2$ at 450°C and lower makes this process CMOS BEOL compatible[64,66]. The large area films can then be transferred to arbitrary substrates using the well-established PMMA-assisted wet transfer process (Methods).

We determined the complex refractive index (Figure 1c) of 17 nm PtSe$_2$ transferred onto 40 nm Al$_2$O$_3$/100 nm Ag/300 nm SiO$_2$/Si by spectroscopic ellipsometry and fitting using Lorentzian oscillators (Methods and supporting information section S1.2.1.). In the complex refractive index, $\eta = n+ik$, $k$ is directly related to absorptivity/extinction while $n$ is related to propagation and power localization. The optical response of PtSe$_2$ in the UV/vis regime is dominated by a high density of interband transitions between Se-$p$ and Pt-$d$ bands near the Fermi level[67], resulting in a large, broadband $k$. The massive $n$ and $k$ in the vis-NIR result in strong broadband absorption. The semimetallic optical behavior of PtSe$_2$ is seen by examining the optical constants: at approximately 470 nm, where the $n$ and $k$ intersect, consistent with prior works[53,54]. Below this critical point, the real part of the relative permittivity (Re($\varepsilon$) = $n^2 - k^2$) is negative, so the optical properties of PtSe$_2$ are more metallic, but the Lorentzian oscillator terms and lack of a Drude term in the optical fitting indicate that the optical properties are still dominated by interband transitions and not free carrier behavior. Re($\varepsilon$) is positive from 470 nm to 900 nm, so PtSe$_2$ behaves like a semiconductor or lossy dielectric in this range.

The extraordinary broadband optical properties of PtSe$_2$ are evident upon transfer from the growth substrate (sapphire wafers) onto a silver mirror coated with 40 nm Al$_2$O$_3$ (Figure 1d, inset), resulting in >85% absorption in the unpatterned case (Figure 1d). Broadband >85% absorption in a 17 nm thick semimetallic film using only a back reflector and a lossless dielectric to recirculate light is remarkable, particularly because it presents a scalable lithography-free approach to broadband light-harvesting in ultra-thin geometries. For the remainder of this study, we explore the enhancement of the optical absorption of this 3-layer structure to near unity by exploiting the PtSe$_2$ absorption and the plasmonic properties of silver.

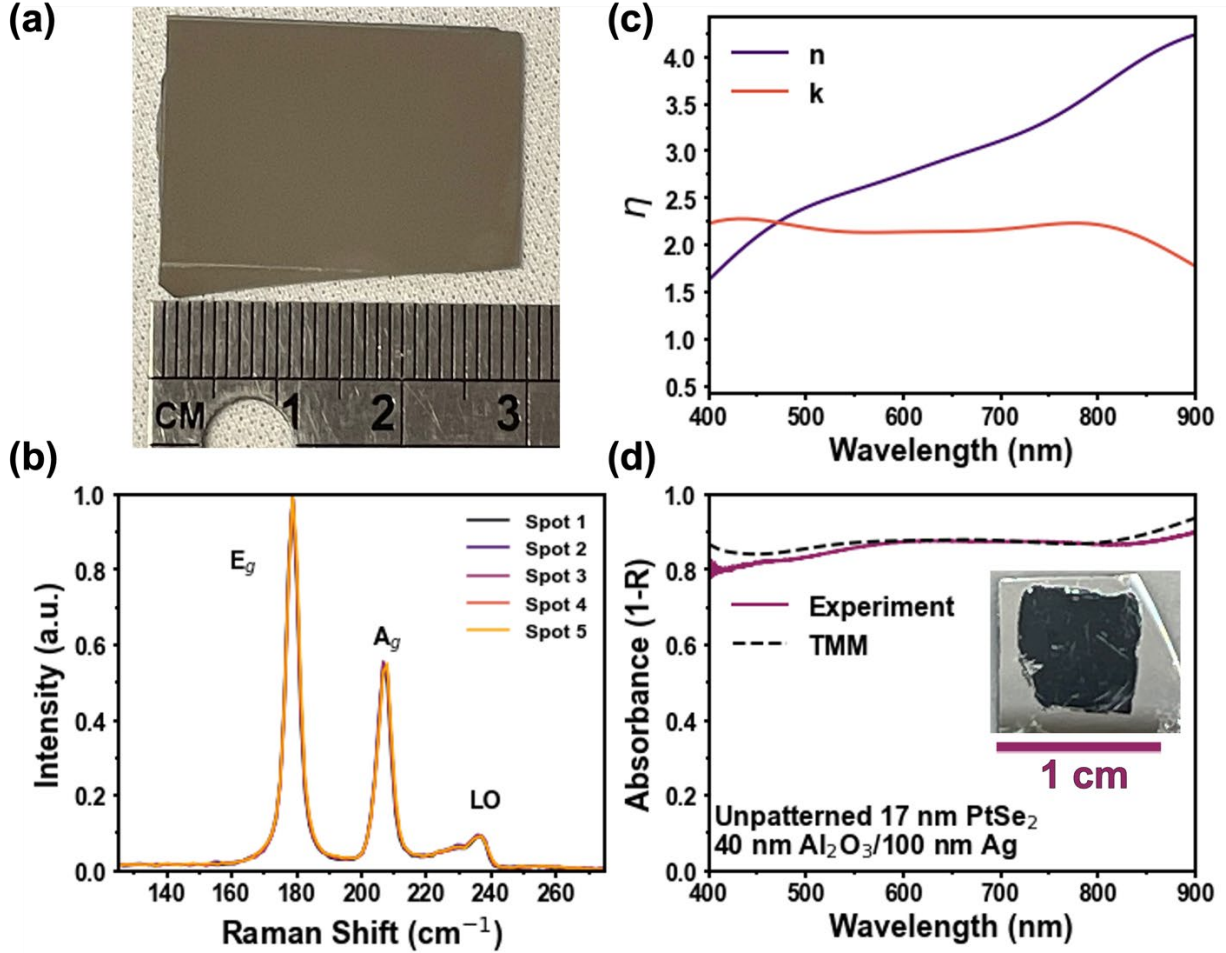

*Figure 1. PtSe$_2$ Film Structural and Optical Characterization.* (a.) A large area PtSe$_2$ film as grown on sapphire (b.) Raman characterization of a PtSe$_2$ film as grown on sapphire, taken at 5 different spots that are randomly and evenly distributed across the film. (c) Real and imaginary parts of the refractive index: $\eta = n + ik$. (d) Experimental and transfer matrix method (TMM) calculated absorbance spectra of an unpatterned PtSe$_2$ film transferred onto 40 nm Al$_2$O$_3$/100 nm Ag/SiO$_2$/Si. Inset: a picture of the sample. Optical constants for PtSe$_2$ were determined by spectroscopic ellipsometry on a transferred film.

## 2.2. Nanoribbon Array Metasurfaces

The first structure considered for broadband absorption is a 1D metasurface based on nanoribbons patterned in the 17 nm thick PtSe$_2$ film by a combination of electron beam lithography and reactive ion etching (further details found in Methods). In this structure, the top metal layer of the common MIM metasurface is replaced with PtSe$_2$. While PtSe$_2$ does not exhibit a LSP resonance, it is highly absorptive. Moreover, nanopatterning of the PtSe$_2$ still excites plasmons[68] and results in GSP-like modes. Figure 2a shows a schematic of the 3-layer structure: the top layer is composed of PtSe$_2$ nanoribbons with width $w$, arranged in an array with period $p$; the middle layer is a 40 nm thick Al$_2$O$_3$ film; and the bottom layer is a sputtered 100 nm thick Ag film. The substrate is Si with a 300 nm thick thermal oxide. There is no transmission through the Ag mirror, so the substrate choice is arbitrary as long as it is flat and smooth. Because there is no transmission, we can approximate the absorption of films and metasurfaces from the reflectance spectra as $A \approx 1 - R$. This is not exact, as there will be some scattering due to surface roughness.

We fabricated nanoribbon arrays with a fixed period $p$ = 300 nm and widths varying from 110 nm to 225 nm by patterning 15x15 micron pixels in PtSe$_2$ (Figure 2b and Figure 2c). The pixels look increasingly dark with increasing width, and the higher fill factor arrays look darker than the unpatterned PtSe$_2$, which itself absorbs over 85% of visible light. The unpolarized reflectance measurements in Figure 2d confirm the absorbance broadening with increasing nanoribbon fill factor. By patterning the PtSe$_2$ into a 1D metasurface, the absorption can be enhanced relative to the5npatternned film across the broadband spectral region between 400 nm and 900 nm.

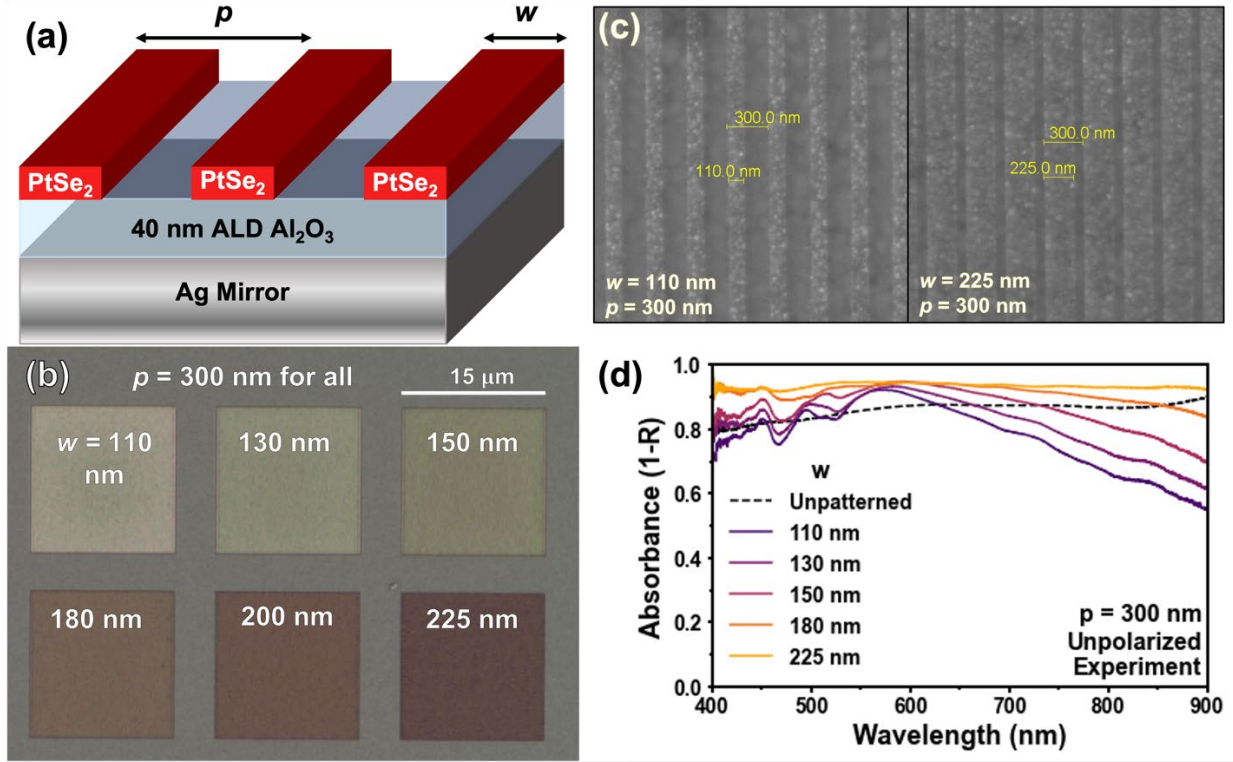

*Figure 2. PtSe$_2$ Nanoribbon Array Structure and Unpolarized Reflectance.* (a) Schematic of a PtSe$_2$ nanoribbon array. The nanoribbons sit on a 40 nm Al$_2$O$_3$ layer on a 100 nm Ag mirror. The nanoribbons have width $w$ and are arranged in a 1D array with period $p$. (b) Unpolarized optical image of a set of array pixels with widths ranging from $w$ = 110 nm to 225 nm. All arrays have period $p$ = 300 nm. The area surrounding the 15 μm x15 μm pixels is the unpatterned PtSe$_2$ film. (c) Scanning electron microscope images of arrays with $w$ = 110 nm (left) and $w$ = 225 nm (right). Both structures have period $p$ = 300 nm. Measurements using the SEM software overlaid on the images confirms the accuracy of the patterning. (d) Unpolarized absorbance spectra for the arrays shown in (b) and the unpatterned film area.

The absorption mechanisms can be better understood by considering separately the transverse electric (TE) and transverse magnetic ™ polarized modes of the nanoribbon array. At normal incidence, we define TE polarized light as linearly polarized plane waves with the electric field component parallel to the nanoribbon length while TM polarized light has its electric field component orthogonal to the nanoribbon length, respectively.

*2.2.1. Transverse Magnetic Modes of PtSe$_2$ Nanoribbons*

For normal incidence TM polarized light, the in-plane wavevector depends only on the array period:

$$k_x(\theta = 0) = 2\pi m / p \qquad m = \text{integer} \qquad (1)$$

The nonzero $k_x$ excites plasmons from the Ag film and results in an enhanced electric field in and around the nanoribbons. To characterize the plasmonic response, we first consider arrays with constant fill factor, $f = w/p \approx 1/3$. Figure 3a is a TM-polarized optical image of a die with various pixels corresponding to different nanoribbon periods ranging from 150 nm to 1500 nm, all with $w$ approximately $p/3$. Structural color changing with period is a clear signature of a visible range plasmonic response. We characterize the plasmon resonances by performing normal-incidence reflectance spectroscopy with TM light (Figure 3b) on the arrays shown in Figure 3a for periods 150 nm to 360 nm. The spectra with increasing period are vertically offset by 0.5 to better visualize the shift of the plasmon resonances, which are taken to be reflectance minima in the spectra.

There are 3 different resonances in the experimental spectra: a lower energy plasmon and a higher energy plasmon – denoted P1 and P2 for brevity – and a waveguide mode that appears in the experimental spectra due to imperfect collimation of the incident beam, resulting in a slight incident angle that causes the peak seen in the simulated spectra to split (supporting information S2.2.3.). Both the P1 and P2 modes are well predicted by simulation. The profiles of the P1 and P2 modes, the effects of the dielectric layer material on the TM modes, and the effects of dielectric layer thickness on the TM modes are further discussed in supporting information section S2.2.

Figure 3c shows the electric field magnitude and power absorbed ($P_{abs}$) profiles for the nanoribbon array structures (seen in Figure 2b) with $p = 300$ nm and $w = 225$ nm under normal incidence TM illumination with wavelength $\lambda = 550$ nm. The plasmonic response results in electric field enhancement in and around the PtSe$_2$. Using this simulated field profile, we can directly calculate the power absorbed in each material using the formula for a nonmagnetic material:

$$P_{abs} = \frac{1}{2}\omega \text{Im}(\varepsilon)|\boldsymbol{E}|^2 \qquad (2)$$

The $P_{abs}$ profiles for the TM modes of the nanoribbon structure are enhanced at the center of the PtSe$_2$ where the electric field is localized. Based on control simulations (supporting information section S2.2.2.) and the relatively small amount of power absorbed in the silver (5.4% on average), we can conclude that the absorption of TM polarized light is almost entirely due to the PtSe$_2$ interband absorption. Moreover, the absorption is at a maximum where the field is enhanced by plasmons, so it is clear that the field localization and enhancement by the plasmonic resonance assist absorption.

In Figure 2d, the absorption spectra increase and slightly redshift with increased $w$ for a fixed $p$ of 300 nm. This trend is observed for TM polarized light in experiment (Figure 3d, left) and simulation (right): the plasmon resonances redshift and broaden with increasing fill factor. The broadening of the resonance and enhancement of the absorption are both enabled by the large extinction of the PtSe$_2$ (supporting information section S2.2.4). The P1 and P2 modes are distinguishable for lower fill factor structures, and each broaden out into a broadband absorbance spectrum with no distinguishable modes as the fill factor increases. The waveguide mode again appears in the experimental spectrum at approximately 510 nm and redshifts slightly with increasing fill factor. The main phenomena of interest– resonance broadening and absorption enhancement– are well predicted by the simulations.

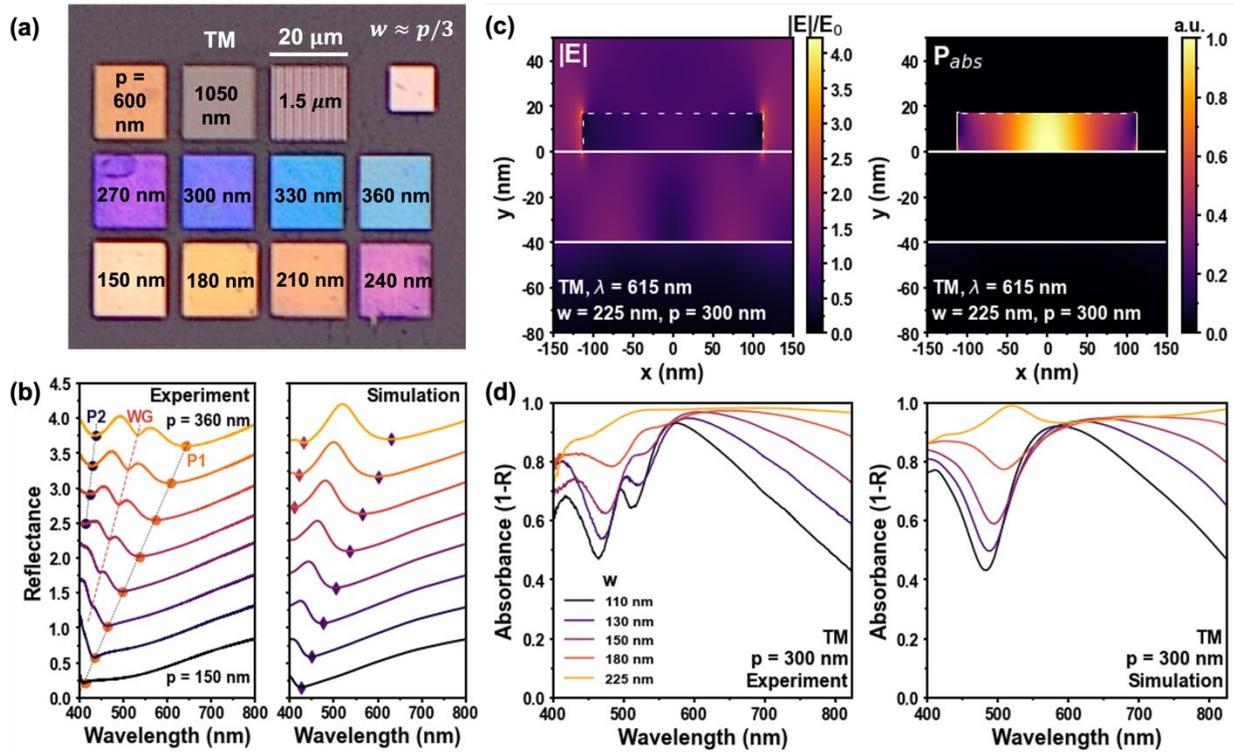

*Figure 3. Nanoribbon Array Transverse Magnetic Modes and Absorption.* (a) A TM-polarized optical image of nanoribbon array pixels with an approximate fill factor of 1/3 and periods ranging from 150 nm to 1500 nm. The area surrounding the pixels is the unpatterned PtSe$_2$ film. (b) Experimentally measured (left) and simulated (right) reflectance spectra for the arrays shown in (a) with *p* ranging from 150 nm to 360 nm in 30 nm increments. Spectra are vertically offset by 0.5 for clarity. The P1, P2, and waveguide (WG) modes are labeled on the experimental spectra, and dotted lines are added to guide the eye. The plasmon resonances on both the experimental and simulated spectra are indicated with dots (experiment) and diamonds (simulation). (c) Electric field magnitude (left) and power absorption (right) profiles for a nanoribbon array (*w* = 225 nm, *p* = 300 nm) illuminated by a TM wave (λ = 615 nm). (d) Experimentally measured (left) and simulated (right) TM-polarized absorbance spectra (taken to be *A = 1-R*) for the arrays shown in Figure 2b with *p* = 300 nm and *w* ranging from 110 nm to 225 nm. The linearly polarized data in (d) extends only up to 825 nm due to the limited bandwidth of the linear polarizer.

### 2.2.2. Broadband Absorption of TE Polarized Light in PtSe$_2$ Nanoribbons

In addition to strong plasmon-enhanced absorption, the PtSe$_2$ nanoribbons exhibit broadband absorption for TE light. Figure 4a shows the experimental (top) and simulated (bottom) TE-polarized absorbance spectra for the structures in Figure 2b with a fixed *p* = 300 nm and varying widths. Like the plot for the TM case, the polarized data is limited from 400 nm to 825 nm by the polarizer bandwidth. The experimental spectra match up well with the corresponding simulated spectra for most of the measured area. Even for TE polarizations, the absorbance is enhanced relative to the unpatterned case for part or all of the measured range, depending on the array dimensions. Here we see the benefit of replacing the top metal layer in a MIM broadband metasurface absorber with a high index, high-*k* material: there is strong absorption even for polarizations that do not excite the plasmon resonance.

Enhanced absorption despite removal of absorbing material indicates that patterning PtSe$_2$ results in the excitement of lossy modes due to the PtSe$_2$ itself, given the lack of a plasmon resonance for TE polarizations. Figure 4b shows the TE $|E|$ field profiles and the $P_{abs}$ profiles for a nanoribbon array with $w$ = 225 nm, $p$ = 300 nm at wavelengths of 405 nm and 825 nm, respectively. The $P_{abs}$ profiles are normalized to the maximum value of $P_{abs}$ across both wavelengths and are not a measure of the fraction of the incoming power absorbed. On average, the $P_{abs}$ in the Ag mirror is only 1.7%. Control simulations further confirm that the absorption is due to the large $k$ of PtSe$_2$ (supporting information S2.2.5.). For $\lambda$ = 405 nm, the electric field is enhanced between the ribbons; we expect that this is due to the Re($\varepsilon$) < 0 of PtSe$_2$ below 470 nm. The corresponding $P_{abs}$ profile shows that power absorption primarily occurs towards the edges of the ribbons. In the spectra shown in Figure 4a, the absorption maximum below 470 nm increases and blue shifts slightly with increasing fill factor as the field is being confined in a narrower area before decreasing for $w$ = 225 nm. In contrast, the electric field and power absorption are more concentrated in the center of the ribbons for $\lambda$ = 825 nm, indicating dielectric confinement due to the large $n$ of PtSe$_2$. At intermediate wavelengths where the material behaves like a lossy dielectric (between 470 nm and 825 nm), we expect that the absorption is also enhanced due to localization of light in the high-index nanoribbons. Overall, dependence of absorbance on fill factor is less drastic than for the TM case, and broadband absorption is seen for all ribbon widths.

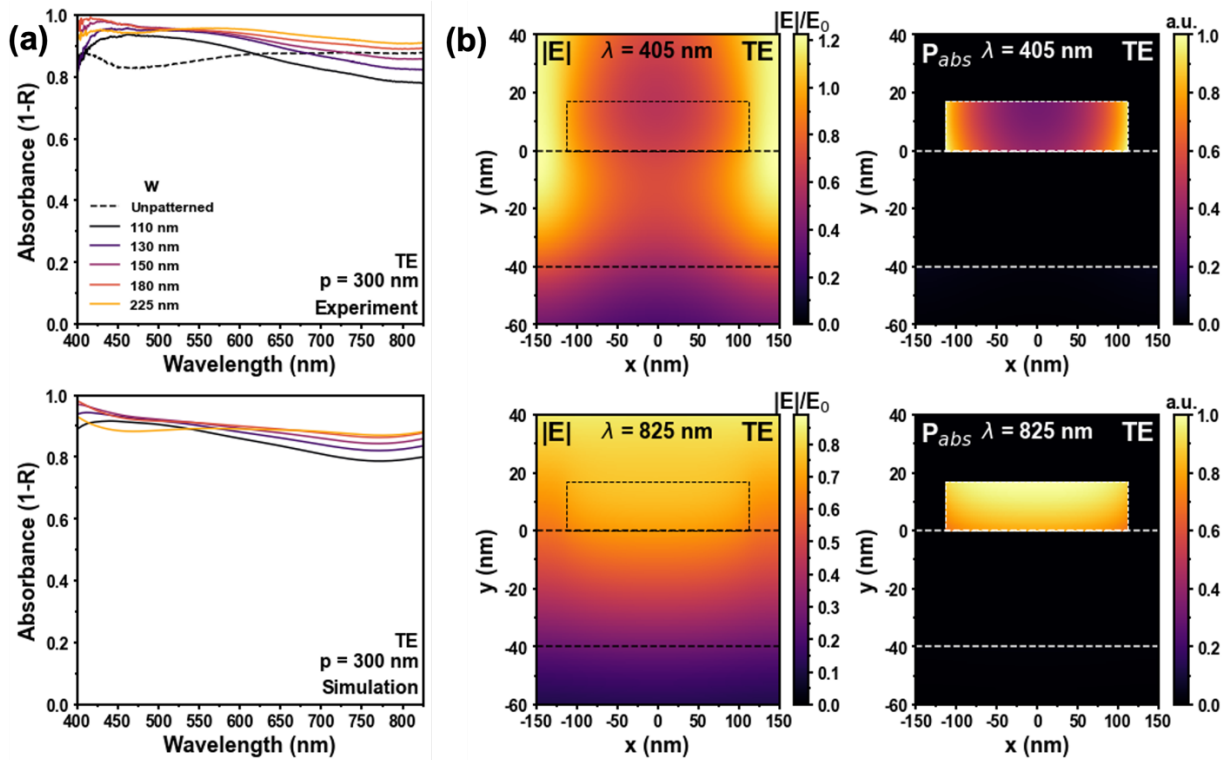

*Figure 4. Nanoribbon Array Transverse Electric Absorption.* (a) Experimentally measured (top) and simulated (bottom) transverse electric (TE)-polarized absorbance spectra for the arrays shown in Figure 2b with $p$ = 300 nm and $w$ ranging from 110 nm to 225 nm. The experimental data (b) $|E|$ field (left) and corresponding $P_{abs}$ (right) profiles for a nanoribbon array ($w$ = 225 nm, $p$ = 300 nm) illuminated with TE-polarized plane waves of wavelength $\lambda$ = 405 nm (top) and $\lambda$ = 825 nm (bottom). The $P_{abs}$ profiles are normalized to the maximum value of $P_{abs}$ to provide a relative comparison.

## 2.3. 2D Array PtSe$_2$ Metasurfaces

Now that we have observed the contributions to absorption of the Ag and PtSe$_2$ in a 1D nanoribbon array structures, we design relatively simple 2D arrays to maximize absorption of unpolarized visible light. We maintain the general 3-layer structure of PtSe$_2$ transferred on 40 nm Al$_2$O$_3$/100 nm Ag, but now fabricate 2D arrays. 2D periodicity results in nonzero in-plane wavevectors in both the $\hat{x}$ and $\hat{y}$ directions, exciting plasmon polaritons for all polarizations. For any given polarization, we expect the simultaneous presence of TM-like absorption modes – for which the Ag plasmons assist absorption by localizing the electric field in and around the PtSe$_2$ – and TE-like lossy dielectric modes due to the strong $n$ and $k$ of PtSe$_2$. Absorber design is therefore flexible and can be tailored to applications as any well-designed periodic, high PtSe$_2$ fill factor pattern will exhibit broadband near-unity absorption. We demonstrate this by fabricating structures with simple, inverted patterns: crossed PtSe$_2$ gratings/nanoribbons (XGs, Figure 5a - top) and a square array of PtSe$_2$ nanosquares (NSQs, Figure 5a - bottom). The XGs can alternatively be thought of as a square array of square nanoholes. The calculated field profiles and power absorption profiles are discussed in detail in Supporting Information section S3.2. The absorption of the XGs is more similar to the TE response of the nanoribbons, i.e., field localization and absorption are largely due to the large refractive index. In contrast, absorption in NSQ arrays is dominated by plasmonic field enhancement in the PtSe$_2$, similar to the TM modes of the nanoribbon arrays.

Dies were fabricated with 3 XGs and 3 NSQ arrays per die. SEM images of representative structures are shown in Figure 5b. The XGs have a fixed $p$ = 300 nm and different $w$'s of 100 nm, 150 nm, and 200 nm, respectively. The NSQ arrays have the following dimensions ($w$, $p$): (150 nm, 200 nm), (150 nm, 300 nm), (250 nm, 300 nm). Figure 5c is an unpolarized optical image of a die of metasurface pixels. We see that all pixels – except for NSQ (150 nm, 300 nm) – look dark or even completely black compared to the unpatterned PtSe$_2$ film, which itself absorbs over 85% of visible light. Unpolarized reflectance measurements (Figure 5d) show broadband near-unity absorption for all metasurfaces except the array of 150 nm NSQs with period 300 nm. We note that the XGs exhibit less dependence on fill factor than the NSQ arrays over the range of dimensions considered. This is well predicted by FDTD simulation of the structures with a broadband plane wave polarized along the $\hat{x}$ direction (supporting information section S3.3.).

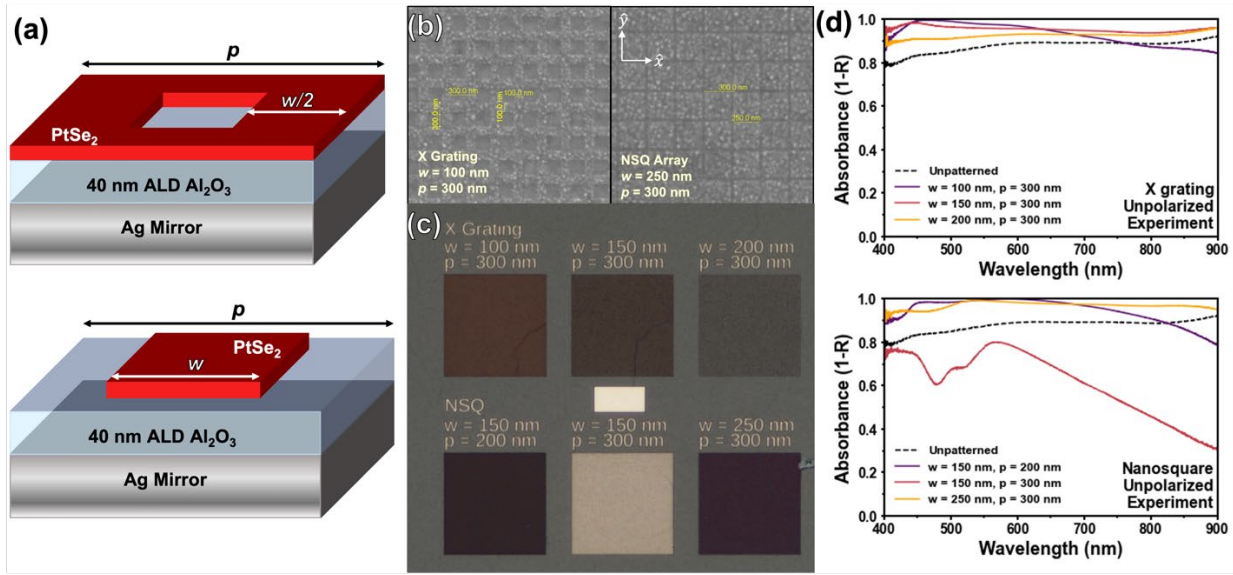

*Figure 5. 2D Array PtSe₂ Broadband Metasurface Superabsorber Structures and Experimental Results.* (a) Schematics of single unit cells of the crossed grating (XG) metasurface (top) and the nanosquare (NSQ) metasurface (bottom). (b) Scanning electron microscope (SEM) images of an XG metasurface (left) with dimensions w = 100 nm and p = 300 nm and NSQ array metasurface (right) with dimensions w = 250 nm and p = 300 nm. Measurements of the feature sizes using SEM software are displayed on the images, confirming accurate dimensions. (c) Unpolarized optical micrograph of a representative die, consisting of six 20 μm by 20 μm pixels patterned into PtSe₂ – the metasurfaces – and a small white rectangular patch where the PtSe₂ is removed to show the alumina-coated Ag substrate. The top row of pixels is a set of XGs while the bottom row is a set of NSQ arrays. The structure dimensions for each pixel are lithographically patterned into the PtSe₂ film. (d) Absorbance spectra of XGs (top) and NSQ arrays (bottom) as determined by normal incidence unpolarized reflectance spectroscopy. The unpatterned PtSe₂ film absorbance spectrum is included for reference. All reflectance data is normalized to a baseline spectrum from substrate and corrected for background noise (methods).

The integrated absorbance of all light in the visible range (400 nm to 700 nm), $A_{vis}$, and the integrated solar absorption efficiency, $A_{solar}$, over the measured range (400 nm to 900 nm) are calculated and tabulated (Table 1) as follows:

$$A_{vis} = \frac{1}{300\ nm} \int_{400\ nm}^{700\ nm} A(\lambda)d\lambda \qquad (3a)$$

$$A_{solar} = \frac{\int_{400\ nm}^{900\ nm} A(\lambda) P_{solar}(\lambda) d\lambda}{\int_{400\ nm}^{900\ nm} P_{solar}(\lambda) d\lambda} \qquad (3b)$$

$A(\lambda) = 1 - R(\lambda)$ is the measured absorbance spectrum for a given structure (Figure 5d), and $P_{solar}(\lambda)$ is the solar irradiance spectrum (ASTM G-0173-03 1.5AM Standard). The results are plotted in Figure 6 for unpatterned films and selected metasurfaces. Up to 97.9% absorption of visible light and 97.0% absorption of solar light (400 nm to 900 nm) is achieved, enhanced from 86.4% of visible light and 87.6% of solar light for the case of unpatterned PtSe₂ on the dielectric coated silver mirror. We reiterate that the absorption is occurring almost entirely within a 17 nm thick film with < 6% power absorbed on average by the Ag mirror from 400-900 nm. Near unity solar absorption concentrated in a large area, 17 nm strongly photo-catalytic[44,58–63] layer is therefore a significant step towards solar photocatalysis in this class of materials.

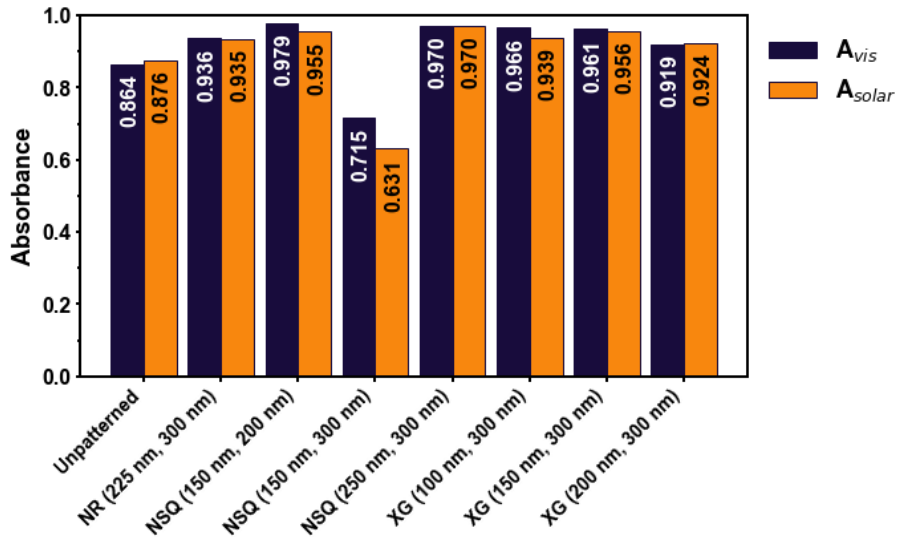

*Figure 6. Integrated Visible and Solar Light Absorption for Selected Structures.* Integrated visible range absorbance ($A_{vis}$) and integrated solar absorbance efficiency ($A_{sol}$) calculated by (3a) and (3b) for unpatterned structures, a selected nanoribbon array (NR) structure, the three different nanosquare array metasurfaces (NSQ), and the three different crossed grating metasurfaces (XG). The dimensions in parentheses are the width and period of the structures, respectively.

We compare the $A_{vis}$ of our best structures for visible range absorption (the NSQ array with $w$ = 150 nm, $p$ = 200 nm) to other broadband absorbers in literature (Figure 7), plotting $A_{vis}$ as a function of total structure thickness and the thickness of the active layer (i.e., the layer(s) in which most or all light is absorbed). Our PtSe$_2$-based metasurface stands out as simultaneously among the thinnest and strongest absorbing broadband absorbers to date. To our knowledge, the only experimental study that has demonstrated higher average visible absorption in a total thickness less than 700 nm was Wang *et al.*[36], which achieved 99.6% average visible absorption in a TiN/AlN nanocomposite active layer (126 nm thick) but required a 100 nm thick anti-reflective coating (ARC). We demonstrate comparable absorption in a thinner overall structure and in a 17 nm thick active layer that is on the surface – i.e., without the need for an ARC. Further, we achieve comparable or greater absorbance as compared to vertical nanowire arrays exceeding 1 micron in thickness[10,11,13].

Broadband near-unity visible light absorption concentrated in a 17 nm thick PtSe$_2$ film is noteworthy result due to its potential for both enhanced photodetection and efficient solar photocatalysis in ultrathin, lightweight devices. While some metals are either good absorbers (Ti, Cr, W) or good catalysts for the hydrogen evolution reaction (Pt, Ir, Rh), they are generally not both[69]. PtSe$_2$ maintains much of the catalytic activity of Pt but is a significantly more efficient absorber as a Dirac semi-metal. Further, the Pt film precursors are extremely thin, lowering the cost relative to pure Pt electrodes. Additionally, like semiconductors, PtSe$_2$ perfect absorbers can be applied to photodetection[70]. However, conventional semiconductors have significantly smaller and more narrowband imaginary refractive indices than PtSe$_2$ for most of the visible-NIR range[71] and are otherwise not suitable for efficient light harvesting or photodetection in ultrathin films with high surface-to-volume ratios due to processing difficulties, oxidation, and the effects of surface states on electrical transport and non-radiative recombination. Van der Waals PtSe$_2$ layers have self-passivated bonds and therefore enable much more practical devices and photocatalysts from ultrathin absorbers.

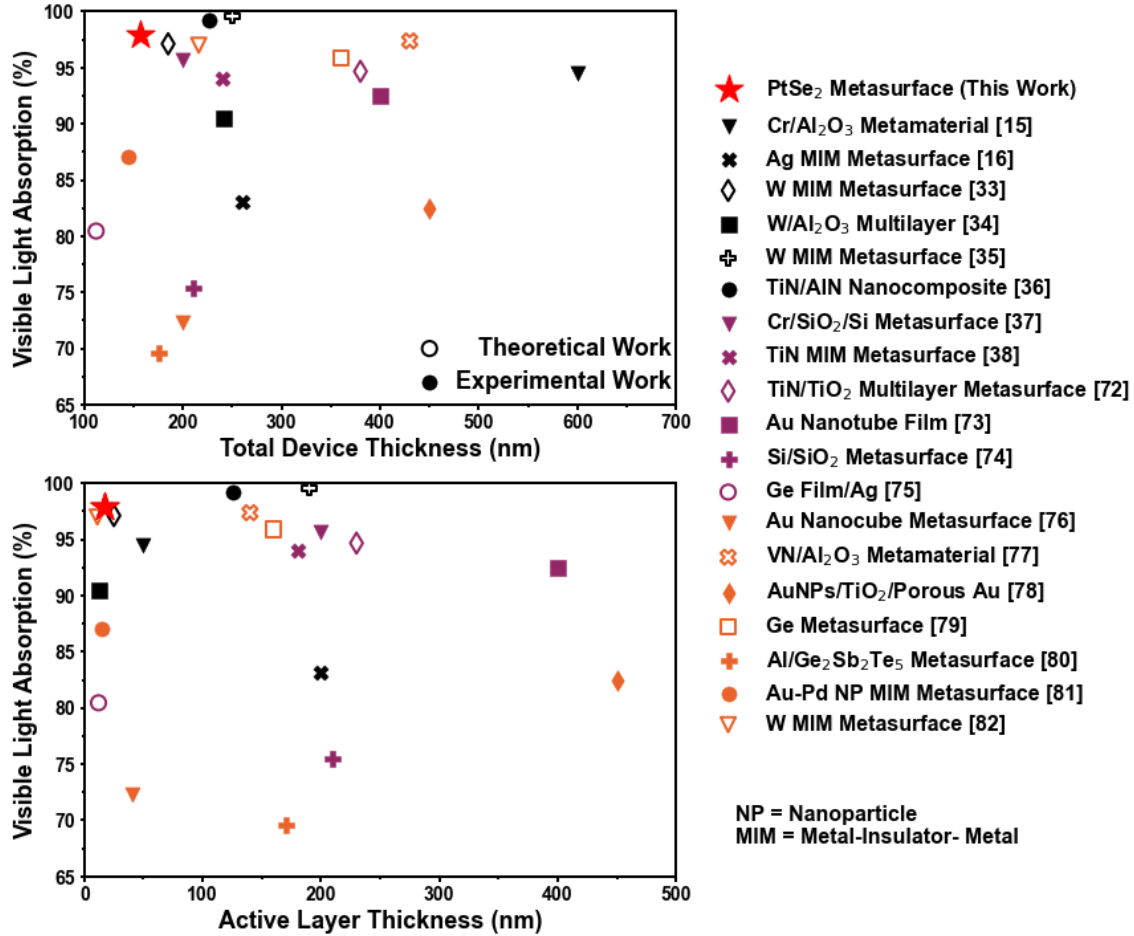

*Figure 7. Benchmarking Visible Absorbance of our PtSe$_2$ Metasurface Against Other Absorbers.* Plot of average visible range (400 nm to 700 nm) absorption as a function of total thickness (top) and active layer thickness (bottom) for the PtSe$_2$ NSQ array metasurface ($w$ = 150 nm, $p$ = 200 nm; denoted by a red star) and other broadband absorbers demonstrated in literature[15,16,33–37,72–82]. We define the active layer to be the layer(s) in which most of the power is absorbed. We estimate the absorbance as a function of wavelength for the best result in each previous work at various points from figures in each publication, then estimate the average light absorption using equation (3a). Experimental works are denoted by solid markers while theoretical works are denoted by unfilled markers.

**Summary and Outlook**

In summary, we have demonstrated the synthesis of 17 nm thick PtSe$_2$ films with large real and imaginary refractive indices. The large, broadband refractive index and extinction were leveraged to realize strong absorption (> 87 %) in the 400-900 nm range for an ultrathin, cm$^2$ scale reflector (Ag)/spacer (alumina)/absorber (PtSe$_2$) structure in a lithography-free approach. Upon using lithography to pattern the PtSe2 into 1D and 2D metasurfaces of nanoresonator arrays, broadband absorption was further enhanced to 97% in the most optimized 2D metasurface comprising nanosquare resonators. Finally, all our experimental observations were verified with electromagnetic simulations which suggest that the enhanced broadband absorption in PtSe2 metasurfaces is a consequence of both: 1. the large optical indices intrinsic to the PtSe2 due to its semi-metallic character as well as 2. nanophotonic effects such as plasmonic and dielectric resonances. This work pioneers the use of ultrathin PtSe$_2$-based metasurfaces for broadband visible and near infrared light absorption. Our approach presented here can be generalized to other existing and to-be-discovered van der Waals semimetallic and semiconducting materials exhibiting high broadband extinction due to interband transitions, presenting an exciting opportunity to design future metasurfaces for photodetection, photochemistry, and solar energy harvesting. We expect that our metasurfaces will be particularly promising for solar photocatalysis given the strong solar absorbance in a catalytically active material, plasmonic field enhancement and hot electron generation.

**Methods**

*$PtSe_2$ Synthesis*

100 mm single-side polished c-plane sapphire substrates (UniversityWafer) were diced with a diamond scribe and subsequently cleaned by sonicating in acetone and isopropyl alcohol before being blow-dried with $N_2$. Samples were then loaded in a custom-built sputtering system with a minimum base pressure of $3 \times 10^{-8}$ Torr. A 2" Pt sputter target was mounted on a 2" Mak sputter source (MeiVac, Inc.) and positioned at an angle of 30° and distance of 8 cm from the substrate. A ~ 2 nm thick Pt precursor layer was deposited at room temperature using a Pinnacle Plus pulsed power supply (Advanced Energy) operated at a nominal power of 90W, pulse frequency of 65 kHz, and a pulse width of 0.4 μs under argon with a pressure of 15 mTorr and Ar flow rate of 25 sccm. The total deposition time was 8 seconds.

Formation of $PtSe_2$ was performed via thermally assisted conversion. The samples were loaded in a tube furnace and held under vacuum for one hour, then pumped under $H_2$ gas (200 sccm) for another hour. The pressure was then increased to 500 Torr under a flow of 10 sccm $H_2$ and 160 sccm $N_2$. The temperature was ramped from room temperature to 550°C over 20 minutes then allowed to equilibrate for 5 minutes before $H_2Se$ gas flow (150 sccm) was turned on. The chamber was held under these conditions for 30 minutes before the tube furnace lid was opened to rapidly cool the chamber. The $H_2Se$ flow was turned off once the temperature decreased below 400°C.

*Vibrational Characterization*

We performed room temperature Raman spectroscopy using a Horiba LabRam HR Evolution Confocal Microscope with a 100x objective lens (Olympus) and a 633 nm laser filtered to have an intensity of <1 mW. Calibration of the spectra was done using a Si reference sample.

*Measurement of Optical Constants*

Spectroscopic ellipsometry (SE) on as-grown films was performed with a Woollam RC2 ellipsometer at wavelengths 210-2500 nm and incident angles of 55°, 65°, and 75°. SE of transferred films and bare ALD $Al_2O_3$/Ag substrates was later performed with a Woollam VASE ellipsometer from 240-900 nm at an incident angle of 65°. The complex permittivity was extracted from the raw Psi-Delta data by fitting the data in the CompleteEASE software (v6.55) with Lorentzian oscillators (See supporting information for details).

*Optical Simulation*

Absorbance of structures with unpatterned films was calculated using the 1D transfer matrix method. Frequency domain optical simulation of the 1D nanoribbon arrays was done using the Wave Optics module in COMSOL Multiphysics. The system was modeled with monochromatic, normal incidence TM or TE plane waves for incident wavelengths of 400-800 nm in 5 nm increments. The field profiles and power absorption profiles for the 1D nanoribbon arrays were calculated using the finite difference time domain (FDTD) solver in Lumerical. The 2D metasurfaces were similarly modeled using the Lumerical FDTD solver. Both systems were modeled using a normal incidence, broadband plane wave source composed of wavelengths 300 nm to 900 nm and measured at 100 frequency points. The optical constants used for $Al_2O_3$ were determined using spectroscopic ellipsometry and the constants for Ag are from Palik[83]. The fraction of light absorbed in the Ag mirror for each structure is determined by integrating the $P_{abs}$ profiles calculated by the Lumerical FDTD solver over the area (for 2D simulations) or volume (3D) of the Ag region. For the 1D nanoribbon arrays, the percent absorption is calculated for TM and TE polarizations. The 2DMSs are both square lattices, so the percent absorption is only calculated for light polarized in the $\hat{x}$ direction.

*Metasurface Fabrication*

A 1 cm x 1 cm p-Si substrate with a 290 nm thermal oxide was cleaned by sonication in acetone and isopropyl alcohol for 5 minutes each before being blow-dried with $N_2$. A 100 nm Ag film was deposited by DC sputtering (Kurt J. Lesker P75). The 40 nm thick $Al_2O_3$ layer was deposited by atomic layer deposition (Cambridge Nanotech) at 150°C at a rate of 0.9Å/cycle for 444 cycles. The precursors for Al and O were trimethyl aluminum (TMA) and $H_2O$, respectively.

The $PtSe_2$ films were transferred onto the dielectric/Ag substrates using a wet transfer method. The $PtSe_2$ films were spin-coated with PMMA A4 (2K rpm, 60s; left to dry overnight) before being removed from the sapphire growth substrate using a ~2 molar potassium hydroxide/water solution. The film was thoroughly cleaned with DI $H_2O$ to remove potassium ions before being transferred to the target substrate. The PMMA was removed by soaking the samples in Remover 1165 on a hot plate set to 80°C for 30 minutes, leaving in room temperature acetone overnight, and then placing the samples in fresh acetone and isopropyl alcohol for 5 minutes each before blow-drying with $N_2$.

The $PtSe_2$ films were patterned by electron beam lithography (Elionix ELS-75) and a subsequent dry etch (Oxford RIE 80). ZEP520A E-beam resist diluted with anisole (1:1 by weight) was spin coated at 2K rpm for 60s and baked at 120°C for 3 minutes. Proximity effect correction with a base dose of 225 $\mu C/cm^2$ was used for the nanoribbon arrays while the 2D metasurfaces used proximity effect correction with a base dose of 190 $\mu C/cm^2$. The samples were then developed using -10°C o-Xylene for 90s before rinsing in room temperature isopropyl alcohol and DI water for 30 seconds each. The film was then etched using reactive ion etching (Oxford RIE 80; 200W RF power, 25 mTorr pressure, 20 sccm $CF_4$ flow, room temperature) for 28 seconds. Finally, the resist was removed by leaving the samples in Remover 1165 on a hot plate set to 80°C for 2 hours, rinsing with acetone and isopropyl alcohol, and blow-drying with $N_2$.

*Reflectance Measurements*

Reflectance measurements were performed at normal incidence with a 50x objective lens (Olympus SLMPLN 50X N.A. = 0.35) in ambient conditions using a confocal microscope (Horiba LabRam HR Evolution) and an external white light source (AvaLight-HAL). The spot size is approximately 4 μm in diameter. For linearly polarized reflectance measurements, a linear polarizer (Edmund Linear Glass Polarizing Filter #43-783) was inserted between the and the sample. Measurements with the linear polarizer were limited to the spectral range of 400 nm to 825 nm due to polarizer bandwidth. Measurements were performed in the dark to minimize noise and the spectra were normalized to a baseline scan taken from the $Al_2O_3$/Ag coated substrate. A background scan was also collected during which the white light source was turned off. The background was subtracted from each scan:

$$R = \frac{R_{device} - R_{background}}{R_{substrate} - R_{background}}$$

*Scanning Electron Microscope (SEM) Characterization*

SEM images of metasurfaces were taken using a FEI Quanta 600 ESEM at an accelerating voltage of 10 kV and with a spot size of 3.0.

**Acknowledgements**

D.J., A.A. and J.L. acknowledge primary support for this work by the Asian Office of Aerospace Research and Development (AOARD) of the Air Force Office of Scientific Research (AFOSR) FA2386-20-1-4074 and FA2386-21-1-4063. D.J. also acknowledges partial support from the University Research Foundation at Penn and the Alfred P. Sloan Foundation for the Sloan Fellowship. D.J., P.K. and H. Z. acknowledge support from the National Science Foundation (NSF) (grant no. DMR-1905853) and support from University of Pennsylvania Materials Research Science and Engineering Center (MRSEC) (grant no. DMR-1720530) in addition to usage of MRSEC supported facilities. The sample fabrication, assembly and characterization were carried out at the Singh Center for Nanotechnology at the University of Pennsylvania, which is supported by the NSF National Nanotechnology Coordinated Infrastructure Program grant no. NNCI-1542153. F.B. is supported by the Vagelos Integrated Program in Energy Research. H.Z. was partially supported by Vagelos Institute of Energy Science and Technology graduate fellowship. J.R.H. acknowledges support from the Air Force Office of Scientific Research (Program Manager Dr. Gernot Pomrenke) under award number FA9550-20RYCOR059. M. S. and N. R. G. acknowledge support from the Air Force Office of Scientific Research under Award No. FA9550-19RYCOR050. Dr. Evan Smith helped with preliminary ellipsometry measurements.

# Supporting Information: Ultrathin Broadband Metasurface Superabsorber Based on a van der Waals Semimetal

Adam D. Alfieri[1], Michael J. Motala[2], Michael Snure[3], Jason Lynch[1], Pawan Kumar[1,4], Huiqin Zhang[1], Susanna Post[5], Christopher Muratore[5], Joshua R. Hendrickson[3], Nicholas R. Glavin*[6], Deep Jariwala*[1].

[1]Department of Electrical and Systems Engineering, University of Pennsylvania, Philadelphia, PA 19104, USA
[2]UES Inc. Dayton, OH, 45432
[3]Air Force Research Laboratory, Sensors Directorate, Wright-Patterson Air Force Base, OH, 45433, USA
[4]Department of Materials Science and Engineering, University of Pennsylvania, Philadelphia, PA 19104, USA
[5]Department of Chemical and Materials Engineering, University of Dayton, Dayton, OH, 45469, USA
[6]Air Force Research Laboratory, Materials and Manufacturing Directorate, Wright-Patterson Air Force Base, OH, 45433, USA

*Corresponding Authors: dmj@seas.upenn.edu; nicholas.glavin.1@afrl.af.mil

## S1. Growth and Characterization

### *S1.1. Growth and Structural Characterization*

In the initial optimization of the growth conditions, we annealed Pt in $H_2Se$ at temperatures ranging from 375°C to 650°C. Temperature dependent Raman and XRD data are shown in Figure S1, showing the synthesis of $PtSe_2$ at temperatures ranging from 375°C to 650°C. The presence of the $E_g$ and $A_{1g}$ Raman peaks at 178 cm$^{-1}$ and 206 cm$^{-1}$ for all growth conditions shows the formation of crystalline $PtSe_2$ at all 3 temperatures. The presence of the $PtSe_2$ peaks and lack of any Pt peaks in the XRD confirm the complete selenization of the Pt film.

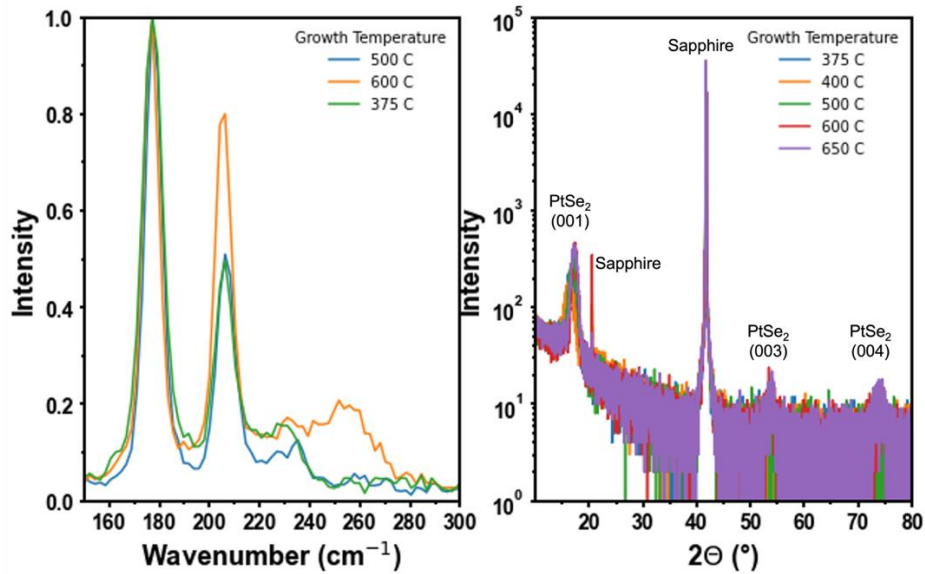

*Figure S1.* Raman spectra (a) and x-ray diffraction scans (b) for different growth temperatures.

Cross-sectional transmission electron microscopy (X-TEM) was performed on a $PtSe_2$ film transferred to a $SiO_2$/Si substrate (Figure S2). The X-TEM sample was fabricated with a dual beam plasma-focus ion beam (FIB) setup. A soft carbon layer (using Sharpie marker) was coated on top of the $PtSe_2$ film to avoid the contamination as well as beam damages while preparing the specimen. 30 keV to 5 keV ion beam were used for the specimen preparation for milling, thinning, and polishing along with an in-situ liftoff processes. The as prepared X-TEM specimen attached to the Omniprobe half grid was analyzed under high angle annular dark field (HAADF) scanning transmission electron microscopy (STEM). Aberration corrected (probe correction) STEM uses a probe size of ~1Å and accelerating voltage of 200kV. A high speed large dual detector was used to get ultra-high resolution EDS mapped images. The van der Waals layered structure of PtSe$_2$ can be easily seen, and the vdW layers are horizontally aligned. The domain size of this nanocrystalline PtSe2 ranges from 50 to 100 nm. Moreover, the HAADF STEM image and elemental mapping confirm complete selenization of the platinum to form PtSe$_2$ layers.

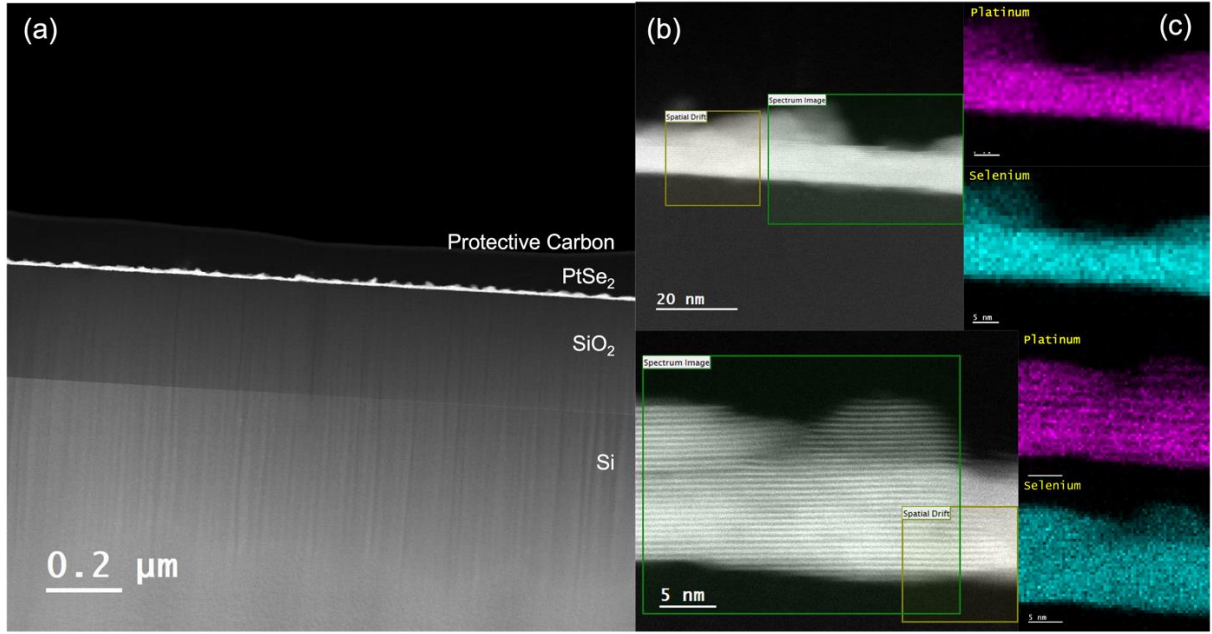

*Figure S2. Cross-Sectional Scanning Transmission Electron Microscope (X-STEM) Analysis.* (a) Low magnification cross sectional high angle annular dark field (HAADF) scanning transmission electron micrograph image of a PtSe$_2$ film transferred onto SiO$_2$/Si and coated with a protective carbon layer. (b) High magnification HAADF X-STEM images of the sample at two different spots and (c) the corresponding elemental mapping. The green boxes in (b) are the range over which the elemental mapping is performed.

### S1.2. Ellipsometry and Optical Constants

#### S1.2.1. Measurement and Fitting
Spectroscopic ellipsometry (SE) measures the ratio of complex reflection coefficients for *s* and *p* polarized light:

$$\frac{r_p}{r_s} = \tan(\psi)e^{-i\Delta} \qquad (1)$$

The measured $\psi$ and $\Delta$ values (Figure S3) are related to the dielectric function of the medium. The dielectric function and complex refractive index are determined by fitting the raw $\psi$ and $\Delta$. We fit the $\psi$ and $\Delta$ values with a background term, $\varepsilon_\infty$, a UV pole, an IR pole, and a series of four Lorentz oscillators:

$$\varepsilon(E) = \varepsilon_\infty + \frac{A_{UV}}{E_{UV}^2 - E^2} - \frac{A_{IR}}{E^2} + \sum_j \frac{A_j \Gamma_j E_j}{E_j^2 - E^2 - i\Gamma_j E_j} \qquad (2)$$

The model parameters are given in Table 1, and the fits are plotted against the raw data in Figure S3 to confirm the validity of the model.

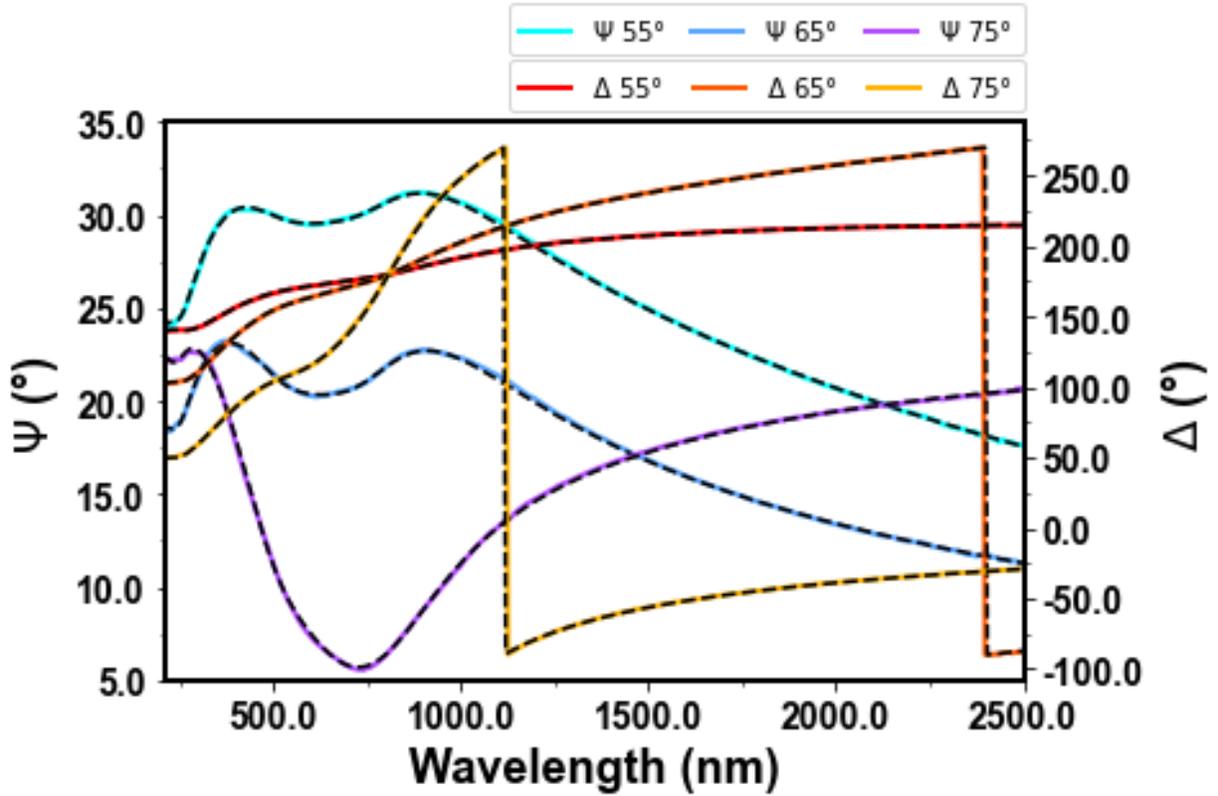

*Figure S3. Experimental Psi-Delta values for the as-grown films on sapphire measured by SE.* Measured data taken at incident angles of 55-75° in 10° increments is shown, and the corresponding fits from the model are overlaid (black dashed lines).

Figure S4 shows the optical constants of as-grown films annealed at different temperatures. We see that the optical properties of the films are highly dependent on the growth temperature. Growth at 500-550°C appears to be necessary to achieve a plasmonic response, as this zero-crossing of $\varepsilon'$ becomes negligible for films grown at higher and lower temperatures. Moreover, the interband absorption is maximized in this temperature range.

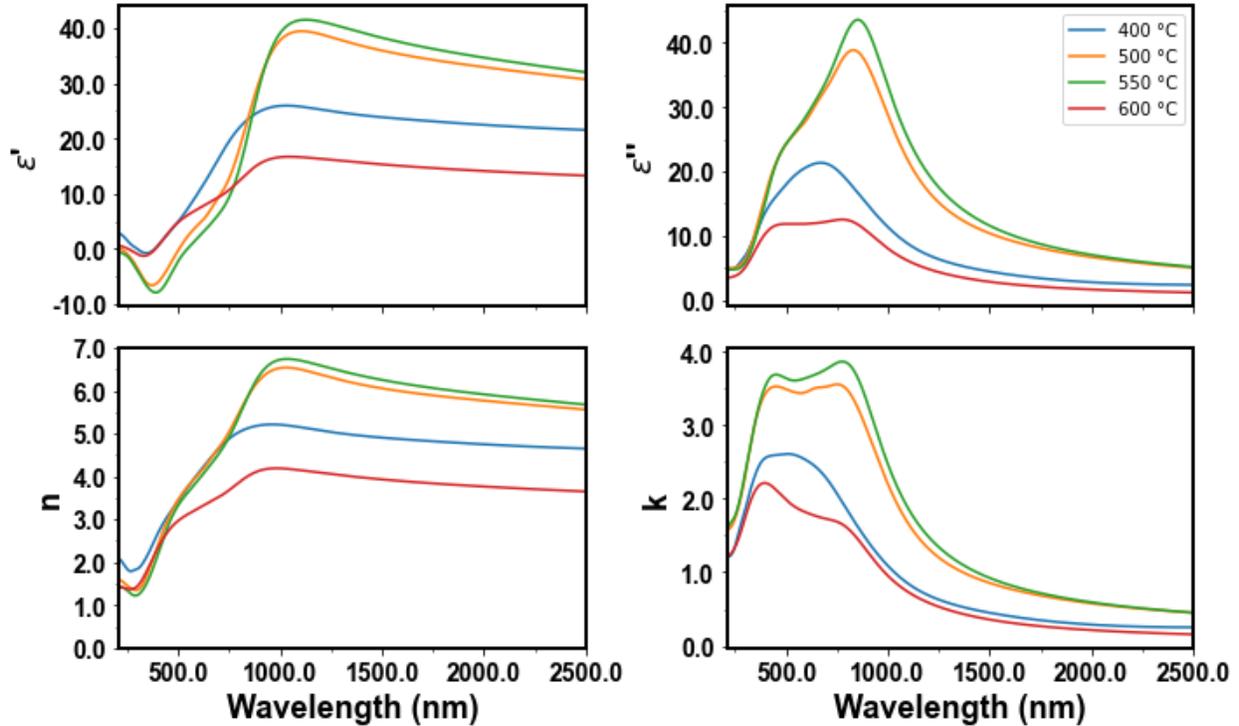

*Figure S4.* Optical constants of as-grown PtSe$_2$ as a function of growth temperature for growth temperatures ranging from 400°C to 600°C. We see that the best optical properties are achieved at growth temperatures of 500-550°C.

### S1.2.2. Transfer-Induced Change in Optical Properties and Thickness

We observed a change in the thickness and optical properties of the PtSe$_2$ after transfer. We expect that this is the result of the nanocrystalline structure "breaking apart" at the grain boundaries, leading to a decreased density. The optical properties of the transferred film (Figures 1c of main MS) differ from those of the as grown film (Figure S4): the magnitude of the constants decreases after transfer. The measured psi-delta data from SE of the transferred film was fit with 4 Lorentzian oscillators (Table 1). The raw data (solid lines) and the fit (dashed lines) are plotted in Figure S5. The discrepancies can likely be explained by the roughness of the underlying substrate, wrinkles of the transferred film, and the roughness of the film itself. The fit is reasonably good for the wavelengths of interest (400 nm to 900 nm).

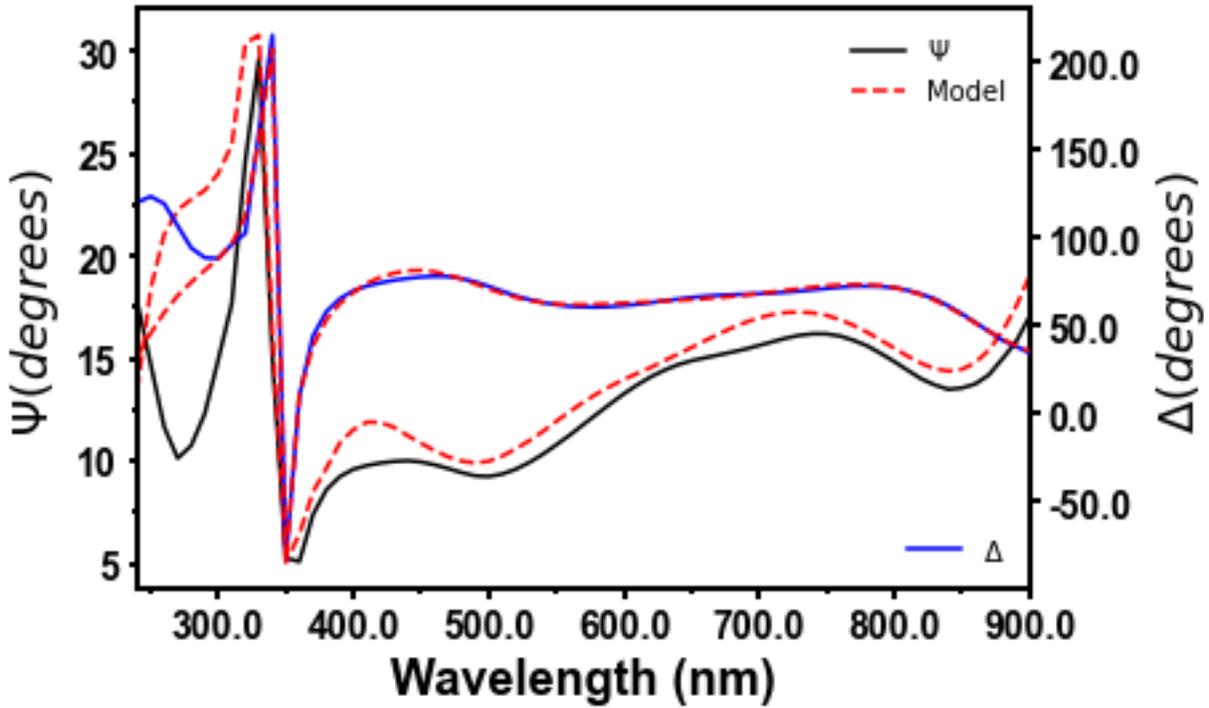

*Figure S5,* Experimental Psi-Delta from ellipsometry taken at an incident angle of 65° and the corresponding fits from the model (red dashed lines).

**S2. Nanoribbon Device Characterization**

*S2.1. Device Images*

In Figures S6a and S6b, we show scanning electron micrographs of devices with a nominal $w$ of 50 nm ($p$ = 150 nm) and $w$ = 80 nm ($p$ = 240 nm), respectively. We show measurements on the SEM images to confirm the dimensions. In Figure S6c, we show an atomic force micrograph of a device with a nominal $w$ of 90 nm ($p$ = 270 nm). While the AFM tip cannot fully resolve the lateral dimensions, we see that the PtSe$_2$ is indeed being removed fully by the etching process.

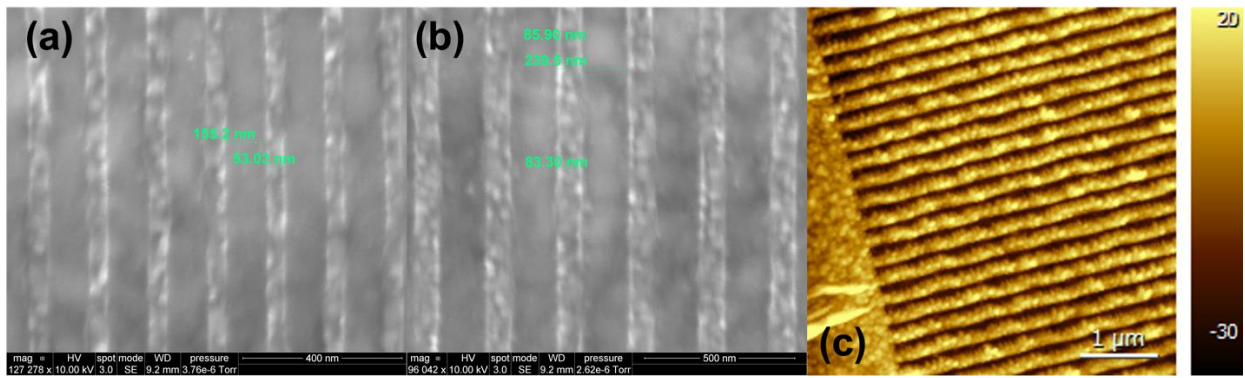

*Figure S6. SEM and AFM Images of nanoribbon devices.*

## S2.2. Characterization of Nanoribbon Array Modes

### S2.2.1. Field Profiles of Modes P1 and P2

Illumination by TM light excites two plasmon modes, the $|E|$, $\text{Re}(E_x)$, and $\text{Re}(E_y)$ field profiles are shown in figure S7. A lower energy plasmon mode (P1, top) is localized above and below the nanoribbon. The higher energy mode (P2, bottom) is localized to the sides of the nanoribbon. Both modes can be considered gap surface plasmon (GSP) like modes. Metal-insulator-metal (MIM) structures for GSP modes can be thought of as Fabry-Perot (FP) cavities with a large mode index[1]. This, of course, requires sufficient reflectance at each interface for multiple reflections to occur. The impedance mismatch between the PtSe$_2$ and alumina layer results in reflectance values of ~0.25 at 640 nm and ~0.3 at 425 nm for light travelling from the alumina incident on the PtSe$_2$. The structure consequently behaves like a lossy FP cavity and supports GSP-like modes.

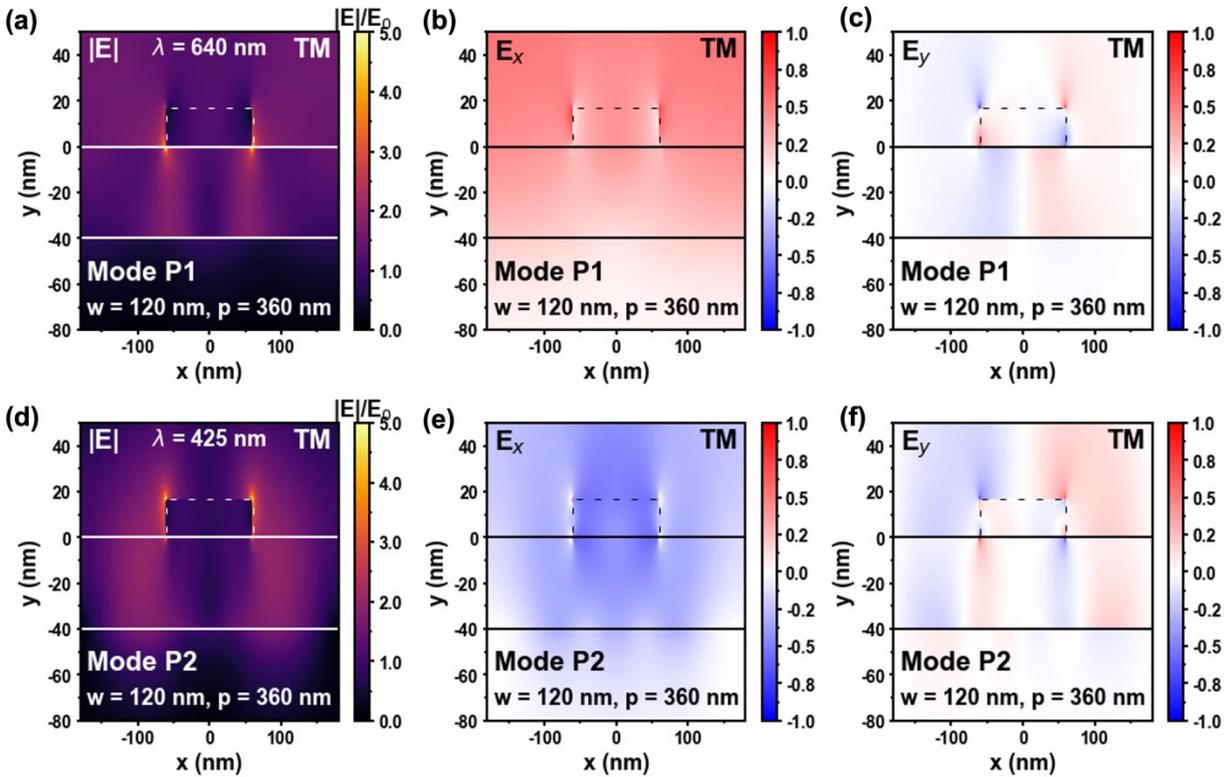

*Figure S7. TM Modes of PtSe$_2$ Nanoribbon Devices.* (a) $|E|$ field profiles, (b) $\text{Re}(E_x)$ field profiles, and (c) $\text{Re}(E_y)$ for the P1 ($\lambda = 640$ nm, top) and P2 ($\lambda = 425$ nm, bottom) modes for a nanoribbon array with $w = 120$ nm and $p = 360$ nm. The $\text{Re}(E_x)$ and $\text{Re}(E_y)$ profiles are normalized to themselves.

### S2.2.2. Origin of Plasmon Resonances

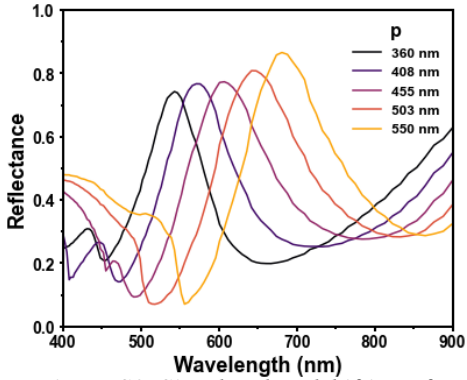

*Figure S8. Simulated Redshifting of P2 Mode Beyond 470 nm.* Simulated reflectance spectra of the PtSe$_2$ nanoribbon structure (17 nm PtSe$_2$/40 nm Al$_2$O$_3$/100 nm Ag) illuminated by TM polarized plane wave. In the simulation, the period $p$ is varied while the fill factor is fixed at 1/3.

To better understand the role of the PtSe$_2$ and Ag, we performed a control simulation for which we replaced the PtSe$_2$ nanoribbons with a fictitious (and non-physical) material that has the same $n$ as PtSe$_2$ but has $k = 0$. Figure S9a shows the calculated TM absorbance spectra for a control device with $w = 120$ nm, $p = 360$ nm. There is a strong absorbance peak at 550 nm due to the P1 mode (Figure S9b) and a small peak at approximately 415 nm due to the P2 mode (Figure S9c). The locations and field profiles of these two modes are in good agreement with the P1 and P2 modes shown in Figure 3 of the main manuscript. We also note that off resonance absorption is minimal, further supporting our main argument that absorbance is due to interband loss in PtSe$_2$, with Ag assisting. Further, while PtSe$_2$ is optically a metal (Re($\varepsilon$) < 0) below 470 nm, it can also be shown by simulation that the P2 mode can be shifted to wavelengths where PtSe$_2$ is optically a dielectric (Figure S8). We can therefore attribute both modes to the silver.

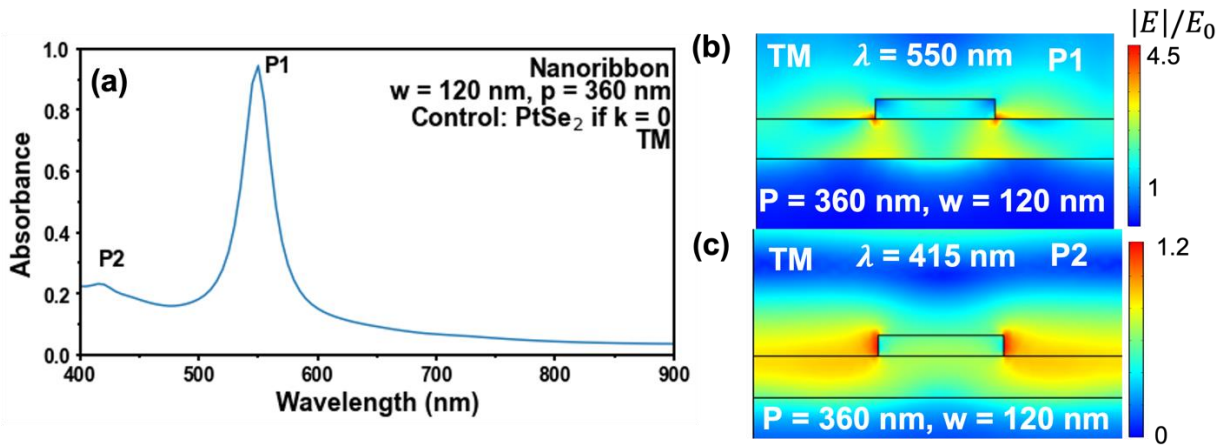

*Figure S9. Control Simulations of Nanoribbons with n = n(PtSe$_2$) and k = 0.* (a) Simulated absorbance spectrum of a control device with $w = 120$ nm, $p = 360$ nm, and the material $k = 0$ under illumination by a TM polarized plane wave. The P1 and P2 modes are labeled in the spectrum, and the $|E|$ field profiles the P1 and P2 modes are shown in (b) and (c), respectively.

### S2.2.3. Waveguide Mode

For TM polarized light, in experimental spectra, we see the emergence of a third peak (labeled 'WG' in Figure 2c) that is not predicted by normal incidence simulations. We find that this is a waveguide-like resonance that emerges at non-zero angles of incidence at ~2.35 eV for a device with $w = 120$ nm and $p = 3w$ (Figure S10). Experimentally, this is the result of the numerical aperture of the microscope being ~0.35, resulting in incident angles up to approximately 20.5°.

### S2.2.4. Effect of Dielectric Layer Permittivity

Plasmons are dependent on the local dielectric environment[2]. For our system, we expect the effective dielectric permittivity of the surrounding environment to be approximately proportional to $(1+\varepsilon_s)$, where $\varepsilon_s$ is the relative permittivity of the dielectric spacer layer. We then expect the plasmon frequency, $\omega_p$, to vary as:

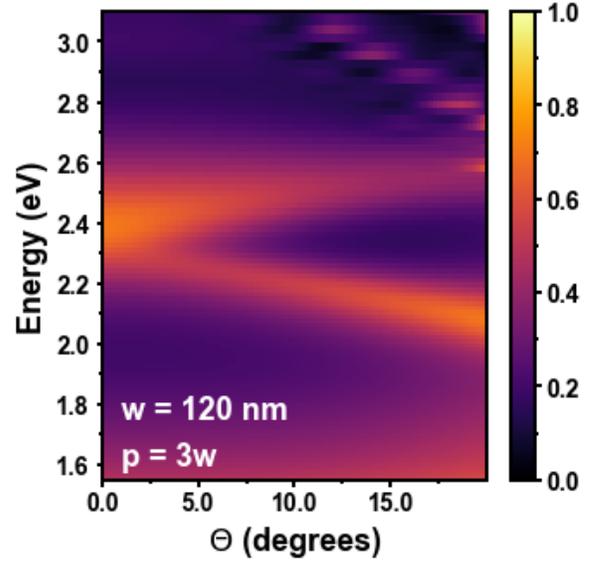

*Figure S10. Waveguide mode.* Simulated reflectance of a device on 40 nm $Al_2O_3$ with $w = 120$ nm and $p = 360$ nm as a function of incident angle and energy. At angles greater than ≈ 5°, the reflectance maximum splits into two peaks, and a reflectance minimum arises.

$$\omega_p \propto (1+\varepsilon_s)^{-1/2} \quad (3)$$

Figure S11a shows the reflectance magnitude as a function of energy and $\varepsilon_s$ for a device with $w = 90$ nm, $p = 270$ nm, and $t = 40$ nm. We observe that both the P1 and P2 modes follow the expected dispersion given by (3).

### S2.2.5. Effect of Dielectric Layer Thickness

As mentioned in *S2.2.1.*, the structure has a GSP-like response that consequently depends on the thickness of the dielectric layer. We would expect a "pure" GSP resonance to redshift with increasing layer thickness $t$. In Figure S11b, we plot the simulated reflectance magnitude as a function of $t$ and energy for a device with an $Al_2O_3$ dielectric, $w = 120$ nm, and $p = 360$ nm. Two phenomena are observed: (i.) modes redshift with increasing $t$ and (ii.) changing $t$ causes different modes to have enhanced absorption for devices with the same width and period. Thinner layers have a higher cavity photon energy and therefore couple to the higher energy P2 mode while thicker layers couple to the P1 mode. The dielectric thickness should be chosen to optimize both plasmon energy and reflectance magnitude.

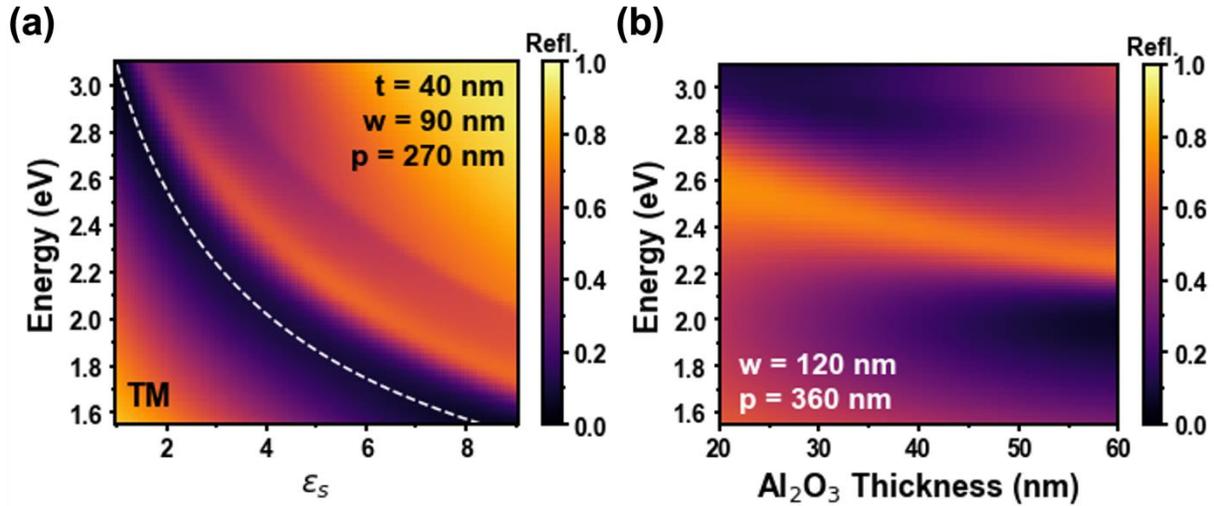

***Figure S11. Effects of (a) Dielectric Permittivity and (b)Thickness on Plasmon Energy and Reflectance Magnitude for PtSe₂ Nanoribbons.*** Results are calculated using the COMSOL Multiphysics wave optics module for a monochromatic, normal incidence, TM polarized plane wave source. The white dotted line in (a) is a fit to equation (3).

### S2.2.5. Confirming the Role of PtSe₂ in Plasmon Broadening

We now fix the period of the lossless control device to 300 nm and sweep the width from 110 nm to 225 nm. The TM dispersion (Figure S11) shows the P1 mode, which redshifts minimally and does not broaden with increasing fill factor, unlike the device with actual PtSe₂ (Figure 3d of main manuscript).

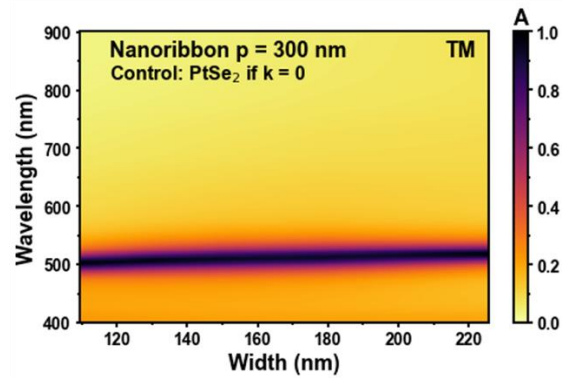

***Figure S12.*** Control $k = 0$ simulation for which the array period is fixed at 300 nm and the width is swept from 110 to 225 nm. Simulated with the COMSOL Multiphysics wave optics module for a monochromatic, normal incidence, TM polarized plane wave source.

### S2.2.6. TE Absorbance due to PtSe₂

Figure S13 shows the transverse electric (TE) dispersion. There is less than 10% absorbance from 400 nm to 900 nm; this is the expected amount of loss due to the silver. We can therefore confirm that nearly all absorbance we see in the TE direction in Figure 4 of the main manuscript is due to the large $k$ of PtSe₂.

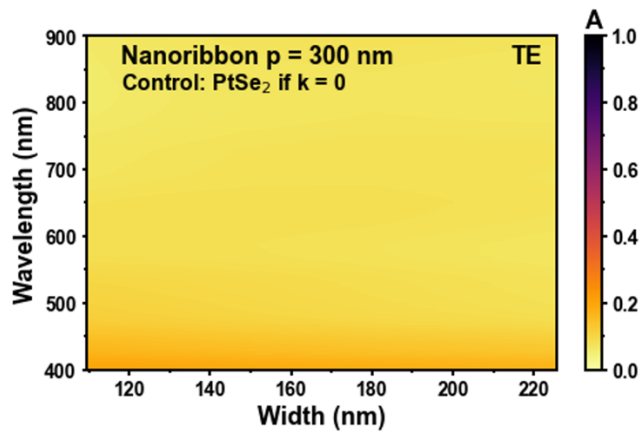

## S3. 2D Metasurface Characterization and Field Profiles

### S3.1. More SEM Images of 2D Metasurfaces

*Figure S13.* Control $k = 0$ simulation for which the array period is fixed at 300 nm and the width is swept from 110 to 225 nm. Simulated with the COMSOL Multiphysics wave optics module for a monochromatic, normal incidence, TE polarized plane wave source.

In Figure 5b of the main manuscript, SEM images of the nanosquare (NSQ) array device with dimensions ($w$ = 250 nm, $p$ = 300 nm) and the crossed grating (XG) device with dimensions (100 nm, 300 nm) are shown. NSQ arrays with dimensions (150 nm, 300 nm) and (150 nm, 200 nm) and XGs with dimensions (150 nm, 300 nm) and (200 nm, 300 nm) were also fabricated and characterized (Figure 5c and 5d of main manuscript). SEM images of these other four devices are shown in Figure S14.

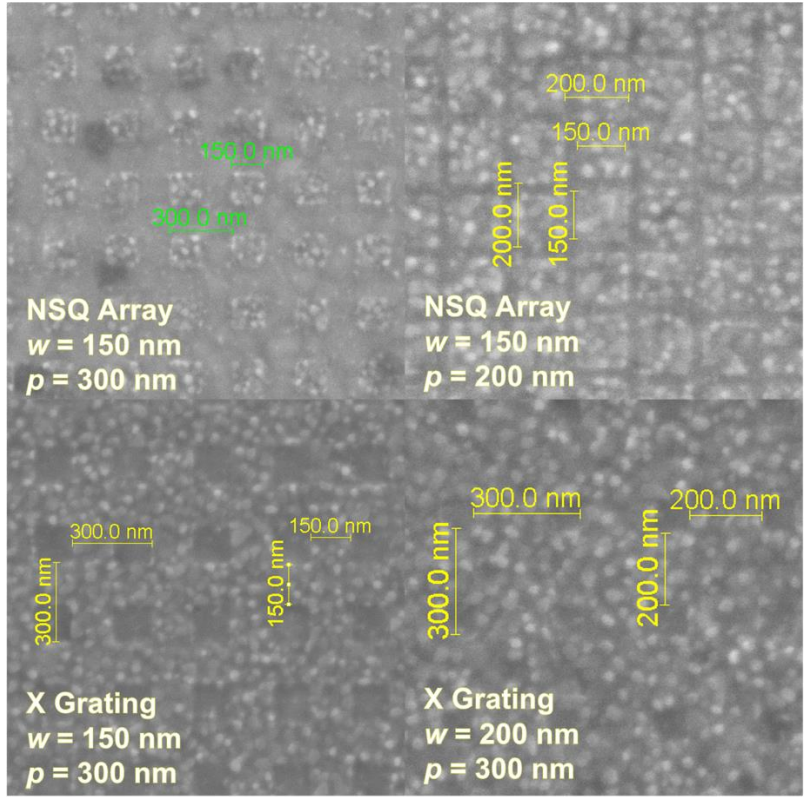

*Figure S14.* SEM Images of 2D Metasurface Devices not included in Figure 5 of main manuscript. SEM images of nanosquare (NSQ) array devices with width 150 nm, period 300 nm (top left) and width 150 nm, period 200 nm (top right). SEM images of crossed grating (X Grating) devices with width 150 nm, period 300 nm (bottom left) and width 200 nm, period 300 nm (bottom right). Dimensions are measured using the SEM software and overlaid.

### S3.2. Field Profiles and Absorption Characteristics of 2D Metasurfaces

The modes of the 2D metasurfaces (2DMSs) were explored using the Lumerical finite difference time domain (FDTD) solver. We simulated the XG and NSQ array structures using normal incident plane waves with linear polarization along the $\hat{x}$-direction (along one of the axes of periodicity for the array). Given the square lattice and square features, the response to light polarized along the $\hat{x}$ direction and $\hat{y}$ direction will be identical. Figure S15a shows the calculated $|E|$ field (top) and corresponding $P_{abs}$ (bottom) profiles in the XZ plane ($y = 0$) for an XG device ($w = 150$ nm, $p = 300$ nm) illuminated by an $\hat{x}$-polarized 550 nm plane wave. The $|E|$ field profile and $P_{abs}$ profile in the XY plane ($z = 8$ nm) for the same device are displayed in Figure S15b. Under this polarization, the modes in the XZ plane can be considered TM modes while the modes in the YZ plane are TE-like. The $P_{abs}$ profile confirms that absorption is dominated by the PtSe2. The plasmonic response observed in the $|E|$ field profile in Figure S15a is similar to the TM P1 mode of the 1D nanoribbon array. In the XY plane $|E|$ field profile (Figure S15b), TE-like field localization in PtSe2 is seen when there is confinement in the $\hat{y}$ direction (i.e., for -75 nm $< x <$ 75 nm). The power absorption is strongest in the region with TE field enhancement (Figure S15b, bottom). Additionally, slight reduced dependence on fill factor and higher energy resonance (absorbance maximum) – relative to the P1 mode – are both phenomena observed for the TE modes of 1D nanoribbon arrays (Figure 4 in main manuscript). Therefore, while plasmonic field enhancement contributes to absorption, the absorption in the XG geometry is more due to TE-like modes.

Figure S15c shows the calculated E field profile (top) and corresponding $P_{abs}$ profile (bottom) in the XZ plane ($y = 0$) for a NSQ array device ($w = 250$ nm, $p = 300$ nm) illuminated by an $\hat{x}$-polarized 550 nm plane wave. Figure S15d shows these profiles for the same structure in the XY ($z = 8$ nm). The role of the plasmon resonance is more obvious for this NSQ device. The electric field is localized near the edges of the nanosquare, with little variation in the $\hat{y}$ direction. The regions for which the plasmon localizes and enhances the field are the regions that exhibit the strongest absorption. In contrast to the XG geometry, absorption in the NSQ geometry is dominated by the plasmonic TM-like modes. Again, negligible absorption can be attributed directly to the Ag, but the plasmon resonance assists absorption in the PtSe2.

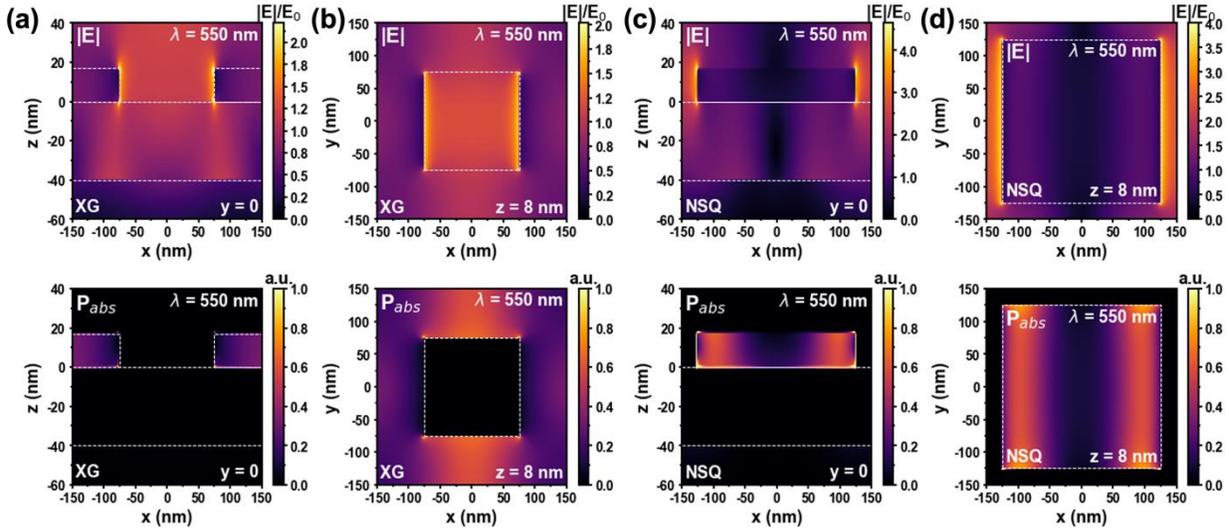

*Figure S15. 2DMS Field and Power Absorption Profiles.* (a) Simulated electric field magnitude ($|E|$, top) and power absorbed ($P_{abs}$, bottom) profiles for (a) an X Grating (XG) device ($w = 150$ nm, $p = 300$ nm) in the XZ plane ($y = 0$); (b) the same XG device in the XY plane ($z = 8$ nm); (c) a nanosquare (NSQ) array device ($w = 250$ nm, $p = 300$ nm) in the XZ plane ($y = 0$); and (d) the same NSQ device in the XY plane ($z = 8$ nm). The field profiles were calculated by the Lumerical FDTD solver for a normal incidence plane wave ($\lambda = 550$ nm) polarized with the $\vec{E}$ component

along the $\hat{x}$ direction. Edges of layers and materials are indicated by a white dashed line. The $P_{abs}$ values are normalized to the maximum $P_{abs}$ value for all plots shown to provide a relative comparison.

We further investigate the XG devices within the two regimes of the optical constants for PtSe$_2$. Figure S16a (top) shows the $|E|$ profile along the YZ plane for an XG device with $w$ = 150 nm, $p$ = 300 nm illuminated by $\hat{x}$-polarized light with $\lambda$ = 425 nm. Like the nanoribbon TE field profiles (Figure 4b of main manuscript), the electric field is localized and enhanced in the area between the PtSe$_2$ parallel gratings, resulting in power absorption being localized near the edges of the PtSe$_2$. For the $\lambda$ = 900 nm case (Figure S16b), absorption is again – like the nanoribbon TE case and the XG device for $\lambda$ = 550 nm – enhanced by the large refractive index of the PtSe$_2$, and power is primarily absorbed in the regions where there is confinement in the $\hat{y}$ direction. The similarities to the TE nanoribbon modes further supports the claim that the primary absorption mechanism for XG devices is field enhancement either between the gratings (below 470 nm) or inside the ribbons (above 470 nm).

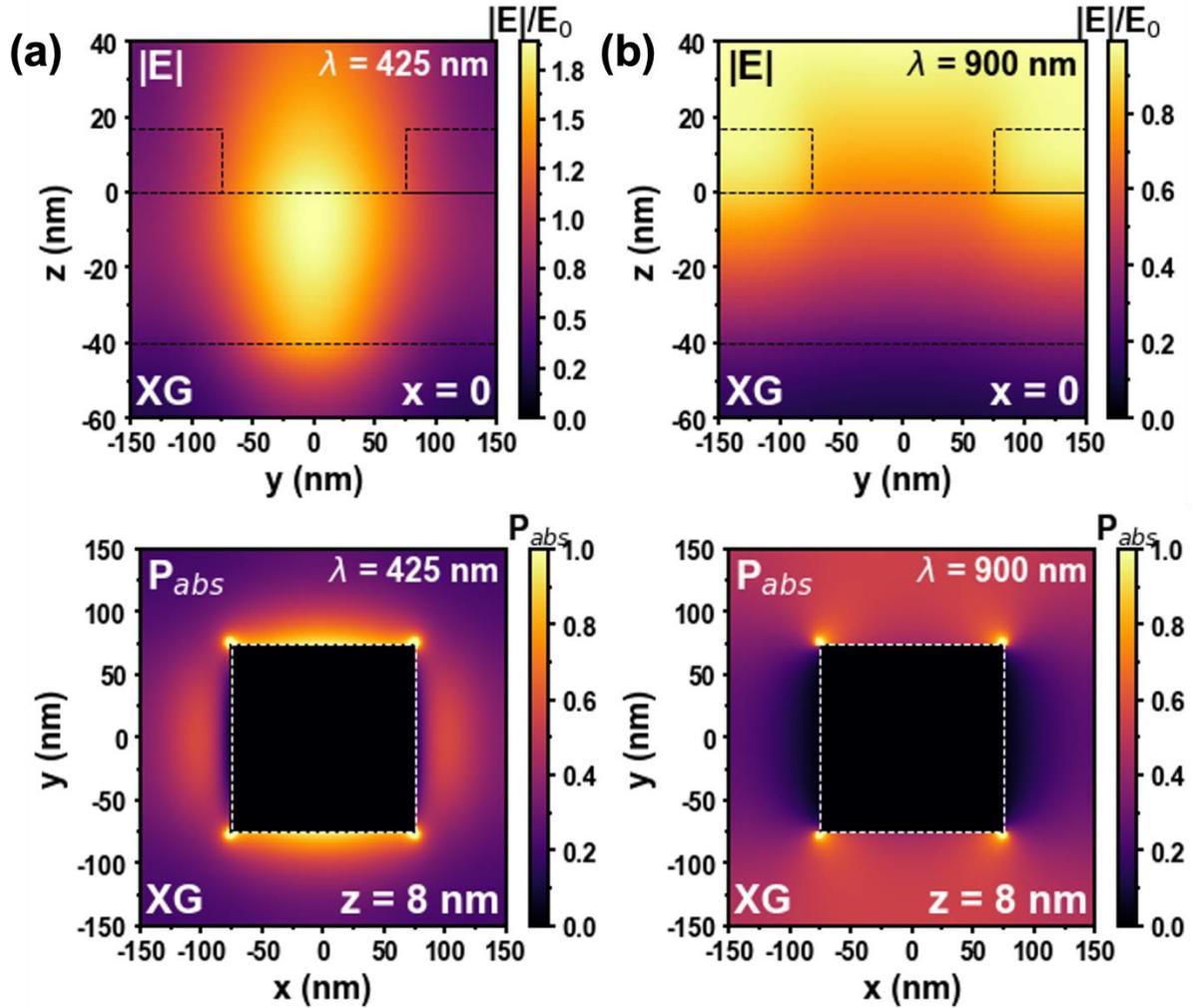

*Figure S16. Simulated Field and Power Absorption Profiles for an X Grating Metasurface Illuminated by $\hat{x}$-Polarized Light with $\lambda$ = 425 nm and 900 nm.* $|E|$ profile along the YZ plane (top) and $P_{abs}$ profile along the XY plane (bottom) for an XG device with $w$ = 150 nm, $p$ = 300 nm illuminated by two other wavelengths: $\lambda$ = 425 nm (a) and $\lambda$ = 900 nm (b).

Similarly, we consider the NSQ array field and power profiles at two more wavelengths. Figure S17 shows the electric field magnitude and $P_{abs}$ profiles for a NSQ array device with $w = 250$ nm, $p = 300$ nm under illumination by an $\hat{x}$-polarized plane wave with wavelengths $\lambda = 450$ nm and 900 nm, both wavelengths away from the main plasmon resonance. For $\lambda = 450$ nm, there is TE-like localization of the electric field between NSQs in the YZ plane (Figure S17a) and TM-like plasmonic field localization towards the edges of the NSQ in the XZ plane (Figure S17b). The combination of these two modes is observed in the XY plane power profile (Figure S17c).

While the plasmon resonance occurs at approximately 550 nm, the off-resonance plasmon assisted absorption is directly observed at 900 nm for the NSQ. The YZ plane $|E|$ profile indicates plasmonic field enhancement in the center of the NSQ (Figure S17d), resulting in power absorption occurring towards the center (in $\hat{x}$) of the NSQ (Figures S17e and S17f). The lack of variation in $\hat{y}$ indicates that absorbance is dominated by the TM-like modes at this wavelength. By using plasmons for field enhancement instead of as the primary absorption mechanism, broadband absorption is achieved without the need for resonance multiplexing.

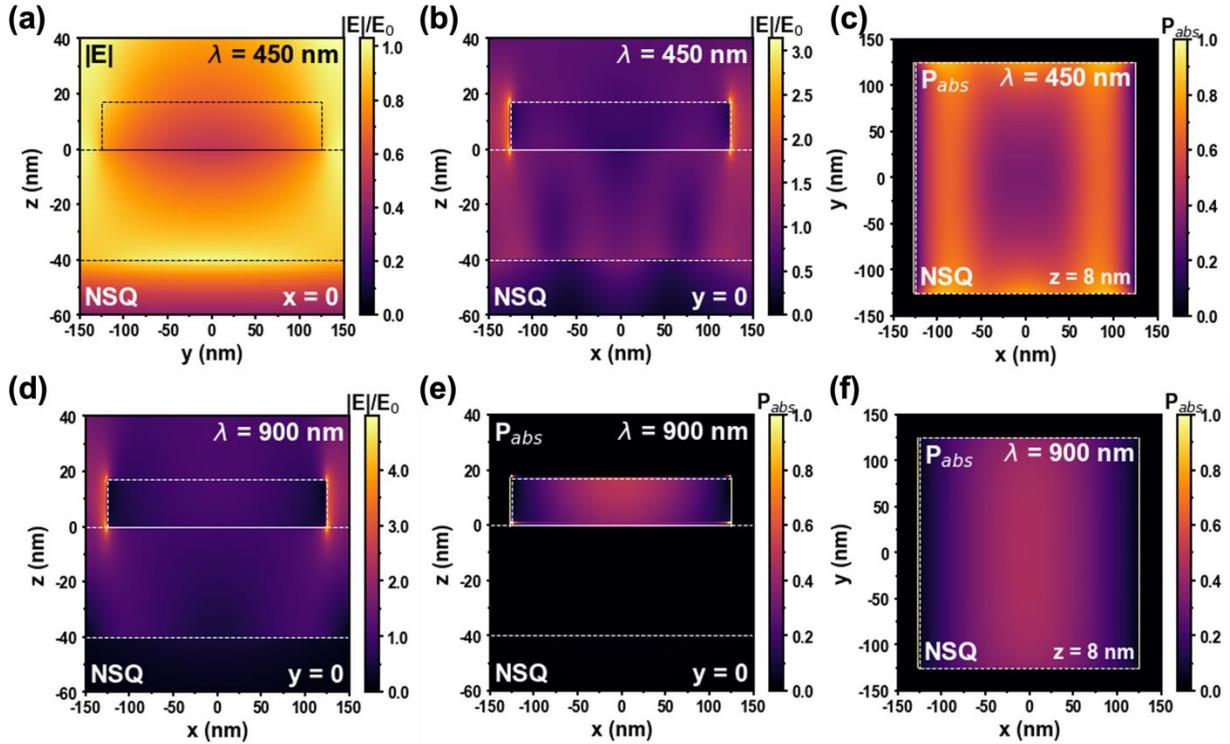

*Figure S17. Simulated Field and Power Absorption Profiles for Nanosquare Array Metasurface Illuminated by $\hat{x}$-Polarized Light with $\lambda = 450$ nm and 900 nm.* $|E|$ profile in the YZ plane (a), $|E|$ profile in the XZ plane (b), and $P_{abs}$ profile in the XY plane (c) for a NSQ array device with $w = 250$ nm, $p = 300$ nm for $\lambda = 450$ nm. $|E|$ profile in the XZ plane (d), $P_{abs}$ profile in the XZ plane (e), and $P_{abs}$ profile in the XY plane (f) for the same NSQ array device for $\lambda = 900$ nm.

### S3.3. Simulated Absorbance Dispersion for 2DMSs

Using the Lumerical FDTD solver, we simulated the 2DMSs shown in Figure 5a of the main manuscript for $\hat{x}$-polarized plane waves at normal incidence using a broadband source (400 nm to

900 nm). Figure S18a is a plot of $A$ in an XG device as a function of wavelength and $p$ for a fixed $w$ of 150 nm. Figure S18b shows $A$ in an XG device as a function of wavelength and $w$ for a fixed $p = 300$ nm. Similarly, Figures S18c and S18d show $A$ spectra as a function of period (fixed $w = 150$ nm) and width (fixed $p = 300$ nm) for a NSQ array. The X Grating devices exhibit significantly less dependence on fill factor than the NSQ arrays. For the NSQ devices, there are distinct plasmonic resonances at low fill factors that broaden with increasing fill factor, resulting in broadband absorption. This is consistent with what is observed for the TM modes of the nanoribbon arrays.

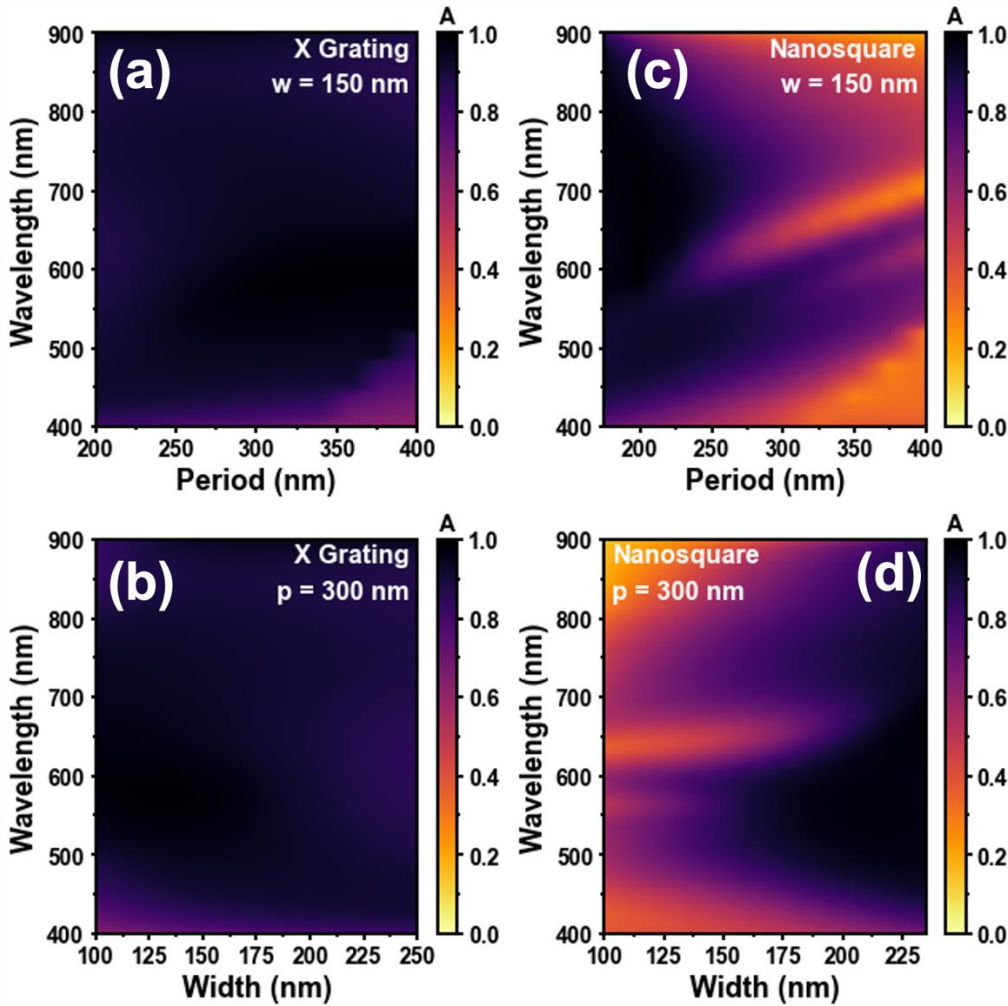

*Figure S18. Dispersions of Absorbance Spectra for 2D Metasurface Devices.* (a) Absorbance as a function of wavelength and period for an X Grating device with fixed $w = 150$ nm. (b) Absorbance as a function of wavelength and width for an X Grating device with fixed $p = 300$ nm. (c) Absorbance as a function of wavelength and period for an NSQ Array device with fixed $w = 150$ nm. (d) Absorbance as a function of wavelength and width for a NSQ array device with fixed $p = 300$ nm.

Table 1. Fit parameters for PtSe₂ grown at 550°C before and after transfer

| PARAMETER | AS GROWN | POST-TRANSFER |
|---|---|---|
| MSE | 2.824 | 2.93 |
| Roughness (nm) | 2.0±0.542 | 1.162 |
| Thickness (nm) | 8.3±0.439 | 17.11 |
| $\varepsilon_\infty$ | 2.041±0.0405 | 1.544 |
| $A_{UV}$ | 15.5362±1.18301 | 20.4668 |
| $E_{UV}$ (eV) | 6.548±0.0406 | 6.750 |
| $A_{IR}$ | 1.0517±0.07634 | 0 |
| $A_1$ | 16.009524±1.3807628 | 7.810893 |
| $\Gamma_1$ (eV) | 1.7411±0.02524 | 1.1840 |
| $E_1$ (eV) | 2.737±0.005758 | 2.779 |
| $A_2$ | 30.05389±2.0949219 | 10.837 |
| $\Gamma_2$ (eV) | 0.6637±0.00412 | 0.5682 |
| $E_2$ (eV) | 1.453±0.002186 | 1.480 |
| $A_3$ | 11.939459±0.9658847 | 4.740844 |
| $\Gamma_3$ (eV) | 1.166±0.04441 | 1.0109 |
| $E_3$ (eV) | 1.945±0.005159 | 1.942 |
| $A_4$ | 3.207054±0.1500741 | 1.536796 |
| $\Gamma_4$ (eV) | 2.4721±0.10155 | 0.4878 |
| $E_4$ (eV) | 5.748±0.0138 | 11.040 |

Table 2. Optical constants for transferred PtSe₂ grown at 550°C for 240 nm to 900 nm.

| Wavelength (nm) | n | k | $\varepsilon'$ | $\varepsilon''$ |
|---|---|---|---|---|
| 240 | 0.805143 | 0.595336 | 0.2938303 | 0.95866123 |
| 250 | 0.737805 | 0.760954 | -0.0346948 | 1.12287133 |
| 260 | 0.711644 | 0.920726 | -0.3412992 | 1.31045827 |
| 270 | 0.712898 | 1.068935 | -0.6343985 | 1.52408325 |
| 280 | 0.732459 | 1.205868 | -0.9176214 | 1.76649774 |
| 290 | 0.765277 | 1.333163 | -1.1916747 | 2.04047796 |
| 300 | 0.808623 | 1.452258 | -1.4551821 | 2.34865844 |
| 310 | 0.860991 | 1.56409 | -1.705072 | 2.69333483 |
| 320 | 0.921505 | 1.669127 | -1.9368135 | 3.07621775 |
| 330 | 0.989585 | 1.767438 | -2.1445586 | 3.49806027 |
| 340 | 1.064755 | 1.858784 | -2.3213747 | 3.95829912 |
| 350 | 1.146506 | 1.94269 | -2.4595684 | 4.45461148 |
| *360* | 1.234205 | 2.018519 | -2.551157 | 4.98253248 |
| *370* | 1.327026 | 2.085558 | -2.5885542 | 5.53517938 |
| *380* | 1.423907 | 2.143112 | -2.5654179 | 6.10318436 |
| *390* | 1.523543 | 2.190606 | -2.4775714 | 6.67496487 |
| *400* | 1.624411 | 2.227693 | -2.323905 | 7.23737803 |
| *410* | 1.724839 | 2.254339 | -2.1069748 | 7.77674365 |
| *420* | 1.823108 | 2.270894 | -1.8332368 | 8.28017004 |
| *430* | 1.917572 | 2.278109 | -1.5126982 | 8.73687606 |
| *440* | 2.006792 | 2.277113 | -1.1580295 | 9.1393843 |
| *450* | 2.089642 | 2.26934 | -0.7833003 | 9.48421635 |
| *460* | 2.16539 | 2.256419 | -0.4025129 | 9.77205428 |
| *470* | 2.233726 | 2.240051 | -0.0282966 | 10.0073203 |
| *480* | 2.294754 | 2.221883 | 0.32913185 | 10.1973498 |
| *490* | 2.348943 | 2.203408 | 0.6625264 | 10.3513596 |
| *500* | 2.397054 | 2.185875 | 0.96781836 | 10.4793208 |
| *510* | 2.440051 | 2.17025 | 1.24386382 | 10.5910414 |
| *520* | 2.479012 | 2.157181 | 1.49207063 | 10.6953552 |
| *530* | 2.515043 | 2.14701 | 1.71578935 | 10.7996449 |
| *540* | 2.549199 | 2.139789 | 1.91971858 | 10.909496 |
| *550* | 2.582414 | 2.135319 | 2.10927484 | 11.0285554 |
| *560* | 2.615452 | 2.133209 | 2.29000853 | 11.1586115 |
| *570* | 2.648867 | 2.132934 | 2.46708894 | 11.299717 |
| *580* | 2.682988 | 2.133914 | 2.64483565 | 11.4505313 |
| *590* | 2.717922 | 2.135579 | 2.82640233 | 11.6086743 |
| *600* | 2.753585 | 2.137444 | 3.0135635 | 11.7712675 |
| *610* | 2.789745 | 2.13916 | 3.20667166 | 11.9354218 |
| *620* | 2.826091 | 2.14055 | 3.40483604 | 12.0987782 |
| *630* | 2.862296 | 2.141622 | 3.6061936 | 12.2599122 |
| *640* | 2.898093 | 2.142554 | 3.80840539 | 12.4186415 |
| *650* | 2.933328 | 2.143661 | 4.00913067 | 12.5761217 |
| *660* | 2.968009 | 2.145343 | 4.20658084 | 12.7347947 |
| *670* | 3.002323 | 2.148021 | 4.39994918 | 12.8981057 |
| *680* | 3.036644 | 2.152072 | 4.58979289 | 13.0701531 |
| *690* | 3.071511 | 2.15777 | 4.77820845 | 13.2552286 |
| *700* | 3.107592 | 2.165226 | 4.96892441 | 13.457278 |
| *710* | 3.145636 | 2.174352 | 5.16721922 | 13.6794399 |
| 720 | 3.186408 | 2.184833 | 5.3797007 | 13.9235387 |
| 730 | 3.230625 | 2.196118 | 5.61400362 | 14.1896674 |
| 740 | 3.278883 | 2.207432 | 5.87831769 | 14.4758225 |
| 750 | 3.331591 | 2.217806 | 6.18083514 | 14.777645 |
| 760 | 3.388913 | 2.226133 | 6.52906319 | 15.0883421 |
| 770 | 3.450721 | 2.231233 | 6.92907472 | 15.3987251 |
| 780 | 3.516577 | 2.231942 | 7.38474871 | 15.6975918 |
| 790 | 3.585735 | 2.2272 | 7.89707565 | 15.972298 |
| 800 | 3.65718 | 2.216142 | 8.46368019 | 16.2096604 |
| 810 | 3.729686 | 2.198174 | 9.07858872 | 16.3969976 |
| 820 | 3.801909 | 2.17302 | 9.73249612 | 16.5232486 |
| 830 | 3.872478 | 2.140748 | 10.4132839 | 16.5799991 |
| 840 | 3.940102 | 2.101752 | 11.1070423 | 16.5622345 |
| 850 | 4.00365 | 2.056709 | 11.7991614 | 16.468686 |
| 860 | 4.062219 | 2.00651 | 12.4755408 | 16.3017661 |
| 870 | 4.115167 | 1.952179 | 13.1235966 | 16.0670852 |
| 880 | 4.162121 | 1.89479 | 13.7330221 | 15.7726905 |
| 890 | 4.202956 | 1.835399 | 14.2961496 | 15.4282025 |
| 900 | 4.237761 | 1.774984 | 14.8080501 | 15.0439159 |

*Table 3. Optical constants for as-grown PtSe$_2$ grown at 550°C for 210 nm to 2500 nm.*

| Wavelength (nm) | n | k | $\varepsilon'$ | $\varepsilon''$ |
|---|---|---|---|---|
| 210 | 1.422846 | 1.537909 | -0.340673 | 4.376415 |
| 211 | 1.420147 | 1.552269 | -0.392721 | 4.408903 |
| 212 | 1.4182 | 1.565147 | -0.438393 | 4.439384 |
| 213 | 1.416832 | 1.576724 | -0.478646 | 4.467905 |
| 214 | 1.415901 | 1.587159 | -0.514297 | 4.49452 |
| 215 | 1.415293 | 1.596589 | -0.546041 | 4.519281 |
| 216 | 1.414912 | 1.605134 | -0.574479 | 4.542249 |
| 217 | 1.414681 | 1.612902 | -0.600131 | 4.563482 |
| 218 | 1.414532 | 1.619985 | -0.623451 | 4.583044 |
| 219 | 1.414413 | 1.626469 | -0.644838 | 4.600999 |
| 220 | 1.414278 | 1.632428 | -0.66464 | 4.617414 |
| 221 | 1.414089 | 1.63793 | -0.683168 | 4.632359 |
| 222 | 1.413815 | 1.643036 | -0.700694 | 4.645901 |
| 223 | 1.413431 | 1.647802 | -0.717463 | 4.658111 |
| 224 | 1.412916 | 1.652278 | -0.733692 | 4.669059 |
| 225 | 1.412251 | 1.65651 | -0.749573 | 4.678818 |
| 226 | 1.411425 | 1.660542 | -0.765279 | 4.687459 |
| 227 | 1.410425 | 1.664411 | -0.780966 | 4.695054 |
| 228 | 1.409244 | 1.668154 | -0.796769 | 4.701673 |
| 229 | 1.407876 | 1.671805 | -0.812815 | 4.707389 |
| 230 | 1.406318 | 1.675393 | -0.829211 | 4.712271 |
| 231 | 1.404567 | 1.678948 | -0.846058 | 4.71639 |
| 232 | 1.402623 | 1.682496 | -0.863441 | 4.719815 |
| 233 | 1.400487 | 1.686062 | -0.88144 | 4.722614 |
| 234 | 1.398161 | 1.689668 | -0.900123 | 4.724854 |
| 235 | 1.395649 | 1.693336 | -0.91955 | 4.726603 |
| 236 | 1.392954 | 1.697085 | -0.939776 | 4.727923 |
| 237 | 1.390083 | 1.700935 | -0.960848 | 4.72888 |
| 238 | 1.387041 | 1.704901 | -0.982805 | 4.729535 |
| 239 | 1.383835 | 1.709 | -1.005682 | 4.729949 |
| 240 | 1.380473 | 1.713247 | -1.029511 | 4.730182 |
| 241 | 1.376962 | 1.717655 | -1.054314 | 4.73029 |
| 242 | 1.373311 | 1.722236 | -1.080114 | 4.730332 |
| 243 | 1.369529 | 1.727002 | -1.106925 | 4.730361 |
| 244 | 1.365626 | 1.731964 | -1.134762 | 4.73043 |
| 245 | 1.361611 | 1.73713 | -1.163634 | 4.730591 |
| 246 | 1.357495 | 1.742509 | -1.193546 | 4.730894 |
| 247 | 1.353286 | 1.74811 | -1.224503 | 4.731386 |
| 248 | 1.348997 | 1.753938 | -1.256505 | 4.732115 |
| 249 | 1.344638 | 1.76 | -1.289551 | 4.733125 |
| 250 | 1.340218 | 1.766302 | -1.323637 | 4.73446 |
| 251 | 1.33575 | 1.772847 | -1.358756 | 4.736162 |
| 252 | 1.331245 | 1.779639 | -1.394902 | 4.73827 |
| 253 | 1.326712 | 1.786681 | -1.432065 | 4.740824 |
| 254 | 1.322164 | 1.793976 | -1.470234 | 4.74386 |
| 255 | 1.31761 | 1.801525 | -1.509397 | 4.747414 |
| 256 | 1.313062 | 1.809329 | -1.54954 | 4.751521 |
| 257 | 1.30853 | 1.817388 | -1.59065 | 4.756212 |
| 258 | 1.304024 | 1.825702 | -1.63271 | 4.76152 |
| 259 | 1.299556 | 1.834271 | -1.675704 | 4.767474 |
| 260 | 1.295134 | 1.843092 | -1.719614 | 4.774103 |
| 261 | 1.29077 | 1.852163 | -1.764423 | 4.781434 |
| 262 | 1.286472 | 1.861484 | -1.810111 | 4.789493 |
| 263 | 1.28225 | 1.871049 | -1.856661 | 4.798305 |
| 264 | 1.278112 | 1.880857 | -1.904051 | 4.807893 |
| 265 | 1.274069 | 1.890903 | -1.952262 | 4.818281 |
| 266 | 1.270127 | 1.901183 | -2.001274 | 4.829489 |
| 267 | 1.266296 | 1.911693 | -2.051065 | 4.841538 |
| 268 | 1.262582 | 1.922428 | -2.101614 | 4.854447 |
| 269 | 1.258994 | 1.933383 | -2.152901 | 4.868235 |
| 270 | 1.255538 | 1.944552 | -2.204904 | 4.882918 |
| 271 | 1.252221 | 1.95593 | -2.257601 | 4.898514 |
| 272 | 1.24905 | 1.96751 | -2.310971 | 4.915037 |
| 273 | 1.246029 | 1.979288 | -2.364992 | 4.932503 |
| 274 | 1.243166 | 1.991257 | -2.419643 | 4.950925 |
| 275 | 1.240464 | 2.00341 | -2.474901 | 4.970316 |
| 276 | 1.237929 | 2.015742 | -2.530746 | 4.99069 |
| 277 | 1.235565 | 2.028244 | -2.587155 | 5.012056 |
| 278 | 1.233377 | 2.040913 | -2.644107 | 5.034428 |
| 279 | 1.231367 | 2.053739 | -2.70158 | 5.057814 |
| 280 | 1.22954 | 2.066718 | -2.759553 | 5.082224 |
| 281 | 1.227898 | 2.079841 | -2.818006 | 5.107668 |
| 282 | 1.226445 | 2.093104 | -2.876916 | 5.134155 |
| 283 | 1.225183 | 2.106498 | -2.936262 | 5.161692 |
| 284 | 1.224113 | 2.120018 | -2.996025 | 5.190286 |
| 285 | 1.223239 | 2.133658 | -3.056182 | 5.219946 |
| 286 | 1.222561 | 2.14741 | -3.116714 | 5.250677 |
| 287 | 1.22208 | 2.161268 | -3.1776 | 5.282485 |
| 288 | 1.221798 | 2.175226 | -3.238819 | 5.315377 |
| 289 | 1.221717 | 2.189279 | -3.300352 | 5.349357 |
| 290 | 1.221835 | 2.20342 | -3.362179 | 5.384431 |
| 291 | 1.222154 | 2.217643 | -3.424278 | 5.420603 |
| 292 | 1.222675 | 2.231942 | -3.486633 | 5.457878 |
| 293 | 1.223396 | 2.246312 | -3.549221 | 5.496259 |
| 294 | 1.224319 | 2.260748 | -3.612024 | 5.53575 |
| 295 | 1.225442 | 2.275243 | -3.675023 | 5.576355 |
| 296 | 1.226765 | 2.289793 | -3.738198 | 5.618076 |
| 297 | 1.228289 | 2.304392 | -3.801531 | 5.660917 |
| 298 | 1.230011 | 2.319036 | -3.865002 | 5.704879 |
| 299 | 1.231932 | 2.33372 | -3.928592 | 5.749966 |
| 300 | 1.23405 | 2.348438 | -3.992284 | 5.796179 |
| 301 | 1.236364 | 2.363187 | -4.056058 | 5.84352 |
| 302 | 1.238874 | 2.377962 | -4.119896 | 5.89199 |
| 303 | 1.241578 | 2.392759 | -4.18378 | 5.941592 |
| 304 | 1.244474 | 2.407573 | -4.247691 | 5.992326 |
| 305 | 1.247563 | 2.4224 | -4.311611 | 6.044194 |
| 306 | 1.250841 | 2.437238 | -4.375523 | 6.097196 |
| 307 | 1.254309 | 2.45208 | -4.439406 | 6.151332 |
| 308 | 1.257964 | 2.466925 | -4.503245 | 6.206604 |
| 309 | 1.261805 | 2.481768 | -4.56702 | 6.263012 |
| 310 | 1.26583 | 2.496606 | -4.630714 | 6.320556 |
| 311 | 1.270038 | 2.511435 | -4.69431 | 6.379235 |
| 312 | 1.274427 | 2.526253 | -4.757788 | 6.43905 |
| 313 | 1.278996 | 2.541055 | -4.821132 | 6.5 |
| 314 | 1.283743 | 2.55584 | -4.884323 | 6.562085 |
| 315 | 1.288667 | 2.570604 | -4.947343 | 6.625304 |
| 316 | 1.293765 | 2.585344 | -5.010176 | 6.689655 |
| 317 | 1.299037 | 2.600057 | -5.072803 | 6.755139 |
| 318 | 1.30448 | 2.614741 | -5.135205 | 6.821754 |
| 319 | 1.310093 | 2.629393 | -5.197366 | 6.889497 |
| 320 | 1.315874 | 2.644011 | -5.259268 | 6.958368 |
| 321 | 1.321822 | 2.658591 | -5.320892 | 7.028366 |
| 322 | 1.327935 | 2.673131 | -5.382221 | 7.099487 |
| 323 | 1.334211 | 2.68763 | -5.443237 | 7.17173 |
| 324 | 1.340649 | 2.702084 | -5.503922 | 7.245092 |
| 325 | 1.347247 | 2.716492 | -5.564257 | 7.319571 |
| 326 | 1.354003 | 2.730851 | -5.624225 | 7.395164 |
| 327 | 1.360917 | 2.74516 | -5.683808 | 7.471868 |
| 328 | 1.367985 | 2.759415 | -5.742988 | 7.549679 |
| 329 | 1.375208 | 2.773615 | -5.801746 | 7.628595 |
| 330 | 1.382582 | 2.787759 | -5.860065 | 7.708612 |
| 331 | 1.390108 | 2.801843 | -5.917925 | 7.789725 |
| 332 | 1.397782 | 2.815866 | -5.975308 | 7.871932 |
| 333 | 1.405603 | 2.829826 | -6.032197 | 7.955225 |
| 334 | 1.413571 | 2.843722 | -6.088573 | 8.039603 |
| 335 | 1.421682 | 2.857551 | -6.144417 | 8.12506 |
| 336 | 1.429937 | 2.871312 | -6.199711 | 8.211589 |
| 337 | 1.438333 | 2.885002 | -6.254436 | 8.299187 |
| 338 | 1.446869 | 2.898621 | -6.308573 | 8.387846 |
| 339 | 1.455542 | 2.912166 | -6.362104 | 8.477561 |
| 340 | 1.464353 | 2.925635 | -6.41501 | 8.568326 |
| 341 | 1.473299 | 2.939028 | -6.467273 | 8.660132 |
| 342 | 1.482379 | 2.952341 | -6.518873 | 8.752975 |
| 343 | 1.49159 | 2.965575 | -6.569792 | 8.846847 |
| 344 | 1.500933 | 2.978726 | -6.62001 | 8.941737 |
| *345* | 1.510405 | 2.991794 | -6.66951 | 9.037642 |
| *346* | 1.520004 | 3.004777 | -6.718271 | 9.134548 |
| *347* | 1.52973 | 3.017673 | -6.766276 | 9.23245 |
| *348* | 1.53958 | 3.030481 | -6.813505 | 9.331338 |
| *349* | 1.549554 | 3.043199 | -6.85994 | 9.431202 |
| *350* | 1.559649 | 3.055825 | -6.905561 | 9.532031 |
| *351* | 1.569865 | 3.068359 | -6.95035 | 9.633817 |
| *352* | 1.580199 | 3.080798 | -6.994288 | 9.736547 |
| *353* | 1.59065 | 3.093141 | -7.037356 | 9.840209 |
| *354* | 1.601216 | 3.105387 | -7.079536 | 9.944794 |
| *355* | 1.611897 | 3.117534 | -7.120809 | 10.050288 |
| *356* | 1.62269 | 3.129581 | -7.161156 | 10.156679 |
| *357* | 1.633594 | 3.141526 | -7.200559 | 10.263954 |
| *358* | 1.644606 | 3.153368 | -7.238999 | 10.372099 |
| *359* | 1.655727 | 3.165105 | -7.276459 | 10.481101 |
| *360* | 1.666954 | 3.176736 | -7.31292 | 10.590944 |
| *361* | 1.678284 | 3.18826 | -7.348365 | 10.701614 |
| *362* | 1.689718 | 3.199675 | -7.382775 | 10.813096 |
| *363* | 1.701252 | 3.21098 | -7.416133 | 10.925374 |
| *364* | 1.712886 | 3.222173 | -7.448421 | 11.038431 |
| *365* | 1.724617 | 3.233254 | -7.479623 | 11.152251 |
| *366* | 1.736445 | 3.24422 | -7.509722 | 11.266816 |
| *367* | 1.748366 | 3.25507 | -7.5387 | 11.38211 |
| *368* | 1.76038 | 3.265804 | -7.56654 | 11.498111 |
| *369* | 1.772484 | 3.27642 | -7.593228 | 11.614803 |
| *370* | 1.784677 | 3.286916 | -7.618747 | 11.732166 |
| *371* | 1.796957 | 3.297292 | -7.643082 | 11.850181 |
| *372* | 1.809321 | 3.307546 | -7.666216 | 11.968826 |
| *373* | 1.821769 | 3.317677 | -7.688136 | 12.088082 |
| *374* | 1.834298 | 3.327683 | -7.708826 | 12.207927 |
| *375* | 1.846907 | 3.337564 | -7.728272 | 12.328339 |
| *376* | 1.859592 | 3.347319 | -7.746461 | 12.449296 |
| *377* | 1.872353 | 3.356946 | -7.76338 | 12.570776 |
| *378* | 1.885187 | 3.366444 | -7.779016 | 12.692755 |
| *379* | 1.898093 | 3.375813 | -7.793355 | 12.815209 |
| *380* | 1.911067 | 3.38505 | -7.806386 | 12.938116 |
| *381* | 1.924109 | 3.394156 | -7.818098 | 13.06145 |
| *382* | 1.937216 | 3.403128 | -7.828479 | 13.185186 |
| *383* | 1.950385 | 3.411968 | -7.83752 | 13.3093 |
| *384* | 1.963615 | 3.420672 | -7.84521 | 13.433765 |
| 385 | 1.976904 | 3.42924 | -7.851541 | 13.558556 |
| 386 | 1.990248 | 3.437672 | -7.856503 | 13.683644 |
| 387 | 2.003647 | 3.445967 | -7.860089 | 13.809006 |
| 388 | 2.017098 | 3.454124 | -7.86229 | 13.934611 |
| 389 | 2.030598 | 3.462142 | -7.863101 | 14.060435 |
| 390 | 2.044145 | 3.470021 | -7.862516 | 14.186448 |
| 391 | 2.057736 | 3.477759 | -7.860528 | 14.312622 |
| 392 | 2.071371 | 3.485357 | -7.857134 | 14.43893 |
| 393 | 2.085045 | 3.492813 | -7.852328 | 14.565341 |
| 394 | 2.098757 | 3.500127 | -7.846109 | 14.691829 |
| 395 | 2.112504 | 3.507299 | -7.838473 | 14.818362 |
| 396 | 2.126283 | 3.514328 | -7.829419 | 14.944913 |
| 397 | 2.140093 | 3.521214 | -7.818945 | 15.071451 |
| 398 | 2.153931 | 3.527956 | -7.807052 | 15.197948 |
| 399 | 2.167794 | 3.534554 | -7.79374 | 15.324372 |
| 400 | 2.18168 | 3.541008 | -7.77901 | 15.450695 |
| 401 | 2.195587 | 3.547318 | -7.762865 | 15.576887 |
| 402 | 2.209511 | 3.553483 | -7.745307 | 15.702918 |
| 403 | 2.22345 | 3.559504 | -7.726341 | 15.828758 |
| 404 | 2.237402 | 3.56538 | -7.70597 | 15.954376 |
| 405 | 2.251364 | 3.571112 | -7.684202 | 16.079744 |
| 406 | 2.265333 | 3.576699 | -7.661041 | 16.20483 |
| 407 | 2.279308 | 3.582142 | -7.636495 | 16.329607 |
| 408 | 2.293285 | 3.58744 | -7.610573 | 16.454044 |
| 409 | 2.307261 | 3.592595 | -7.583282 | 16.57811 |
| 410 | 2.321235 | 3.597606 | -7.554633 | 16.701778 |
| 411 | 2.335204 | 3.602473 | -7.524636 | 16.82502 |
| 412 | 2.349164 | 3.607198 | -7.493303 | 16.947803 |
| 413 | 2.363115 | 3.61178 | -7.460645 | 17.070103 |
| 414 | 2.377052 | 3.616221 | -7.426676 | 17.191889 |
| 415 | 2.390973 | 3.62052 | -7.39141 | 17.313133 |
| 416 | 2.404877 | 3.624678 | -7.354862 | 17.433811 |
| 417 | 2.41876 | 3.628697 | -7.317046 | 17.553892 |
| 418 | 2.432619 | 3.632577 | -7.277978 | 17.673351 |
| 419 | 2.446453 | 3.636318 | -7.237677 | 17.792162 |
| 420 | 2.460258 | 3.639922 | -7.19616 | 17.910299 |
| 421 | 2.474034 | 3.643389 | -7.153445 | 18.027737 |
| 422 | 2.487776 | 3.646722 | -7.109551 | 18.144449 |
| 423 | 2.501482 | 3.649919 | -7.064499 | 18.260414 |
| 424 | 2.515151 | 3.652984 | -7.018309 | 18.375608 |
| 425 | 2.528779 | 3.655916 | -6.971003 | 18.490009 |
| 426 | 2.542365 | 3.658718 | -6.922601 | 18.603592 |
| 427 | 2.555905 | 3.661391 | -6.873128 | 18.716335 |
| 428 | 2.569399 | 3.663935 | -6.822606 | 18.82822 |
| 429 | 2.582843 | 3.666352 | -6.77106 | 18.939228 |
| 430 | 2.596236 | 3.668645 | -6.718512 | 19.049334 |
| 431 | 2.609575 | 3.670813 | -6.664989 | 19.158524 |
| 432 | 2.622858 | 3.67286 | -6.610515 | 19.266777 |
| 433 | 2.636083 | 3.674786 | -6.555117 | 19.374079 |
| 434 | 2.649248 | 3.676593 | -6.498821 | 19.48041 |
| 435 | 2.662351 | 3.678283 | -6.441653 | 19.585756 |
| 436 | 2.67539 | 3.679857 | -6.383641 | 19.690104 |
| 437 | 2.688363 | 3.681319 | -6.324812 | 19.793438 |
| 438 | 2.701268 | 3.682668 | -6.265195 | 19.895746 |
| 439 | 2.714104 | 3.683908 | -6.204816 | 19.997015 |
| 440 | 2.726868 | 3.68504 | -6.143705 | 20.097233 |
| 441 | 2.73956 | 3.686065 | -6.081891 | 20.196392 |
| 442 | 2.752177 | 3.686987 | -6.019401 | 20.294481 |
| 443 | 2.764717 | 3.687808 | -5.956265 | 20.391491 |
| 444 | 2.77718 | 3.688529 | -5.892513 | 20.487415 |
| 445 | 2.789564 | 3.689152 | -5.828174 | 20.582247 |
| 446 | 2.801867 | 3.689679 | -5.763276 | 20.675978 |
| 447 | 2.814088 | 3.690114 | -5.697849 | 20.768606 |
| 448 | 2.826225 | 3.690457 | -5.631923 | 20.860125 |
| 449 | 2.838278 | 3.690711 | -5.565527 | 20.950533 |
| 450 | 2.850246 | 3.69088 | -5.498691 | 21.039827 |
| 451 | 2.862126 | 3.690963 | -5.431444 | 21.128004 |
| 452 | 2.873919 | 3.690965 | -5.363814 | 21.215067 |
| 453 | 2.885623 | 3.690887 | -5.295831 | 21.301014 |
| 454 | 2.897236 | 3.690732 | -5.227525 | 21.385845 |
| 455 | 2.908759 | 3.690502 | -5.158922 | 21.469563 |
| 456 | 2.920191 | 3.690199 | -5.090053 | 21.552172 |
| 457 | 2.93153 | 3.689826 | -5.020944 | 21.633675 |
| 458 | 2.942777 | 3.689385 | -4.951625 | 21.714073 |
| 459 | 2.95393 | 3.688879 | -4.882122 | 21.793377 |
| 460 | 2.964989 | 3.688309 | -4.812463 | 21.87159 |
| 461 | 2.975953 | 3.687679 | -4.742675 | 21.948717 |
| 462 | 2.986823 | 3.68699 | -4.672784 | 22.024769 |
| 463 | 2.997597 | 3.686245 | -4.602816 | 22.099752 |
| 464 | 3.008275 | 3.685447 | -4.532798 | 22.173676 |
| 465 | 3.018858 | 3.684597 | -4.462753 | 22.24655 |
| 466 | 3.029345 | 3.683699 | -4.392708 | 22.318384 |
| 467 | 3.039735 | 3.682754 | -4.322687 | 22.389191 |
| 468 | 3.050029 | 3.681765 | -4.252712 | 22.458979 |
| 469 | 3.060227 | 3.680734 | -4.182808 | 22.527763 |
| 470 | 3.070329 | 3.679663 | -4.112998 | 22.595556 |
| 471 | 3.080335 | 3.678555 | -4.043302 | 22.662369 |
| 472 | 3.090246 | 3.677413 | -3.973743 | 22.728216 |
| 473 | 3.10006 | 3.676237 | -3.904342 | 22.793114 |
| 474 | 3.10978 | 3.675031 | -3.83512 | 22.857077 |
| 475 | 3.119405 | 3.673797 | -3.766096 | 22.920116 |
| 476 | 3.128935 | 3.672536 | -3.697289 | 22.982254 |
| 477 | 3.138371 | 3.671252 | -3.628719 | 23.043503 |
| 478 | 3.147714 | 3.669946 | -3.560402 | 23.103878 |
| 479 | 3.156963 | 3.66862 | -3.492356 | 23.163399 |
| 480 | 3.166121 | 3.667277 | -3.424599 | 23.222082 |
| 481 | 3.175186 | 3.665918 | -3.357145 | 23.279945 |
| 482 | 3.184161 | 3.664545 | -3.29001 | 23.337006 |
| 483 | 3.193046 | 3.663161 | -3.22321 | 23.393282 |
| 484 | 3.201841 | 3.661767 | -3.156758 | 23.448793 |
| 485 | 3.210547 | 3.660366 | -3.090667 | 23.503555 |
| 486 | 3.219166 | 3.658959 | -3.02495 | 23.55759 |
| 487 | 3.227698 | 3.657547 | -2.959619 | 23.610916 |
| 488 | 3.236144 | 3.656134 | -2.894686 | 23.663551 |
| 489 | 3.244505 | 3.65472 | -2.830162 | 23.715515 |
| 490 | 3.252783 | 3.653307 | -2.766056 | 23.766827 |
| 491 | 3.260977 | 3.651897 | -2.702378 | 23.817507 |
| 492 | 3.26909 | 3.650491 | -2.639136 | 23.867573 |
| 493 | 3.277123 | 3.649092 | -2.576339 | 23.917046 |
| 494 | 3.285076 | 3.647701 | -2.513995 | 23.965946 |
| 495 | 3.292951 | 3.646318 | -2.452111 | 24.014292 |
| 496 | 3.300749 | 3.644946 | -2.390691 | 24.062103 |
| 497 | 3.308471 | 3.643586 | -2.329743 | 24.109398 |
| 498 | 3.316118 | 3.64224 | -2.269271 | 24.156199 |
| 499 | 3.323693 | 3.640908 | -2.20928 | 24.202522 |
| 500 | 3.331195 | 3.639593 | -2.149774 | 24.248388 |
| 501 | 3.338627 | 3.638294 | -2.090756 | 24.293818 |
| 502 | 3.34599 | 3.637015 | -2.032228 | 24.338827 |
| 503 | 3.353284 | 3.635754 | -1.974193 | 24.383438 |
| 504 | 3.360513 | 3.634515 | -1.916653 | 24.427666 |
| 505 | 3.367676 | 3.633297 | -1.859608 | 24.471531 |
| 506 | 3.374775 | 3.632102 | -1.803059 | 24.515053 |
| 507 | 3.381812 | 3.630931 | -1.747007 | 24.558249 |
| 508 | 3.388788 | 3.629784 | -1.69145 | 24.601135 |
| 509 | 3.395704 | 3.628663 | -1.636388 | 24.643732 |
| 510 | 3.402563 | 3.627568 | -1.581819 | 24.686056 |
| 511 | 3.409364 | 3.626501 | -1.527742 | 24.728123 |
| 512 | 3.416111 | 3.625461 | -1.474155 | 24.769953 |
| 513 | 3.422803 | 3.62445 | -1.421053 | 24.811558 |
| 514 | 3.429444 | 3.623468 | -1.368436 | 24.852959 |
| 515 | 3.436033 | 3.622516 | -1.316297 | 24.894171 |
| 516 | 3.442573 | 3.621594 | -1.264635 | 24.935207 |
| 517 | 3.449065 | 3.620704 | -1.213445 | 24.976088 |
| 518 | 3.455511 | 3.619844 | -1.162721 | 25.016823 |
| 519 | 3.461911 | 3.619017 | -1.112459 | 25.05743 |
| 520 | 3.468267 | 3.618222 | -1.062653 | 25.097923 |
| 521 | 3.474582 | 3.61746 | -1.013298 | 25.138319 |
| 522 | 3.480855 | 3.61673 | -0.964388 | 25.178627 |
| 523 | 3.487089 | 3.616034 | -0.915916 | 25.218864 |
| 524 | 3.493284 | 3.615371 | -0.867876 | 25.259041 |
| 525 | 3.499443 | 3.614743 | -0.820261 | 25.299173 |
| 526 | 3.505567 | 3.614147 | -0.773063 | 25.339272 |
| 527 | 3.511656 | 3.613586 | -0.726276 | 25.379347 |
| 528 | 3.517713 | 3.61306 | -0.679892 | 25.419415 |
| 529 | 3.523739 | 3.612567 | -0.633904 | 25.459482 |
| 530 | 3.529734 | 3.612108 | -0.588302 | 25.499563 |
| 531 | 3.535701 | 3.611684 | -0.543079 | 25.539667 |
| 532 | 3.54164 | 3.611294 | -0.498227 | 25.579805 |
| 533 | 3.547553 | 3.610938 | -0.453737 | 25.619986 |
| 534 | 3.553441 | 3.610615 | -0.409601 | 25.660219 |
| 535 | 3.559305 | 3.610327 | -0.36581 | 25.700516 |
| 536 | 3.565147 | 3.610073 | -0.322354 | 25.740883 |
| 537 | 3.570967 | 3.609852 | -0.279225 | 25.78133 |
| 538 | 3.576767 | 3.609665 | -0.236415 | 25.821863 |
| 539 | 3.582549 | 3.60951 | -0.193912 | 25.862494 |
| 540 | 3.588312 | 3.609389 | -0.15171 | 25.903225 |
| 541 | 3.594058 | 3.6093 | -0.109797 | 25.944067 |
| 542 | 3.599788 | 3.609244 | -0.068165 | 25.985023 |
| 543 | 3.605504 | 3.609219 | -0.026804 | 26.026104 |
| 544 | 3.611206 | 3.609226 | 0.014294 | 26.067312 |
| 545 | 3.616895 | 3.609264 | 0.05514 | 26.108656 |
| 546 | 3.622572 | 3.609333 | 0.095743 | 26.150137 |
| 547 | 3.628238 | 3.609432 | 0.136112 | 26.191763 |
| 548 | 3.633895 | 3.609562 | 0.176257 | 26.233538 |
| 549 | 3.639543 | 3.609721 | 0.216187 | 26.275467 |
| 550 | 3.645182 | 3.609909 | 0.255912 | 26.317553 |
| 551 | 3.650815 | 3.610126 | 0.29544 | 26.359798 |
| 552 | 3.65644 | 3.61037 | 0.334781 | 26.402208 |
| 553 | 3.66206 | 3.610643 | 0.373944 | 26.444784 |
| 554 | 3.667675 | 3.610942 | 0.412939 | 26.487532 |
| 555 | 3.673287 | 3.611269 | 0.451773 | 26.530449 |
| 556 | 3.678894 | 3.611621 | 0.490457 | 26.573544 |
| 557 | 3.684499 | 3.611999 | 0.528999 | 26.616812 |
| 558 | 3.690102 | 3.612401 | 0.567407 | 26.660259 |
| 559 | 3.695703 | 3.612829 | 0.605691 | 26.703884 |
| 560 | 3.701304 | 3.61328 | 0.643859 | 26.74769 |
| 561 | 3.706904 | 3.613754 | 0.681919 | 26.791677 |
| 562 | 3.712504 | 3.614251 | 0.71988 | 26.835846 |
| 563 | 3.718106 | 3.61477 | 0.757749 | 26.880196 |
| 564 | 3.723709 | 3.615311 | 0.795534 | 26.924728 |
| 565 | 3.729313 | 3.615872 | 0.833244 | 26.96944 |
| 566 | 3.73492 | 3.616454 | 0.870886 | 27.014336 |
| 567 | 3.74053 | 3.617056 | 0.908468 | 27.059412 |
| 568 | 3.746142 | 3.617677 | 0.945996 | 27.104668 |
| 569 | 3.751759 | 3.618317 | 0.983478 | 27.150103 |
| 570 | 3.757379 | 3.618974 | 1.020922 | 27.195715 |
| 571 | 3.763003 | 3.619649 | 1.058333 | 27.241505 |
| 572 | 3.768632 | 3.620341 | 1.095718 | 27.28747 |
| 573 | 3.774266 | 3.62105 | 1.133085 | 27.333609 |
| 574 | 3.779905 | 3.621773 | 1.170438 | 27.379919 |
| 575 | 3.785549 | 3.622513 | 1.207785 | 27.426399 |
| 576 | 3.791199 | 3.623266 | 1.24513 | 27.473047 |
| 577 | 3.796854 | 3.624034 | 1.282481 | 27.519859 |
| 578 | 3.802516 | 3.624815 | 1.319842 | 27.566837 |
| 579 | 3.808184 | 3.62561 | 1.357218 | 27.613976 |
| 580 | 3.813858 | 3.626416 | 1.394615 | 27.66127 |
| 581 | 3.819538 | 3.627235 | 1.432037 | 27.708723 |
| 582 | 3.825225 | 3.628065 | 1.469489 | 27.756329 |
| 583 | 3.830918 | 3.628906 | 1.506976 | 27.804085 |
| 584 | 3.836618 | 3.629757 | 1.544502 | 27.851988 |
| 585 | 3.842325 | 3.630619 | 1.582071 | 27.900036 |
| 586 | 3.848039 | 3.63149 | 1.619687 | 27.948225 |
| 587 | 3.853759 | 3.63237 | 1.657353 | 27.996555 |
| 588 | 3.859487 | 3.633258 | 1.695073 | 28.045019 |
| 589 | 3.865221 | 3.634155 | 1.732851 | 28.093616 |
| 590 | 3.870961 | 3.635059 | 1.770688 | 28.142344 |
| 591 | 3.876709 | 3.635971 | 1.808589 | 28.191198 |
| 592 | 3.882463 | 3.636889 | 1.846556 | 28.240177 |
| 593 | 3.888224 | 3.637815 | 1.884591 | 28.289276 |
| 594 | 3.893991 | 3.638746 | 1.922696 | 28.338491 |
| 595 | 3.899765 | 3.639683 | 1.960874 | 28.387823 |
| 596 | 3.905546 | 3.640626 | 1.999127 | 28.437265 |
| 597 | 3.911332 | 3.641575 | 2.037456 | 28.486816 |
| 598 | 3.917125 | 3.642528 | 2.075863 | 28.536474 |
| 599 | 3.922924 | 3.643485 | 2.11435 | 28.586235 |
| 600 | 3.928729 | 3.644448 | 2.152916 | 28.636095 |
| 601 | 3.93454 | 3.645414 | 2.191565 | 28.686052 |
| 602 | 3.940357 | 3.646384 | 2.230296 | 28.736105 |
| 603 | 3.946179 | 3.647358 | 2.269109 | 28.786251 |
| 604 | 3.952006 | 3.648335 | 2.308007 | 28.836485 |
| 605 | 3.957839 | 3.649315 | 2.346988 | 28.886808 |
| 606 | 3.963677 | 3.650299 | 2.386055 | 28.937216 |
| 607 | 3.96952 | 3.651286 | 2.425205 | 28.987705 |
| 608 | 3.975368 | 3.652275 | 2.46444 | 29.038275 |
| 609 | 3.981221 | 3.653267 | 2.503759 | 29.088924 |
| 610 | 3.987078 | 3.654262 | 2.543162 | 29.13965 |
| 611 | 3.992939 | 3.655259 | 2.582649 | 29.190453 |
| 612 | 3.998805 | 3.656258 | 2.622219 | 29.241327 |
| 613 | 4.004675 | 3.65726 | 2.661872 | 29.292273 |
| 614 | 4.010549 | 3.658264 | 2.701607 | 29.343288 |
| 615 | 4.016426 | 3.65927 | 2.741422 | 29.394375 |
| 616 | 4.022307 | 3.660278 | 2.781318 | 29.445528 |
| 617 | 4.028192 | 3.661289 | 2.821293 | 29.49675 |
| 618 | 4.03408 | 3.662302 | 2.861346 | 29.548037 |
| 619 | 4.039971 | 3.663317 | 2.901477 | 29.599388 |
| 620 | 4.045866 | 3.664334 | 2.941683 | 29.650806 |
| 621 | 4.051763 | 3.665354 | 2.981964 | 29.702288 |
| 622 | 4.057663 | 3.666376 | 3.022319 | 29.753834 |
| 623 | 4.063566 | 3.6674 | 3.062745 | 29.805445 |
| 624 | 4.069471 | 3.668427 | 3.103243 | 29.857119 |
| *625* | 4.07538 | 3.669456 | 3.143809 | 29.908857 |
| *626* | 4.08129 | 3.670489 | 3.184444 | 29.960661 |
| *627* | 4.087204 | 3.671524 | 3.225145 | 30.012531 |
| *628* | 4.093119 | 3.672562 | 3.265912 | 30.064466 |
| *629* | 4.099037 | 3.673603 | 3.306742 | 30.11647 |
| *630* | 4.104957 | 3.674648 | 3.347634 | 30.168541 |
| *631* | 4.110879 | 3.675696 | 3.388588 | 30.22068 |
| *632* | 4.116804 | 3.676747 | 3.429601 | 30.272892 |
| *633* | 4.12273 | 3.677803 | 3.470672 | 30.325174 |
| *634* | 4.128659 | 3.678862 | 3.5118 | 30.377533 |
| *635* | 4.13459 | 3.679925 | 3.552984 | 30.429964 |
| *636* | 4.140523 | 3.680993 | 3.594222 | 30.482475 |
| *637* | 4.146459 | 3.682065 | 3.635513 | 30.535065 |
| *638* | 4.152397 | 3.683143 | 3.676856 | 30.587738 |
| *639* | 4.158337 | 3.684225 | 3.718251 | 30.640495 |
| *640* | 4.164279 | 3.685312 | 3.759695 | 30.69334 |
| *641* | 4.170225 | 3.686405 | 3.801189 | 30.746273 |
| *642* | 4.176172 | 3.687504 | 3.842731 | 30.799301 |
| *643* | 4.182123 | 3.688608 | 3.884321 | 30.852425 |
| *644* | 4.188076 | 3.689719 | 3.925958 | 30.905647 |
| *645* | 4.194033 | 3.690836 | 3.967641 | 30.958971 |
| *646* | 4.199992 | 3.691959 | 4.009371 | 31.012403 |
| *647* | 4.205956 | 3.69309 | 4.051147 | 31.065943 |
| *648* | 4.211922 | 3.694228 | 4.092968 | 31.119596 |
| *649* | 4.217892 | 3.695373 | 4.134835 | 31.173368 |
| *650* | 4.223866 | 3.696525 | 4.176748 | 31.227259 |
| *651* | 4.229845 | 3.697686 | 4.218708 | 31.281277 |
| *652* | 4.235828 | 3.698854 | 4.260713 | 31.335423 |
| *653* | 4.241816 | 3.700032 | 4.302765 | 31.389702 |
| *654* | 4.247808 | 3.701217 | 4.344866 | 31.44412 |
| *655* | 4.253806 | 3.702411 | 4.387015 | 31.49868 |
| *656* | 4.259809 | 3.703615 | 4.429213 | 31.553387 |
| *657* | 4.265819 | 3.704828 | 4.471462 | 31.608246 |
| *658* | 4.271834 | 3.70605 | 4.513763 | 31.66326 |
| *659* | 4.277856 | 3.707282 | 4.556118 | 31.718435 |
| *660* | 4.283885 | 3.708523 | 4.598528 | 31.773777 |
| *661* | 4.289922 | 3.709775 | 4.640996 | 31.82929 |
| *662* | 4.295966 | 3.711038 | 4.683522 | 31.884979 |
| *663* | 4.302018 | 3.71231 | 4.72611 | 31.940849 |
| *664* | 4.308078 | 3.713594 | 4.768762 | 31.996906 |
| 665 | 4.314148 | 3.714888 | 4.81148 | 32.053154 |
| 666 | 4.320227 | 3.716193 | 4.854268 | 32.109596 |
| 667 | 4.326315 | 3.71751 | 4.897128 | 32.166241 |
| 668 | 4.332415 | 3.718837 | 4.940064 | 32.223091 |
| 669 | 4.338524 | 3.720177 | 4.983078 | 32.280155 |
| 670 | 4.344646 | 3.721528 | 5.026176 | 32.337437 |
| 671 | 4.350778 | 3.72289 | 5.06936 | 32.394939 |
| 672 | 4.356924 | 3.724264 | 5.112635 | 32.452671 |
| 673 | 4.363081 | 3.725651 | 5.156005 | 32.510632 |
| 674 | 4.369253 | 3.727049 | 5.199474 | 32.568836 |
| 675 | 4.375438 | 3.728459 | 5.243047 | 32.627277 |
| 676 | 4.381637 | 3.729881 | 5.286729 | 32.68597 |
| 677 | 4.387852 | 3.731316 | 5.330525 | 32.744919 |
| 678 | 4.394082 | 3.732762 | 5.374441 | 32.804123 |
| 679 | 4.400328 | 3.734221 | 5.418481 | 32.86359 |
| 680 | 4.406591 | 3.735692 | 5.462652 | 32.923328 |
| 681 | 4.412871 | 3.737174 | 5.50696 | 32.983337 |
| 682 | 4.419169 | 3.738669 | 5.551411 | 33.043625 |
| 683 | 4.425486 | 3.740176 | 5.596011 | 33.104195 |
| 684 | 4.431822 | 3.741695 | 5.640766 | 33.165054 |
| 685 | 4.438178 | 3.743226 | 5.685685 | 33.226204 |
| 686 | 4.444554 | 3.744768 | 5.730772 | 33.287651 |
| 687 | 4.450952 | 3.746322 | 5.776038 | 33.349396 |
| 688 | 4.45737 | 3.747888 | 5.821487 | 33.411449 |
| 689 | 4.463812 | 3.749465 | 5.867128 | 33.473812 |
| 690 | 4.470276 | 3.751053 | 5.91297 | 33.536484 |
| 691 | 4.476764 | 3.752652 | 5.959019 | 33.599476 |
| 692 | 4.483276 | 3.754262 | 6.005284 | 33.662788 |
| 693 | 4.489814 | 3.755882 | 6.051775 | 33.726425 |
| 694 | 4.496377 | 3.757513 | 6.098498 | 33.790386 |
| 695 | 4.502966 | 3.759154 | 6.145463 | 33.854683 |
| 696 | 4.509582 | 3.760804 | 6.19268 | 33.919312 |
| 697 | 4.516226 | 3.762464 | 6.240157 | 33.984276 |
| 698 | 4.522898 | 3.764133 | 6.287904 | 34.04958 |
| 699 | 4.529599 | 3.765811 | 6.33593 | 34.115227 |
| 700 | 4.536329 | 3.767498 | 6.384245 | 34.181221 |
| 701 | 4.54309 | 3.769192 | 6.432859 | 34.247559 |
| 702 | 4.549882 | 3.770894 | 6.481783 | 34.314247 |
| 703 | 4.556705 | 3.772604 | 6.531026 | 34.381283 |
| 704 | 4.563561 | 3.77432 | 6.580598 | 34.448673 |
| 705 | 4.570449 | 3.776042 | 6.630512 | 34.516415 |
| 706 | 4.577371 | 3.777771 | 6.680776 | 34.584515 |
| 707 | 4.584327 | 3.779504 | 6.731402 | 34.652966 |
| 708 | 4.591318 | 3.781243 | 6.782401 | 34.721775 |
| 709 | 4.598344 | 3.782986 | 6.833784 | 34.790939 |
| 710 | 4.605406 | 3.784733 | 6.885563 | 34.860462 |
| 711 | 4.612505 | 3.786483 | 6.937748 | 34.93034 |
| 712 | 4.619641 | 3.788236 | 6.990352 | 35.000576 |
| 713 | 4.626815 | 3.78999 | 7.043386 | 35.071167 |
| 714 | 4.634027 | 3.791747 | 7.096861 | 35.142113 |
| 715 | 4.641278 | 3.793504 | 7.15079 | 35.213409 |
| 716 | 4.648569 | 3.795261 | 7.205185 | 35.285061 |
| 717 | 4.655899 | 3.797017 | 7.260057 | 35.357059 |
| 718 | 4.66327 | 3.798773 | 7.315419 | 35.429409 |
| 719 | 4.670683 | 3.800526 | 7.371282 | 35.502106 |
| 720 | 4.678137 | 3.802277 | 7.42766 | 35.575146 |
| 721 | 4.685634 | 3.804024 | 7.484564 | 35.648525 |
| 722 | 4.693172 | 3.805767 | 7.542007 | 35.72224 |
| 723 | 4.700755 | 3.807505 | 7.600001 | 35.796295 |
| 724 | 4.708381 | 3.809237 | 7.658559 | 35.870674 |
| 725 | 4.71605 | 3.810963 | 7.717693 | 35.945385 |
| 726 | 4.723764 | 3.812681 | 7.777417 | 36.020416 |
| 727 | 4.731524 | 3.814391 | 7.837741 | 36.095764 |
| 728 | 4.739328 | 3.816092 | 7.898679 | 36.171421 |
| 729 | 4.747179 | 3.817782 | 7.960244 | 36.247391 |
| 730 | 4.755075 | 3.819462 | 8.022449 | 36.323658 |
| 731 | 4.763018 | 3.82113 | 8.085304 | 36.400219 |
| 732 | 4.771008 | 3.822785 | 8.148824 | 36.477074 |
| 733 | 4.779044 | 3.824427 | 8.21302 | 36.554207 |
| 734 | 4.787128 | 3.826054 | 8.277906 | 36.631618 |
| 735 | 4.795259 | 3.827665 | 8.343494 | 36.709297 |
| 736 | 4.803439 | 3.82926 | 8.409797 | 36.787235 |
| 737 | 4.811667 | 3.830837 | 8.476825 | 36.865425 |
| 738 | 4.819943 | 3.832396 | 8.544591 | 36.943863 |
| 739 | 4.828268 | 3.833935 | 8.613109 | 37.022533 |
| 740 | 4.836641 | 3.835454 | 8.68239 | 37.101433 |
| 741 | 4.845064 | 3.836951 | 8.752445 | 37.180546 |
| 742 | 4.853535 | 3.838426 | 8.823288 | 37.259869 |
| 743 | 4.862056 | 3.839877 | 8.894929 | 37.33939 |
| 744 | 4.870625 | 3.841303 | 8.967381 | 37.419098 |
| 745 | 4.879245 | 3.842704 | 9.040654 | 37.498985 |
| 746 | 4.887913 | 3.844078 | 9.114761 | 37.579037 |
| 747 | 4.896631 | 3.845424 | 9.189713 | 37.659241 |
| 748 | 4.905398 | 3.846741 | 9.265519 | 37.73959 |
| 749 | 4.914215 | 3.848028 | 9.342193 | 37.820068 |
| 750 | 4.923081 | 3.849283 | 9.419743 | 37.900669 |
| 751 | 4.931996 | 3.850507 | 9.498181 | 37.981373 |
| 752 | 4.94096 | 3.851697 | 9.577518 | 38.062172 |
| 753 | 4.949974 | 3.852854 | 9.657763 | 38.143051 |
| 754 | 4.959036 | 3.853974 | 9.738926 | 38.223995 |
| 755 | 4.968147 | 3.855058 | 9.821017 | 38.304993 |
| 756 | 4.977307 | 3.856104 | 9.904045 | 38.386028 |
| 757 | 4.986515 | 3.857111 | 9.988021 | 38.467087 |
| 758 | 4.99577 | 3.858079 | 10.072952 | 38.548157 |
| 759 | 5.005075 | 3.859005 | 10.158848 | 38.629219 |
| 760 | 5.014426 | 3.85989 | 10.245716 | 38.710258 |
| 761 | 5.023824 | 3.860731 | 10.333566 | 38.791264 |
| 762 | 5.033269 | 3.861528 | 10.422404 | 38.872215 |
| 763 | 5.042761 | 3.862279 | 10.512239 | 38.953098 |
| 764 | 5.052299 | 3.862984 | 10.603079 | 39.033897 |
| 765 | 5.061882 | 3.863641 | 10.694929 | 39.114594 |
| 766 | 5.071511 | 3.86425 | 10.787797 | 39.195168 |
| 767 | 5.081184 | 3.864809 | 10.881689 | 39.275608 |
| 768 | 5.090902 | 3.865316 | 10.976612 | 39.355896 |
| 769 | 5.100663 | 3.865772 | 11.07257 | 39.436008 |
| 770 | 5.110468 | 3.866176 | 11.169569 | 39.515934 |
| 771 | 5.120315 | 3.866525 | 11.267613 | 39.595654 |
| 772 | 5.130205 | 3.866819 | 11.36671 | 39.675144 |
| 773 | 5.140135 | 3.867057 | 11.46686 | 39.754391 |
| 774 | 5.150106 | 3.867238 | 11.568069 | 39.833374 |
| 775 | 5.160118 | 3.867361 | 11.670341 | 39.912075 |
| 776 | 5.170169 | 3.867424 | 11.773678 | 39.990475 |
| 777 | 5.180259 | 3.867428 | 11.878083 | 40.06855 |
| 778 | 5.190386 | 3.86737 | 11.983558 | 40.14629 |
| 779 | 5.200551 | 3.86725 | 12.090105 | 40.223667 |
| 780 | 5.210752 | 3.867068 | 12.197725 | 40.300667 |
| 781 | 5.220989 | 3.866821 | 12.30642 | 40.377266 |
| 782 | 5.231261 | 3.86651 | 12.416191 | 40.453445 |
| 783 | 5.241567 | 3.866133 | 12.527037 | 40.529182 |
| 784 | 5.251905 | 3.865689 | 12.638957 | 40.604465 |
| 785 | 5.262276 | 3.865178 | 12.751951 | 40.679264 |
| 786 | 5.272678 | 3.864598 | 12.866018 | 40.753563 |
| 787 | 5.283111 | 3.863949 | 12.981157 | 40.827339 |
| 788 | 5.293572 | 3.86323 | 13.097365 | 40.900574 |
| 789 | 5.304063 | 3.86244 | 13.214641 | 40.973248 |
| 790 | 5.31458 | 3.861578 | 13.332978 | 41.045338 |
| 791 | 5.325125 | 3.860645 | 13.452377 | 41.116825 |
| 792 | 5.335694 | 3.859637 | 13.572832 | 41.187691 |
| 793 | 5.346288 | 3.858557 | 13.694339 | 41.257908 |
| 794 | 5.356905 | 3.857401 | 13.816893 | 41.327461 |
| 795 | 5.367545 | 3.85617 | 13.940488 | 41.396332 |
| 796 | 5.378205 | 3.854863 | 14.065119 | 41.464493 |
| 797 | 5.388885 | 3.85348 | 14.190779 | 41.531925 |
| 798 | 5.399585 | 3.85202 | 14.317462 | 41.598614 |
| 799 | 5.410302 | 3.850481 | 14.445161 | 41.664536 |
| 800 | 5.421036 | 3.848865 | 14.573867 | 41.729668 |
| 801 | 5.431785 | 3.847169 | 14.703573 | 41.793991 |
| 802 | 5.442548 | 3.845394 | 14.83427 | 41.857487 |
| 803 | 5.453324 | 3.84354 | 14.965948 | 41.920135 |
| 804 | 5.464112 | 3.841604 | 15.098599 | 41.981918 |
| 805 | 5.474911 | 3.839588 | 15.232212 | 42.042812 |
| 806 | 5.485719 | 3.837491 | 15.366776 | 42.102798 |
| 807 | 5.496535 | 3.835312 | 15.502282 | 42.161861 |
| 808 | 5.507359 | 3.833051 | 15.638717 | 42.219978 |
| 809 | 5.518188 | 3.830708 | 15.776069 | 42.27713 |
| 810 | 5.529021 | 3.828282 | 15.914326 | 42.333302 |
| 811 | 5.539857 | 3.825773 | 16.053474 | 42.38847 |
| 812 | 5.550695 | 3.823181 | 16.193504 | 42.442623 |
| 813 | 5.561534 | 3.820505 | 16.334398 | 42.495739 |
| 814 | 5.572372 | 3.817746 | 16.476143 | 42.547798 |
| 815 | 5.583208 | 3.814903 | 16.618727 | 42.598785 |
| 816 | 5.59404 | 3.811975 | 16.762131 | 42.648685 |
| 817 | 5.604868 | 3.808964 | 16.906343 | 42.697479 |
| 818 | 5.61569 | 3.805868 | 17.051348 | 42.745152 |
| 819 | 5.626505 | 3.802688 | 17.197126 | 42.791683 |
| 820 | 5.637311 | 3.799423 | 17.343664 | 42.837063 |
| 821 | 5.648108 | 3.796074 | 17.490946 | 42.881275 |
| 822 | 5.658893 | 3.792641 | 17.63895 | 42.924297 |
| 823 | 5.669666 | 3.789123 | 17.787663 | 42.966122 |
| 824 | 5.680425 | 3.785521 | 17.937067 | 43.006733 |
| 825 | 5.691169 | 3.781834 | 18.087141 | 43.046116 |
| 826 | 5.701897 | 3.778064 | 18.237867 | 43.084259 |
| 827 | 5.712607 | 3.774209 | 18.389231 | 43.121147 |
| 828 | 5.723299 | 3.77027 | 18.541206 | 43.156765 |
| 829 | 5.733969 | 3.766248 | 18.693781 | 43.191101 |
| 830 | 5.744619 | 3.762142 | 18.84693 | 43.224148 |
| 831 | 5.755246 | 3.757954 | 19.000635 | 43.25589 |
| 832 | 5.765848 | 3.753682 | 19.154877 | 43.28632 |
| 833 | 5.776425 | 3.749327 | 19.309635 | 43.315418 |
| 834 | 5.786976 | 3.744891 | 19.464888 | 43.343185 |
| 835 | 5.797499 | 3.740372 | 19.620613 | 43.369606 |
| 836 | 5.807993 | 3.735771 | 19.776793 | 43.394669 |
| 837 | 5.818457 | 3.73109 | 19.933405 | 43.418369 |
| 838 | 5.828888 | 3.726327 | 20.090425 | 43.440697 |
| 839 | 5.839288 | 3.721485 | 20.247837 | 43.461643 |
| 840 | 5.849654 | 3.716562 | 20.405613 | 43.481201 |
| 841 | 5.859984 | 3.71156 | 20.563736 | 43.499367 |
| 842 | 5.870278 | 3.706479 | 20.722179 | 43.516129 |
| 843 | 5.880535 | 3.70132 | 20.880926 | 43.531483 |
| 844 | 5.890753 | 3.696083 | 21.039949 | 43.545422 |
| 845 | 5.900932 | 3.690768 | 21.199228 | 43.557945 |
| 846 | 5.91107 | 3.685377 | 21.358742 | 43.569042 |
| 847 | 5.921166 | 3.67991 | 21.518465 | 43.578716 |
| 848 | 5.931219 | 3.674368 | 21.678377 | 43.586956 |
| 849 | 5.941227 | 3.668751 | 21.838455 | 43.593765 |
| 850 | 5.951191 | 3.663059 | 21.998674 | 43.599136 |
| 851 | 5.961109 | 3.657295 | 22.159016 | 43.603069 |
| 852 | 5.97098 | 3.651458 | 22.319454 | 43.60556 |
| 853 | 5.980802 | 3.645549 | 22.479965 | 43.606609 |
| 854 | 5.990575 | 3.639569 | 22.640532 | 43.606216 |
| 855 | 6.000298 | 3.633518 | 22.801125 | 43.604382 |
| 856 | 6.00997 | 3.627398 | 22.961725 | 43.601105 |
| 857 | 6.01959 | 3.621209 | 23.122311 | 43.596386 |
| 858 | 6.029157 | 3.614952 | 23.282858 | 43.590225 |
| 859 | 6.03867 | 3.608628 | 23.443342 | 43.58263 |
| 860 | 6.048129 | 3.602238 | 23.603746 | 43.573593 |
| 861 | 6.057532 | 3.595782 | 23.764044 | 43.563126 |
| 862 | 6.066878 | 3.589262 | 23.924215 | 43.551228 |
| 863 | 6.076168 | 3.582678 | 24.084236 | 43.537899 |
| 864 | 6.085399 | 3.576031 | 24.244087 | 43.523151 |
| 865 | 6.094572 | 3.569322 | 24.403746 | 43.506981 |
| 866 | 6.103685 | 3.562553 | 24.563189 | 43.489399 |
| 867 | 6.112738 | 3.555723 | 24.722397 | 43.470406 |
| 868 | 6.12173 | 3.548835 | 24.881348 | 43.450012 |
| 869 | 6.13066 | 3.541888 | 25.040022 | 43.428219 |
| 870 | 6.139528 | 3.534884 | 25.198397 | 43.405037 |
| 871 | 6.148333 | 3.527824 | 25.356455 | 43.38047 |
| 872 | 6.157074 | 3.520709 | 25.514172 | 43.354527 |
| 873 | 6.165751 | 3.513539 | 25.67153 | 43.327213 |
| 874 | 6.174363 | 3.506316 | 25.828508 | 43.298542 |
| 875 | 6.18291 | 3.499042 | 25.985088 | 43.268517 |
| 876 | 6.191391 | 3.491715 | 26.141249 | 43.237152 |
| 877 | 6.199806 | 3.484339 | 26.296972 | 43.204449 |
| 878 | 6.208153 | 3.476913 | 26.45224 | 43.170422 |
| 879 | 6.216434 | 3.46944 | 26.607033 | 43.135082 |
| 880 | 6.224646 | 3.461919 | 26.761333 | 43.098434 |
| 881 | 6.23279 | 3.454352 | 26.915121 | 43.060493 |
| 882 | 6.240865 | 3.446739 | 27.06838 | 43.021271 |
| 883 | 6.248871 | 3.439083 | 27.221094 | 42.980774 |
| 884 | 6.256807 | 3.431384 | 27.373243 | 42.939018 |
| 885 | 6.264674 | 3.423643 | 27.524813 | 42.896011 |
| 886 | 6.27247 | 3.41586 | 27.675785 | 42.851768 |
| 887 | 6.280197 | 3.408038 | 27.826145 | 42.806301 |
| 888 | 6.287852 | 3.400177 | 27.975876 | 42.759621 |
| 889 | 6.295436 | 3.392278 | 28.124962 | 42.711742 |
| 890 | 6.302948 | 3.384342 | 28.273388 | 42.662674 |
| 891 | 6.31039 | 3.376371 | 28.421141 | 42.612434 |
| 892 | 6.31776 | 3.368365 | 28.568203 | 42.561035 |
| 893 | 6.325057 | 3.360324 | 28.714563 | 42.508488 |
| 894 | 6.332282 | 3.352252 | 28.860205 | 42.454807 |
| 895 | 6.339435 | 3.344147 | 29.005117 | 42.400009 |
| 896 | 6.346516 | 3.336012 | 29.149284 | 42.344105 |
| 897 | 6.353524 | 3.327847 | 29.292696 | 42.287109 |
| 898 | 6.360459 | 3.319654 | 29.435339 | 42.229042 |
| 899 | 6.367321 | 3.311432 | 29.5772 | 42.16991 |
| 900 | 6.374111 | 3.303185 | 29.718267 | 42.109734 |
| 901 | 6.380828 | 3.294911 | 29.85853 | 42.048523 |
| 902 | 6.387472 | 3.286613 | 29.997978 | 41.986298 |
| 903 | 6.394043 | 3.278291 | 30.136599 | 41.923073 |
| 904 | 6.400542 | 3.269947 | 30.274384 | 41.85886 |
| *905* | 6.406967 | 3.26158 | 30.41132 | 41.793678 |
| *906* | 6.41332 | 3.253194 | 30.547401 | 41.727539 |
| *907* | 6.4196 | 3.244787 | 30.682615 | 41.660461 |
| *908* | 6.425807 | 3.236361 | 30.816956 | 41.592461 |
| *909* | 6.431941 | 3.227918 | 30.950413 | 41.523552 |
| *910* | 6.438003 | 3.219457 | 31.082977 | 41.453747 |
| *911* | 6.443992 | 3.210981 | 31.214642 | 41.383068 |
| *912* | 6.44991 | 3.202489 | 31.345398 | 41.311527 |
| *913* | 6.455755 | 3.193983 | 31.475241 | 41.239143 |
| *914* | 6.461528 | 3.185464 | 31.60416 | 41.165924 |
| *915* | 6.467229 | 3.176932 | 31.732151 | 41.091896 |
| *916* | 6.472858 | 3.168389 | 31.859209 | 41.017071 |
| *917* | 6.478417 | 3.159835 | 31.985325 | 40.94146 |
| *918* | 6.483904 | 3.151272 | 32.110493 | 40.865086 |
| *919* | 6.48932 | 3.142699 | 32.234711 | 40.78796 |
| *920* | 6.494665 | 3.134119 | 32.357971 | 40.710098 |
| *921* | 6.499939 | 3.125531 | 32.48027 | 40.631519 |
| *922* | 6.505144 | 3.116936 | 32.601604 | 40.552238 |
| *923* | 6.510278 | 3.108336 | 32.721966 | 40.472271 |
| *924* | 6.515343 | 3.099732 | 32.841354 | 40.391628 |
| *925* | 6.520338 | 3.091123 | 32.959766 | 40.310333 |
| *926* | 6.525264 | 3.082511 | 33.077198 | 40.228397 |
| *927* | 6.530121 | 3.073897 | 33.193645 | 40.145836 |
| *928* | 6.53491 | 3.06528 | 33.309105 | 40.062664 |
| *929* | 6.53963 | 3.056664 | 33.423576 | 39.978897 |
| *930* | 6.544283 | 3.048046 | 33.537056 | 39.894554 |
| *931* | 6.548868 | 3.039429 | 33.649544 | 39.809647 |
| *932* | 6.553386 | 3.030814 | 33.76104 | 39.72419 |
| *933* | 6.557837 | 3.022201 | 33.871536 | 39.638199 |
| *934* | 6.562222 | 3.01359 | 33.981041 | 39.551689 |
| *935* | 6.566541 | 3.004982 | 34.089542 | 39.46468 |
| *936* | 6.570794 | 2.996379 | 34.197052 | 39.377174 |
| *937* | 6.574982 | 2.98778 | 34.303558 | 39.2892 |
| *938* | 6.579105 | 2.979187 | 34.409069 | 39.20076 |
| *939* | 6.583163 | 2.970599 | 34.51358 | 39.111877 |
| *940* | 6.587158 | 2.962018 | 34.617096 | 39.022564 |
| *941* | 6.591088 | 2.953445 | 34.719612 | 38.932831 |
| *942* | 6.594956 | 2.944879 | 34.821133 | 38.842697 |
| *943* | 6.598761 | 2.936322 | 34.921658 | 38.752171 |
| *944* | 6.602503 | 2.927773 | 35.021191 | 38.661266 |
| 945 | 6.606184 | 2.919235 | 35.119728 | 38.570004 |
| 946 | 6.609802 | 2.910707 | 35.217278 | 38.47839 |
| 947 | 6.61336 | 2.902189 | 35.313835 | 38.38644 |
| 948 | 6.616858 | 2.893682 | 35.409409 | 38.29417 |
| 949 | 6.620295 | 2.885188 | 35.503994 | 38.201588 |
| 950 | 6.623672 | 2.876706 | 35.597599 | 38.108711 |
| 951 | 6.62699 | 2.868236 | 35.690224 | 38.015549 |
| 952 | 6.63025 | 2.85978 | 35.781872 | 37.922115 |
| 953 | 6.633451 | 2.851338 | 35.872547 | 37.828423 |
| 954 | 6.636594 | 2.84291 | 35.96225 | 37.734482 |
| 955 | 6.63968 | 2.834497 | 36.050983 | 37.640308 |
| 956 | 6.64271 | 2.826099 | 36.138756 | 37.54591 |
| 957 | 6.645682 | 2.817717 | 36.225567 | 37.451302 |
| 958 | 6.648599 | 2.809351 | 36.311417 | 37.356495 |
| 959 | 6.651461 | 2.801001 | 36.396317 | 37.261501 |
| 960 | 6.654267 | 2.792669 | 36.480267 | 37.166328 |
| 961 | 6.657019 | 2.784354 | 36.563271 | 37.070992 |
| 962 | 6.659717 | 2.776057 | 36.645336 | 36.975502 |
| 963 | 6.662361 | 2.767777 | 36.726463 | 36.879868 |
| 964 | 6.664952 | 2.759517 | 36.806656 | 36.7841 |
| 965 | 6.667491 | 2.751275 | 36.885921 | 36.68821 |
| 966 | 6.669978 | 2.743053 | 36.964264 | 36.592209 |
| 967 | 6.672413 | 2.73485 | 37.041691 | 36.496105 |
| 968 | 6.674797 | 2.726668 | 37.118198 | 36.39991 |
| 969 | 6.677131 | 2.718505 | 37.193802 | 36.303631 |
| 970 | 6.679414 | 2.710364 | 37.268501 | 36.207283 |
| 971 | 6.681648 | 2.702243 | 37.3423 | 36.11087 |
| 972 | 6.683833 | 2.694144 | 37.415207 | 36.014408 |
| 973 | 6.685968 | 2.686065 | 37.487225 | 35.9179 |
| 974 | 6.688056 | 2.67801 | 37.558361 | 35.821354 |
| 975 | 6.690096 | 2.669976 | 37.62862 | 35.724785 |
| 976 | 6.69209 | 2.661964 | 37.698009 | 35.628201 |
| 977 | 6.694036 | 2.653975 | 37.766529 | 35.531605 |
| 978 | 6.695936 | 2.646009 | 37.83419 | 35.435013 |
| 979 | 6.69779 | 2.638066 | 37.900993 | 35.338425 |
| 980 | 6.699599 | 2.630147 | 37.966953 | 35.241856 |
| 981 | 6.701363 | 2.622251 | 38.032066 | 35.145309 |
| 982 | 6.703083 | 2.614379 | 38.09634 | 35.048798 |
| 983 | 6.704759 | 2.606531 | 38.159786 | 34.952328 |
| 984 | 6.706391 | 2.598708 | 38.222404 | 34.8559 |
| 985 | 6.707981 | 2.590909 | 38.284203 | 34.759533 |
| 986 | 6.709528 | 2.583134 | 38.345192 | 34.663223 |
| 987 | 6.711034 | 2.575385 | 38.405369 | 34.566986 |
| 988 | 6.712498 | 2.56766 | 38.464745 | 34.470825 |
| 989 | 6.713921 | 2.559961 | 38.523327 | 34.374744 |
| 990 | 6.715302 | 2.552287 | 38.58112 | 34.278759 |
| 991 | 6.716645 | 2.544638 | 38.63813 | 34.182865 |
| 992 | 6.717947 | 2.537016 | 38.694363 | 34.087078 |
| 993 | 6.71921 | 2.529419 | 38.749828 | 33.991398 |
| 994 | 6.720435 | 2.521848 | 38.804523 | 33.895832 |
| 995 | 6.721621 | 2.514303 | 38.858467 | 33.800388 |
| 996 | 6.722769 | 2.506785 | 38.911655 | 33.70507 |
| 997 | 6.72388 | 2.499293 | 38.964096 | 33.60989 |
| 998 | 6.724954 | 2.491827 | 39.0158 | 33.514843 |
| 999 | 6.725991 | 2.484388 | 39.066772 | 33.419941 |
| 1000 | 6.726992 | 2.476975 | 39.117016 | 33.325191 |
| 1006 | 6.732265 | 2.433067 | 39.403576 | 32.760098 |
| 1012 | 6.736343 | 2.390139 | 39.665558 | 32.201591 |
| 1018 | 6.739315 | 2.348203 | 39.904305 | 31.650555 |
| 1024 | 6.741261 | 2.307264 | 40.12114 | 31.107738 |
| 1030 | 6.742262 | 2.267323 | 40.317348 | 30.573769 |
| 1036 | 6.742391 | 2.228375 | 40.494183 | 30.049158 |
| 1042 | 6.741718 | 2.190414 | 40.652851 | 29.534313 |
| 1048 | 6.740309 | 2.153429 | 40.794506 | 29.029558 |
| 1054 | 6.738225 | 2.117407 | 40.920261 | 28.535126 |
| 1060 | 6.735523 | 2.082332 | 41.03117 | 28.051186 |
| 1066 | 6.732258 | 2.048186 | 41.128235 | 27.577839 |
| 1072 | 6.72848 | 2.014953 | 41.21241 | 27.115135 |
| 1078 | 6.724235 | 1.98261 | 41.284588 | 26.663074 |
| 1084 | 6.719566 | 1.951139 | 41.345623 | 26.221615 |
| 1090 | 6.714514 | 1.920517 | 41.396309 | 25.790678 |
| 1096 | 6.709115 | 1.890723 | 41.437393 | 25.370157 |
| 1102 | 6.703405 | 1.861735 | 41.469585 | 24.959921 |
| 1108 | 6.697415 | 1.833529 | 41.493538 | 24.559814 |
| 1114 | 6.691175 | 1.806085 | 41.509869 | 24.169662 |
| 1120 | 6.684711 | 1.77938 | 41.519157 | 23.789284 |
| 1126 | 6.678048 | 1.753392 | 41.521938 | 23.418476 |
| 1132 | 6.67121 | 1.7281 | 41.518715 | 23.057037 |
| 1138 | 6.664218 | 1.703482 | 41.509956 | 22.704748 |
| 1144 | 6.657092 | 1.679517 | 41.496098 | 22.361395 |
| *1150* | 6.649849 | 1.656185 | 41.477543 | 22.026756 |
| *1156* | 6.642506 | 1.633465 | 41.454674 | 21.700605 |
| *1162* | 6.635077 | 1.611339 | 41.427837 | 21.382721 |
| *1168* | 6.627578 | 1.589788 | 41.397362 | 21.072882 |
| *1174* | 6.620019 | 1.568792 | 41.363548 | 20.770864 |
| *1180* | 6.612414 | 1.548334 | 41.326683 | 20.476452 |
| *1186* | 6.604773 | 1.528397 | 41.287025 | 20.189425 |
| *1192* | 6.597105 | 1.508963 | 41.24482 | 19.909573 |
| *1198* | 6.589418 | 1.490017 | 41.200287 | 19.636686 |
| *1204* | 6.581723 | 1.471542 | 41.153641 | 19.370564 |
| *1210* | 6.574025 | 1.453524 | 41.105072 | 19.110998 |
| *1216* | 6.566331 | 1.435946 | 41.054764 | 18.857801 |
| *1222* | 6.558648 | 1.418797 | 41.00288 | 18.610779 |
| *1228* | 6.550981 | 1.402061 | 40.949574 | 18.369745 |
| *1234* | 6.543334 | 1.385725 | 40.894993 | 18.13452 |
| *1240* | 6.535714 | 1.369776 | 40.839264 | 17.904928 |
| *1246* | 6.528122 | 1.354203 | 40.782513 | 17.680799 |
| *1252* | 6.520564 | 1.338992 | 40.724854 | 17.461967 |
| *1258* | 6.513042 | 1.324133 | 40.666393 | 17.248272 |
| *1264* | 6.50556 | 1.309615 | 40.607224 | 17.039557 |
| *1270* | 6.49812 | 1.295426 | 40.547436 | 16.835672 |
| *1276* | 6.490725 | 1.281557 | 40.487118 | 16.636473 |
| *1282* | 6.483376 | 1.267998 | 40.426342 | 16.441816 |
| *1288* | 6.476075 | 1.254739 | 40.365181 | 16.251564 |
| *1294* | 6.468824 | 1.24177 | 40.303699 | 16.065588 |
| *1300* | 6.461626 | 1.229084 | 40.241959 | 15.883758 |
| *1306* | 6.454479 | 1.216671 | 40.180012 | 15.705949 |
| *1312* | 6.447386 | 1.204522 | 40.117916 | 15.532043 |
| *1318* | 6.440348 | 1.192632 | 40.055714 | 15.361926 |
| *1324* | 6.433365 | 1.18099 | 39.99345 | 15.195483 |
| *1330* | 6.426438 | 1.169591 | 39.931168 | 15.032609 |
| *1336* | 6.419568 | 1.158427 | 39.8689 | 14.873198 |
| *1342* | 6.412754 | 1.147491 | 39.806683 | 14.71715 |
| *1348* | 6.405998 | 1.136776 | 39.744545 | 14.564366 |
| *1354* | 6.399298 | 1.126276 | 39.682522 | 14.414755 |
| *1360* | 6.392656 | 1.115986 | 39.620632 | 14.268224 |
| *1366* | 6.386072 | 1.105898 | 39.558903 | 14.124685 |
| *1372* | 6.379544 | 1.096007 | 39.497353 | 13.984055 |
| *1378* | 6.373074 | 1.086308 | 39.436008 | 13.84625 |
| *1384* | 6.366661 | 1.076796 | 39.374882 | 13.711191 |
| 1390 | 6.360304 | 1.067465 | 39.313992 | 13.578802 |
| 1396 | 6.354005 | 1.05831 | 39.253357 | 13.44901 |
| 1402 | 6.347761 | 1.049326 | 39.192982 | 13.321741 |
| 1408 | 6.341573 | 1.040509 | 39.132889 | 13.196928 |
| 1414 | 6.33544 | 1.031854 | 39.073082 | 13.074504 |
| 1420 | 6.329363 | 1.023358 | 39.013573 | 12.954403 |
| 1426 | 6.323339 | 1.015015 | 38.954369 | 12.836563 |
| 1432 | 6.31737 | 1.006821 | 38.895481 | 12.720924 |
| 1438 | 6.311455 | 0.998774 | 38.836914 | 12.607428 |
| 1444 | 6.305593 | 0.990868 | 38.778675 | 12.496017 |
| 1450 | 6.299782 | 0.9831 | 38.720768 | 12.386638 |
| 1456 | 6.294024 | 0.975468 | 38.6632 | 12.279236 |
| 1462 | 6.288317 | 0.967967 | 38.605968 | 12.173761 |
| 1468 | 6.28266 | 0.960593 | 38.549084 | 12.070164 |
| 1474 | 6.277054 | 0.953345 | 38.492546 | 11.968395 |
| 1480 | 6.271498 | 0.946218 | 38.436356 | 11.868409 |
| 1486 | 6.26599 | 0.93921 | 38.380516 | 11.770162 |
| 1492 | 6.260531 | 0.932318 | 38.325031 | 11.673608 |
| 1498 | 6.255119 | 0.925538 | 38.269894 | 11.578706 |
| 1504 | 6.249754 | 0.918869 | 38.215111 | 11.485415 |
| 1510 | 6.244436 | 0.912308 | 38.160679 | 11.393695 |
| 1516 | 6.239164 | 0.905851 | 38.106598 | 11.303509 |
| 1522 | 6.233937 | 0.899497 | 38.052872 | 11.214817 |
| 1528 | 6.228754 | 0.893243 | 37.999493 | 11.127586 |
| 1534 | 6.223615 | 0.887087 | 37.946461 | 11.04178 |
| 1540 | 6.21852 | 0.881027 | 37.89378 | 10.957363 |
| 1546 | 6.213467 | 0.875059 | 37.841442 | 10.874304 |
| 1552 | 6.208457 | 0.869183 | 37.789452 | 10.792571 |
| 1558 | 6.203487 | 0.863396 | 37.737801 | 10.712133 |
| 1564 | 6.198559 | 0.857696 | 37.686489 | 10.63296 |
| 1570 | 6.193671 | 0.852081 | 37.635517 | 10.555022 |
| 1576 | 6.188822 | 0.84655 | 37.584877 | 10.478292 |
| 1582 | 6.184013 | 0.8411 | 37.534569 | 10.402741 |
| 1588 | 6.179242 | 0.835729 | 37.484589 | 10.328344 |
| 1594 | 6.174509 | 0.830436 | 37.434937 | 10.255074 |
| 1600 | 6.169813 | 0.82522 | 37.385609 | 10.182905 |
| 1606 | 6.165154 | 0.820078 | 37.336601 | 10.111814 |
| 1612 | 6.160532 | 0.815009 | 37.28791 | 10.041778 |
| 1618 | 6.155944 | 0.810012 | 37.239532 | 9.972772 |
| 1624 | 6.151392 | 0.805084 | 37.191463 | 9.904773 |
| *1630* | 6.146874 | 0.800225 | 37.143703 | 9.837761 |
| *1636* | 6.142391 | 0.795433 | 37.096249 | 9.771714 |
| *1642* | 6.13794 | 0.790706 | 37.049095 | 9.706613 |
| *1648* | 6.133523 | 0.786044 | 37.002239 | 9.642435 |
| *1654* | 6.129138 | 0.781445 | 36.955673 | 9.579163 |
| *1660* | 6.124784 | 0.776907 | 36.909401 | 9.516776 |
| *1666* | 6.120462 | 0.77243 | 36.863415 | 9.455257 |
| *1672* | 6.116171 | 0.768012 | 36.817711 | 9.394588 |
| *1678* | 6.111911 | 0.763652 | 36.772289 | 9.33475 |
| *1684* | 6.10768 | 0.75935 | 36.727142 | 9.275728 |
| *1690* | 6.103478 | 0.755103 | 36.68227 | 9.217505 |
| *1696* | 6.099306 | 0.75091 | 36.637665 | 9.160064 |
| *1702* | 6.095161 | 0.746772 | 36.593327 | 9.103389 |
| *1708* | 6.091045 | 0.742686 | 36.549252 | 9.047465 |
| *1714* | 6.086957 | 0.738651 | 36.505436 | 8.992278 |
| *1720* | 6.082895 | 0.734668 | 36.461876 | 8.937814 |
| *1726* | 6.07886 | 0.730734 | 36.418568 | 8.884056 |
| *1732* | 6.074851 | 0.726848 | 36.375507 | 8.830993 |
| *1738* | 6.070868 | 0.723011 | 36.332695 | 8.77861 |
| *1744* | 6.06691 | 0.719221 | 36.290123 | 8.726894 |
| *1750* | 6.062977 | 0.715476 | 36.247791 | 8.675834 |
| *1756* | 6.059069 | 0.711777 | 36.205692 | 8.625415 |
| *1762* | 6.055185 | 0.708123 | 36.163826 | 8.575626 |
| *1768* | 6.051324 | 0.704512 | 36.122189 | 8.526456 |
| *1774* | 6.047487 | 0.700943 | 36.08078 | 8.477892 |
| *1780* | 6.043673 | 0.697417 | 36.039589 | 8.429924 |
| *1786* | 6.039881 | 0.693933 | 35.998619 | 8.38254 |
| *1792* | 6.036111 | 0.690488 | 35.957867 | 8.33573 |
| *1798* | 6.032363 | 0.687084 | 35.917324 | 8.289482 |
| *1804* | 6.028637 | 0.683719 | 35.876991 | 8.243787 |
| *1810* | 6.024932 | 0.680392 | 35.836868 | 8.198635 |
| *1816* | 6.021247 | 0.677104 | 35.796947 | 8.154016 |
| *1822* | 6.017583 | 0.673852 | 35.757225 | 8.109922 |
| *1828* | 6.013938 | 0.670637 | 35.717701 | 8.066341 |
| *1834* | 6.010314 | 0.667458 | 35.678375 | 8.023266 |
| *1840* | 6.006709 | 0.664314 | 35.639236 | 7.980687 |
| *1846* | 6.003123 | 0.661206 | 35.600288 | 7.938596 |
| *1852* | 5.999555 | 0.658131 | 35.561527 | 7.896984 |
| *1858* | 5.996006 | 0.65509 | 35.522945 | 7.855843 |
| *1864* | 5.992475 | 0.652081 | 35.484547 | 7.815164 |
| *1870* | 5.988962 | 0.649106 | 35.446327 | 7.774941 |
| *1876* | 5.985466 | 0.646162 | 35.408279 | 7.735165 |
| *1882* | 5.981987 | 0.64325 | 35.370407 | 7.695828 |
| *1888* | 5.978526 | 0.640369 | 35.332703 | 7.656924 |
| *1894* | 5.975081 | 0.637518 | 35.295162 | 7.618445 |
| *1900* | 5.971652 | 0.634697 | 35.25779 | 7.580384 |
| *1906* | 5.968239 | 0.631906 | 35.220577 | 7.542735 |
| *1912* | 5.964842 | 0.629144 | 35.183525 | 7.505489 |
| *1918* | 5.961461 | 0.62641 | 35.146626 | 7.468641 |
| *1924* | 5.958095 | 0.623705 | 35.109886 | 7.432184 |
| *1930* | 5.954743 | 0.621027 | 35.073296 | 7.396112 |
| *1936* | 5.951406 | 0.618376 | 35.036854 | 7.36042 |
| *1942* | 5.948084 | 0.615753 | 35.000557 | 7.325099 |
| *1948* | 5.944777 | 0.613156 | 34.964409 | 7.290145 |
| *1954* | 5.941483 | 0.610584 | 34.928402 | 7.255553 |
| *1960* | 5.938202 | 0.608039 | 34.892532 | 7.221315 |
| *1966* | 5.934935 | 0.605519 | 34.856804 | 7.187428 |
| *1972* | 5.931682 | 0.603023 | 34.821209 | 7.153884 |
| *1978* | 5.928441 | 0.600552 | 34.785748 | 7.120679 |
| *1984* | 5.925213 | 0.598106 | 34.750416 | 7.087808 |
| *1990* | 5.921998 | 0.595683 | 34.715214 | 7.055265 |
| *1996* | 5.918794 | 0.593283 | 34.680141 | 7.023046 |
| *2002* | 5.915603 | 0.590907 | 34.645191 | 6.991145 |
| *2008* | 5.912424 | 0.588554 | 34.610363 | 6.959558 |
| *2014* | 5.909256 | 0.586223 | 34.575657 | 6.92828 |
| *2020* | 5.906101 | 0.583914 | 34.541069 | 6.897306 |
| *2026* | 5.902956 | 0.581627 | 34.506596 | 6.866632 |
| *2032* | 5.899822 | 0.579361 | 34.47224 | 6.836254 |
| *2038* | 5.896699 | 0.577117 | 34.437996 | 6.806166 |
| *2044* | 5.893587 | 0.574893 | 34.403862 | 6.776365 |
| *2050* | 5.890485 | 0.57269 | 34.369839 | 6.746847 |
| *2056* | 5.887393 | 0.570508 | 34.335922 | 6.717607 |
| *2062* | 5.884312 | 0.568345 | 34.302113 | 6.688642 |
| *2068* | 5.88124 | 0.566203 | 34.268402 | 6.659947 |
| *2074* | 5.878179 | 0.564079 | 34.234798 | 6.631519 |
| *2080* | 5.875126 | 0.561975 | 34.20129 | 6.603354 |
| *2086* | 5.872083 | 0.55989 | 34.167881 | 6.575447 |
| *2092* | 5.869049 | 0.557824 | 34.134571 | 6.547796 |
| *2098* | 5.866024 | 0.555776 | 34.101353 | 6.520396 |
| *2104* | 5.863008 | 0.553747 | 34.06823 | 6.493245 |
| 2110 | 5.860001 | 0.551735 | 34.035198 | 6.466339 |
| 2116 | 5.857002 | 0.549742 | 34.002254 | 6.439674 |
| 2122 | 5.854012 | 0.547765 | 33.969402 | 6.413247 |
| 2128 | 5.851029 | 0.545806 | 33.936634 | 6.387055 |
| 2134 | 5.848054 | 0.543864 | 33.903954 | 6.361095 |
| 2140 | 5.845088 | 0.541939 | 33.871357 | 6.335363 |
| 2146 | 5.842129 | 0.540031 | 33.83884 | 6.309856 |
| 2152 | 5.839178 | 0.538138 | 33.806404 | 6.284572 |
| 2158 | 5.836234 | 0.536263 | 33.774048 | 6.259507 |
| 2164 | 5.833297 | 0.534403 | 33.741772 | 6.234658 |
| 2170 | 5.830368 | 0.532558 | 33.709568 | 6.210023 |
| 2176 | 5.827445 | 0.53073 | 33.677441 | 6.185598 |
| 2182 | 5.824529 | 0.528917 | 33.645386 | 6.161382 |
| 2188 | 5.82162 | 0.527119 | 33.613407 | 6.137371 |
| 2194 | 5.818717 | 0.525336 | 33.581493 | 6.113562 |
| 2200 | 5.815821 | 0.523568 | 33.549652 | 6.089952 |
| 2206 | 5.812932 | 0.521814 | 33.517879 | 6.066541 |
| 2212 | 5.810048 | 0.520075 | 33.486176 | 6.043324 |
| 2218 | 5.80717 | 0.518351 | 33.454533 | 6.020299 |
| 2224 | 5.804298 | 0.51664 | 33.422958 | 5.997464 |
| 2230 | 5.801432 | 0.514943 | 33.391441 | 5.974816 |
| 2236 | 5.798571 | 0.51326 | 33.359993 | 5.952353 |
| 2242 | 5.795716 | 0.511591 | 33.328602 | 5.930073 |
| 2248 | 5.792866 | 0.509935 | 33.297268 | 5.907974 |
| 2254 | 5.790022 | 0.508293 | 33.265995 | 5.886052 |
| 2260 | 5.787183 | 0.506663 | 33.234776 | 5.864306 |
| 2266 | 5.784348 | 0.505047 | 33.203617 | 5.842734 |
| 2272 | 5.781519 | 0.503443 | 33.172508 | 5.821333 |
| 2278 | 5.778695 | 0.501852 | 33.141457 | 5.800102 |
| 2284 | 5.775875 | 0.500274 | 33.110455 | 5.779038 |
| 2290 | 5.773059 | 0.498708 | 33.079506 | 5.75814 |
| 2296 | 5.770249 | 0.497154 | 33.048607 | 5.737405 |
| 2302 | 5.767442 | 0.495612 | 33.017757 | 5.716831 |
| 2308 | 5.76464 | 0.494083 | 32.986958 | 5.696417 |
| 2314 | 5.761842 | 0.492565 | 32.956203 | 5.67616 |
| 2320 | 5.759048 | 0.491058 | 32.925495 | 5.656059 |
| 2326 | 5.756258 | 0.489564 | 32.894833 | 5.636111 |
| 2332 | 5.753471 | 0.488081 | 32.864212 | 5.616316 |
| 2338 | 5.750689 | 0.486609 | 32.833637 | 5.596671 |
| 2344 | 5.74791 | 0.485148 | 32.803101 | 5.577174 |
| *2350* | 5.745135 | 0.483698 | 32.77261 | 5.557823 |
| *2356* | 5.742362 | 0.482259 | 32.742157 | 5.538618 |
| *2362* | 5.739594 | 0.480832 | 32.711742 | 5.519556 |
| *2368* | 5.736829 | 0.479414 | 32.681366 | 5.500635 |
| *2374* | 5.734066 | 0.478008 | 32.651028 | 5.481854 |
| *2380* | 5.731308 | 0.476611 | 32.620728 | 5.463212 |
| *2386* | 5.728551 | 0.475225 | 32.590462 | 5.444707 |
| *2392* | 5.725798 | 0.47385 | 32.56023 | 5.426336 |
| *2398* | 5.723048 | 0.472484 | 32.530033 | 5.4081 |
| *2404* | 5.7203 | 0.471129 | 32.499866 | 5.389996 |
| *2410* | 5.717555 | 0.469783 | 32.469734 | 5.372022 |
| *2416* | 5.714812 | 0.468447 | 32.439632 | 5.354177 |
| *2422* | 5.712072 | 0.467121 | 32.409561 | 5.336461 |
| *2428* | 5.709334 | 0.465805 | 32.37952 | 5.318871 |
| *2434* | 5.706599 | 0.464498 | 32.34951 | 5.301406 |
| *2440* | 5.703866 | 0.4632 | 32.319527 | 5.284064 |
| *2446* | 5.701134 | 0.461912 | 32.28957 | 5.266845 |
| *2452* | 5.698405 | 0.460633 | 32.25964 | 5.249746 |
| *2458* | 5.695678 | 0.459363 | 32.229733 | 5.232768 |
| *2464* | 5.692953 | 0.458102 | 32.199856 | 5.215907 |
| *2470* | 5.690229 | 0.45685 | 32.169998 | 5.199163 |
| *2476* | 5.687508 | 0.455607 | 32.140167 | 5.182536 |
| *2482* | 5.684788 | 0.454373 | 32.110355 | 5.166023 |
| *2488* | 5.682069 | 0.453147 | 32.08057 | 5.149622 |
| *2494* | 5.679352 | 0.45193 | 32.050804 | 5.133335 |
| *2500* | 5.676637 | 0.450721 | 32.021057 | 5.117158 |